\documentclass[manuscript]{aastex}

\shorttitle{}
\shortauthors{K. Lakhchaura et al.}
\usepackage{psfig}
\usepackage{epsfig}
\usepackage{rotating}
\usepackage{graphicx}
\usepackage{rotate}
\usepackage{subfigure}
\usepackage{textcomp}
\usepackage{threeparttable}
\usepackage[abs]{overpic}
\usepackage[dvips]{color}
\usepackage{amsmath}
\usepackage{longtable}
\usepackage{capt-of}

\begin{document}
\title{Dynamics of ten clusters of galaxies with substructures}

\author{Kiran Lakhchaura\altaffilmark{1} and K. P. Singh}
\affil{Department of Astronomy and Astrophysics, Tata Institute of Fundamental Research, \\1 Homi Bhabha Road, Mumbai 400 005, India}

\altaffiltext{1}{e-mail : kiran\_astro@tifr.res.in}

\begin{abstract}
We present a detailed \textit{Chandra} study of a sample of ten clusters of galaxies selected based on the 
presence of substructures in their optical images. 
The X-ray surface brightness maps of most of these clusters show anisotropic morphologies, especially in the central regions. 
A total of 22 well resolved significantly bright X-ray peaks (corresponding with high-density regions) are seen 
in the central parts (within r$_{\rm c}/2$) of the clusters. Multiple 
peaks are seen in central parts of six clusters. 
Eleven peaks are found to have 
optical counterparts (10 coinciding with the BCGs of the 10 clusters and one coinciding with the second brightest galaxy in A539). 
For most of the clusters, the optical substructures detected in the previous studies are found to be outside the field 
of view of \textit{Chandra}. In the spectroscopically produced 2-D temperature maps,  
significantly lower temperatures are seen at the location of three peaks (two in A539 and one in A376). 
The centers of five clusters in our sample also host regions of higher temperature compared to the ambient medium, 
indicating the presence of galaxy scale mergers. 
The X-ray luminosity, gas mass and central cooling time estimates for all the 
clusters are presented. The radial X-ray surface-brightness profiles of all but one of the 
clusters are found to be best-fitted with a double-$\beta$ model, pointing towards the presence of double-phased central gas due 
to cool-cores. The cooling time estimates of all the clusters, however, indicate that none of them hosts a strong cool-core, although the 
possibility of weak cool-cores cannot be ruled out.
\end{abstract}


\section{Introduction}
\label{sec:Intro}
Clusters of galaxies, the largest known gravitationally bound objects, are believed to grow by the 
mergers of smaller groups of galaxies \citep{GeBe:1982,DressShect:1988,Girardi:1997,KriessBe:1997,JonFor:1999,Schuecker:2001,Burgett:2004}. 
The intracluster medium (ICM), i.e. the hot (T$\sim10^{7}-10^{8}$K) and tenuous (n$\sim10^{-3}$ cm$^{-3}$) ionized plasma that fills the space 
between the galaxies in a cluster, emits in the X-rays mainly through bremsstrahlung and line emission 
\citep{Kellogg:1972,Mitchell:1979,Sarazin:1988,McNamaNul2007}.

   Based on their X-ray morphologies, clusters of galaxies can be classified as regular (or evolved clusters) and irregular \citep[or early 
type clusters; see][]{Sarazin:1988}. In general, the X-ray surface-brightness profiles of the rich, evolved and relaxed clusters are smooth in 
the outer parts and sharply peaked at the centres \citep{ForJon:1982,Sarazin:1988}. The central sharp peaks characterize a very 
important phenomenon in the relaxed clusters, viz. the \textquotedblleft{}central cooling flows\textquotedblright{}. Until very recently, 
the central gas in clusters was thought to be a \textquotedblleft{}single phase\textquotedblright{} rapidly cooling dense gas, that is pushed 
towards the cluster centre by the pressure exerted by the outer layers, thereby, constituting the \textquotedblleft{}cooling 
flows\textquotedblright{} \citep[see][for a review]{Sarazin:1988,Fabian:1994}. However, the absence of large amounts of expected cooled 
gas at the centre, the inadequacy of the single-$\beta$ model to fit the X-ray surface brightness 
profiles of the clusters (specially at the centres) and the requirement of an additional cool component in the model while fitting the X-ray 
spectrum of the central region, led to the \textquotedblleft{}two phase\textquotedblright{} model of central gas in the clusters 
\citep{Fukazawa:1994,Makishima:2001,Takahashi:2009,Gu:2012}. 
In this model, the central region of the clusters is assumed to be filled with  
a mixture of a hotter and a cooler X-ray gas phase. 
Unlike the rich and relaxed clusters, the X-ray surface brightness distributions of the relatively poorer, younger and 
unrelaxed clusters are characterized by significant substructures and relatively smaller and multiple central peaks 
\citep[see][]{ForJon:1990,JonFor:1992,JonFor:1999,Schuecker:2001}.

  Mergers in the clusters are characterized by shocks 
that compress the intracluster gas and result in an increase in the temperature (and hence, entropy and pressure) of the ICM. This creates large 
scale anisotropies in the thermodynamic maps of the clusters \citep{Roettiger:1996}. Mergers generate bulk gas flows and turbulence in the ICM,
 and therefore, can result in the disruption of central cooling flows in the unrelaxed clusters 
\citep[see][]{Fabian:1994,MarkeVikh:2007,Owers:2011,Maurogordato:2011}. In addition, mergers can also lead to metal enrichment of the ICM 
through enhanced ram-pressure stripping \citep{Domainko:2005}. Observations of the clusters with the \textit{Chandra X-ray Observatory} (CXO) 
have led to a better understanding of various small-scale phenomena in the clusters, such as central morphologies, shock and cold fronts, 
\textquotedblleft{}cavities\textquotedblright{} due to radio sources and the cooling flow problem.

    In this paper, we present a study of the X-ray properties of the ICM in a sample of ten low-redshift (0.028 $\leq$ z $\leq$ 0.76) clusters 
of galaxies using the publicly available \textit{Chandra} X-ray data. Our sample contains A193, A376, A539, A970, A1377, A1831B, A2124, A2457, 
A2665, and A3822, selected on the basis of the presence of significant substructures in their optical 
images, as revealed in \citet{FlinKry:2006} and \citet{Ramella:2007}. The study aims at finding evidence for X-ray substructures and/or mergers 
in these clusters. X-ray morphology and the thermodynamic maps of the clusters are the two main tools used for this purpose.

 The paper is organized as follows. Details of the \textit{Chandra} observations used in this paper, and the data reduction methods employed 
are presented in \S\ref{sec:sample_sel_and_data_red}. The resultant X-ray maps overlaid on the optical maps, the radial profiles of X-ray 
surface-brightness, the global spectral parameters, the azimuthally averaged profiles of the projected spectral parameters and the 
two-dimensional projected thermodynamic maps are 
presented in \S\ref{sec:xray_analy}. A discussion based on the general properties of the clusters derived on the basis of these results is 
given in \S\ref{sec:gen_discussion}. A comparison of the values of the general L$_{X}$-kT scaling relation for the clusters of galaxies with 
the values obtained for our sample, and its implications are discussed in \S\ref{sec:Lx_kT_rela}. Special features and other properties of 
individual clusters are summarized in \S\ref{sec:spec_discussion}. A lambda cold dark matter cosmology with $H_{0}$ = 70 km s$^{-1}$ 
Mpc$^{-1}$ and $\Omega_{M}$ = 0.3 ($\Omega_{\Lambda}$ = 0.7) has been assumed throughout.

\section{X-ray observations and Data Reduction}

\label{sec:sample_sel_and_data_red}
  Redshifts and celestial coordinates of the ten clusters studied here along with a log of their \textit{Chandra} observations (observation IDs, 
dates of observations and exposure times) are given in Table \ref{tab:observation_table}. The data were analyzed with the CIAO version 4.3 and 
CALDB version 4.4.0. X-ray images for all clusters were made in the 0.3-8.0 keV band using the ciao task \textit{dmcopy} with a pixel size of 
$4.0^{\prime\prime}$. The diffuse X-ray emission maps of the clusters were made as follows: (a) Point sources were detected using the CIAO 
task \textit{wavdetect}, and were removed from both the image and event files. The holes created in the X-ray images due to the removal of 
point sources were filled with the average value of the counts in the neighboring pixels, using the ciao task \textit{dmfilth}. (b) Exposure 
maps were created for all the clusters, which were used to normalize the X-ray images. (c) The normalized images were smoothed using 
Gaussian kernels of appropriate width (see the caption of Figure \ref{fig:cont1}), using the ciao task \textit{aconvolve}. One of the clusters, 
A539, was observed twice with the \textit{CXO}, therefore, combined event files and exposure maps from both the observations were used to 
create the exposure corrected images of the cluster. Radial profiles of X-ray surface brightness of all the clusters were produced from 
the normalized and unsmoothed diffuse X-ray emission maps obtained from step (b). To make these profiles we calculated the average 
surface-brightness 
for a number of circular annuli, and the total number of and size of the annuli were chosen on the basis of total number of 
photon counts in that cluster (see \S\ref{sec:Total-Spec-Analy}). The last annulus was fixed such that the surface brightness calculated for 
that annulus is at least 1.25 $\sigma$ above the mean 
local background.

  Using the point source removed event files, average spectra of all the clusters were extracted in the energy band of 0.5-7.0 keV, and 
used in the analysis described in \S\ref{sec:Total-Spec-Analy}. To make radial profiles of various thermodynamical quantities (described in 
\S\ref{sec:Azimuth_spec_analys}), spectra were extracted in a number of circular annuli in each cluster (except in A970 and A1377). The total 
number of the circular annuli was chosen such that each annulus has sufficient ($\ge$600) counts in it. For each cluster, the brightest peak 
of the X-ray emission was fixed as a common centre for all the annuli, and spectra were extracted in the energy band 
of 0.5-7.0 keV. Spectra were also extracted from a number of box-shaped regions in each cluster in the energy range of 0.5-7.0 keV. All spectra 
were analyzed using XSPEC (\S\ref{sec:Total-Spec-Analy}) and resultant spectral parameters were used to make maps of thermodynamic quantities 
(\S\ref{sec:box_thermodynamic_maps}). The number and sizes of boxes were chosen carefully such that each box had sufficient ($\ge$600) counts 
in it. In the inner brighter parts of the cluster small sized boxes were chosen and in the outer fainter parts larger boxes were chosen. 
Spectral information for A539 was extracted separately from the two independent exposures but the spectral data were analyzed 
simultaneously with a common spectral model.

  It was found that local background subtraction was not sufficient for removing the particle and cosmic background components 
present in the average X-ray spectra of the clusters, and large residuals were seen in the high-energy-end of the spectrum of all clusters. 
Additionally, as most of these sources have low redshifts, their emission filled nearly the entire field of view of the detector and it was 
impossible to find nearby emission-free regions. Therefore, we used the \textit{Chandra} blank-sky background observations to create the 
background spectrum for all the spectral analyses. For this we followed the standard procedure described in the \textit{CXO science threads}. 
No background subtraction was done for the imaging analysis used to make the surface brightness maps. 
Local background subtraction was, however, used while making the radial profiles of X-ray surface-brightness. For this purpose, 
 background was estimated from an annular region just outside the outermost annulus used in making the surface-brightness profiles.

\section{Analysis and Results}
\label{sec:xray_analy}

\subsection{X-ray and Optical Morphology}
\label{sec:X-ray_morphology}
  The exposure-corrected, point source removed and smoothed X-ray images of all the ten clusters, produced in 
\S\ref{sec:sample_sel_and_data_red}, are shown in the left-hand-side of Figure \ref{fig:cont1}. Six of the ten clusters showed significant 
substructures or multiple peaks in their central regions. Therefore, zoomed-in images of their central regions have been 
shown in the left-hand-side of Figure \ref{fig:cont2}. The right-hand-sides of Figures \ref{fig:cont1} and \ref{fig:cont2} show the 
overlays of the X-ray intensity contours on the optical images of the clusters from the SuperCOSMOS survey in the B$_{J}$ band. The Brightest 
Cluster Galaxies (BCGs) of some of the clusters are found to be shifted away from their X-ray peaks, and the shift is maximum for A970 
($\sim33^{\prime\prime}=38$ kpc). The departure from spherically symmetric X-ray morphology is found to be maximum for A1377. A detailed 
discussion on the X-ray and optical morphologies of the individual clusters is given in \S\ref{sec:spec_discussion}. 

\subsection{X-ray Surface Brightness Profiles}
\label{sec:xray_sb_prof}
  The radial profiles of X-ray surface brightness of the ten clusters, produced in \S\ref{sec:sample_sel_and_data_red}, are shown in Figure 
\ref{fig:xsb_prof_db_fit}. The profiles were fitted with both single-$\beta$ model (S(r)$\rm =S_{01} (1+(r/r_{c1})^{2})^{-3\beta+0.5}$) and 
double-$\beta$ model (S(r)$\rm =\sum\limits_{i=1}^2 S_{0i} (1+(r/r_{ci})^{2})^{-3\beta+0.5}$). 
Here, S(r) is the total surface-brightness at a distance r from the centre of the cluster; $S_{0i}$ and r$_{ci}$ are values of peak 
surface-brightness and core radius for the i$^{\rm th}$ component of the model (i$=$1 for single-$\beta$ model and i$=$1,2 for 
double-$\beta$ model), and $\beta$ is the ratio of the specific kinetic-energy of the dark-matter particles to that of the intracluster 
gas \citep{Rosati:2002}.

All the ten clusters were fitted with the single-$\beta$ model using chi-square minimization. The average value of minimum reduced 
chi-square ($(\chi^{2}_{\nu})_{min}$) thus obtained was $\sim$2.8, resulting in a poor fit in all cases. The best-fit parameter value 
for the core radii (r$_{c1}$) were in the range of 70-270 kpc. We, therefore, employed the double-$\beta$ model to all the clusters. 
As a result, the value of $(\chi^{2}_{\nu})_{min}$ reduced to $\sim$1 for all the clusters, except A1377. The improvement in the fit 
is $>$99\% based on F-statistic. The cluster A1377 showed a very poor fit for both the single-$\beta$ and double-$\beta$ models with 
the value of $(\chi^{2}_{\nu})_{min}\sim2.4$ 
in both cases. For this cluster, the results did not improve even when the values of S$_{02}$, r$_{c2}$ and $\beta$ parameters, single or 
combined, were frozen to the values of S$_{01}$, r$_{c1}$ and $\beta$, obtained from the single-$\beta$ model fit, respectively. 
The results of the double-$\beta$ model fitting for the nine clusters and single-$\beta$ model fitting for A1377, are given in 
Table \ref{tab:xsb_db_fit_results}. All the fits are shown as dashed lines in Figure \ref{fig:xsb_prof_db_fit}. 
For seven clusters viz. A376, A539, A970, A2124, A2457 and A2665, A3822, that showed large central residuals in their single-$\beta$ 
fits, the radii of the 
smaller core components (r$_{\rm c1}$) obtained from the double-$\beta$ fits were found to lie in the range of $\sim$18-46 
kpc. For A193 and A1831B, the best-fit values of r$_{\rm c1}$ were found to be 64 kpc and 118 kpc respectively, 
which are significantly larger than those obtained for the other seven clusters.

  Considering the possibility of X-ray emission from the cluster's hot gas being affected by emission from the central 
bright galaxies, we removed the central galaxy regions (the central 20-25 kpc regions) from the surface-brightness profiles of all 
the clusters. The resultant profiles were then fitted with the single-$\beta$ model. Despite the absence of large central excess emission, 
the improvement in $(\chi^{2}_{\nu})_{min}$ obtained from the new single-$\beta$ model fit was not significant 
relative to the initial single-$\beta$ fits, for all the clusters except 
A376 and A539. For these two clusters, the value of $(\chi^{2}_{\nu})_{min}$ for the new single-$\beta$ fits was found to be close to 1. 
This was expected for A376 and A539, as the core radius r$_{c1}$ from their double-$\beta$ model fits was found to be $\sim$ 20 kpc (see 
Table \ref{tab:xsb_db_fit_results}), which is comparable to the zone of exclusion. For the other eight clusters (with r$_{c1}>30$ kpc), the 
average 
value of $(\chi^{2}_{\nu})_{min}$ for the new single-$\beta$ model fits was found similar to that obtained in the initial 
single-$\beta$ fits. A discussion based on all these results and their implications, is given in \S\ref{sec:gen_discussion}.

\subsection{Average X-ray Spectral Analysis}
\label{sec:Total-Spec-Analy}

X-ray spectral fitting package XSPEC was used for all the spectral analyses done in this paper. Average 
spectra for all clusters were extracted as described in \S\ref{sec:sample_sel_and_data_red}. 
For the spectral fits, the single-temperature \textit{apec} 
plasma emission model \citep{Smith:2001} (with the AtomDB code v2.0) along with the \textit{wabs} photoelectric absorption model 
\citep{MorrMcCam:1983}, was used. Relative elemental abundances used in the \textit{apec} model are from \citet{AndGre:1989} and 
those for the \textit{wabs} model are from \citet{AndEbi:1982}. Due to the poor spectral 
statistics multi-temperature fits could not be constrained. 
The neutral hydrogen column densities 
along the line of sight to the clusters (N$_{\rm H}$) were frozen to the values (see Table \ref{tab:total_regions_spectral_results}) obtained 
from the Leiden/Argentine/Bonn (LAB) Galactic HI survey \citep{Kalberla:2005}. The redshifts (Table \ref{tab:observation_table}) were frozen 
to the values obtained from the SIMBAD astronomical database, for all the clusters, except A1831B, for which the redshift obtained by 
\citet{Kopylov:2010} was used. The 
resulting spectra of all the clusters, along with the histograms of best fit model spectra are shown in Figure~\ref{fig:tot_spec}. The best fit 
values of the temperature, abundance and \textit{apec} normalizations are given in Table~\ref{tab:total_regions_spectral_results}. 
The maximum average temperature of $5.3\pm0.3$ keV is seen for A3822, while A1377 is found to have the minimum average temperature of 
$1.7\pm0.2$ keV. The estimated average elemental abundance for the clusters ranges from $0.09\pm0.04$ (for A1377) to 
$0.5\pm0.1$ (for A2665) times the solar value. We have also estimated the X-ray luminosity in the energy range 0.5-7.0 keV 
(L$_{\rm X}$) and the bolometric X-ray luminosity (L$^{bol}_{\rm X}$) in the energy range 0.1-100 keV (using the 
\textquoteleft{}$dummyrsp$\textquoteright{} command in XSPEC) for all the clusters. The L$^{bol}_{\rm X}$ values of all the clusters were scaled 
up to a radius of r$_{200}$ (radius within which mean density of the cluster equals 200 times the critical density at the redshift of the 
cluster), using the double $\beta$-model fits obtained for their X-ray surface brightness in \S\ref{sec:xray_sb_prof}. The R$_{200}$ values 
were obtained from the literature and for A2665, for which R$_{200}$ could not be found in the literature, a typical value of 
1.5 Mpc was assumed. The values of L$_{\rm X}$ and L$^{bol}_{\rm X}$, along with the R$_{200}$ values used and their references, are given in 
Table \ref{tab:cluster_properties}. The results show that A3822, the hottest cluster of the sample, is also the most X-ray luminous cluster 
while A1377, the coolest cluster, is the least X-ray luminous.

\subsection{Radial Profiles of Thermodynamic Quantities : Cooling time and Gas Mass}
\label{sec:Azimuth_spec_analys}
  We have produced the azimuthally averaged profiles of temperature (kT), electron number density (n$_{e}$), entropy (S), and pressure (P) for 
all the clusters of the sample, except for A970 and A1377. These two clusters were dropped in this analysis because their X-ray surface 
brightness maps were found to be the most asymmetrical/irregular in the whole sample. Spectra were extracted as described in 
\S\ref{sec:sample_sel_and_data_red}. 
The details of the spectral analyses performed for this section are the same as given in \S\ref{sec:Total-Spec-Analy}, except that here 
the elemental abundances of all the annuli belonging to a specific cluster have been frozen to the respective average abundance value obtained 
for that cluster in \S\ref{sec:Total-Spec-Analy}. The temperature (kT) of each annulus was obtained as a direct result of the spectral analyses 
and the temperature profiles obtained for all the clusters are shown in Figure \ref{fig:anu_temp_prof}. The values of the normalization 
constants (K) obtained for the \textit{apec} model fitted to data from each annulus, and the relation 
$K=10^{-14}EI/(4\pi[D_{A}(1+z)]^{2}$) from \citet{Henry:2004} were used to derive the electron density (n$_{e}$) for that annulus. 
Here, $EI$ is the emission integral $\int n_{e} n_{p}dV$. The units assumed for K, n$_{\rm e}$ (n$_{\rm p}$) and D$_{\rm A}$ in the above 
relation are cm$^{-5}$, cm$^{-3}$ and cm, respectively. Density is assumed to be constant within each annulus. 
Using $n_{p}= 0.855 n_{e}$ \citep{Henry:2004}, we obtained, $EI=0.855n_{e}^{2}$V, where $V$ is the volume of non-deprojected 
spherical shell (if $r_{o}$ and $r_{i}$ are the radii of the 
inner and outer annuli of the annular shell (in angular units) then $V=4/3 D_{A}^3 \pi \Omega (r_{o}^{2}-r_{i}^{2})$, where 
$\Omega$ is the solid angle subtended by 
the annular shell). The resulting density profiles are shown in Figure 
\ref{fig:anu_dens_prof}. To calculate the entropy ($S$) and the pressure ($P$) for an annulus, we used the relations 
$S=kTn_{e}^{-2/3}$ and $P=n_{e}kT$, respectively \citep{Gitti:2010}. The resulting entropy and pressure profiles are shown in Figures 
\ref{fig:anu_entr_prof} and \ref{fig:anu_pres_prof}, respectively.

 The density, entropy and pressure profiles of all the eight clusters show an average decrease, increase and decrease, respectively, from the 
centre outwards. The temperature profiles of all the clusters, however, seem to be almost isothermal, though a significantly low temperature 
is found at the centre of A2457. A high temperature in the outermost annulus of A2665 and in the fourth 
(counting from the centre outwards) annulus of A3822 is also found. The projection effects 
along the line of sight to a cluster tend to smooth out the spatial variations of the thermodynamic quantities. Therefore, we also carried out 
a deprojection analysis of the annuli spectra for each cluster, using the techniques described in 
\cite{Lakhchaura:2011,Lakhchaura:2013}. The resulting deprojected profiles of thermodynamic quantities (although in 
agreement with the projected profiles) had very large errors and, therefore, are not shown in this paper. However, we have used the central 
gas temperatures and densities derived from the deprojection analysis and the following equations from Sarazin (1988) to calculate the cooling 
times for the eight clusters studied in this section.
\begin{equation}
\rm t_{\rm cool}= 8.5 \times 10^{10} \rm yr \left[ \frac{\rm n}{10^{-3} \rm cm^{-3}}\right]^{-1} \left[\frac{\rm T_{\rm g}}{10^{8}K} \right]^{1/2}
\end{equation}
The estimated cooling times of the eight clusters are given in Table \ref{tab:cluster_properties}. 
Cooling times of all the clusters are found to be much greater than the Hubble time therefore, none of them seems to be a cool-core-cluster. 
The poor statistics and large errors in the estimated cooling times, however, do not preclude the possibility of a weak cooling flow in 
some of the clusters. Deeper \textit{Chandra} observations of the central parts of the clusters are required 
for a better estimation of the cooling times.

  We have estimated the gas mass for each cluster by fitting the projected gas densities using the single $\beta$-model i.e., 
\begin{equation}
\rm n_{\rm e}(\rm r)=\rm n_{\rm e}\left(0\right)\left(1+\frac{\rm r^{2}}{\rm r_{\rm c}^{2}}\right)^{(3/2)\rm \beta},
\end{equation}
 where $\rm n_{\rm e}(0)$ is the central density and $\rm r_{\rm c}$ is the 
core radius. The following formula from \cite{Donnelly:2001} was used to estimate the gas mass ($\rm M_{\rm gas}(r)$) of all the eight 
clusters studied in this section out to radii 1 Mpc and R$_{200}$ : 
\begin{equation}
\rm M_{\rm gas}(\rm r)=4 \pi \rho_{0} \int_{0}^{\rm r} \rm s^{2} \left[1+ \left( \frac{\rm s}{\rm r_{\rm c}} \right)^{2} \right] ^{(3/2)\rm \beta} ds
\end{equation} 
 $\rho_{0}=\rm \mu \rm n_{\rm e}(0) \rm m_{\rm p}$ ; where $\rm m_{\rm p}$ and $\mu$ ($=$0.609 from \cite{Gu:2010}) are the mass of a proton 
 and the average molecular weight for a fully ionized gas. We also tried fitting the double-$\beta$ model to the density profiles but due to 
 very few data points available, the results obtained were not significant. The estimated values of the gas mass of eight clusters are given 
 in Table \ref{tab:cluster_properties}.

\subsection{2D Projected Thermodynamic Maps}
\label{sec:box_thermodynamic_maps}

  For a better resolved estimation of the spatial variations of the thermodynamic quantities we have produced projected temperature (kT), 
abundance, density (n$_{e}$), entropy (S), and pressure (P) maps for all the ten clusters of our sample. Spectra were extracted from box 
shaped regions as described in \S\ref{sec:sample_sel_and_data_red}. Details of the spectral fitting, and the relations used to calculate the 
electron density (n$_{e}$), entropy (S), and pressure (P) are given in \S\ref{sec:Total-Spec-Analy} and \S\ref{sec:Azimuth_spec_analys}, 
respectively. The volume of each box was calculated 
using the relation given in \citet{Henry:2004} and \citet{Ehlert:2011}. The resulting temperature (kT) maps are shown in Figure 
\ref{fig:2D_temp_map}, the electron density (n$_{e}$) maps are shown in Figure \ref{fig:2D_dens_map}, the entropy (S) maps are shown in Figure 
\ref{fig:2D_entr_map}, and the pressure (P) maps are shown in Figure \ref{fig:2D_pres_map}. The temperature maps of most of the clusters show 
anisotropies and high temperature regions in many parts of the cluster 
except A193 which seems to be mostly isothermal except at the centre. The maps of density, entropy and pressure of all the clusters show an 
average decrease, increase and decrease from the central outwards, as seen in the azimuthally averaged profiles. The various features 
observed in the thermodynamic maps are discussed in detail in \S\ref{sec:spec_discussion} for the individual clusters.

\section{Discussion}
\label{sec:gen_discussion}
\subsection{Global properties}
The 2-D X-ray substructure and thermodynamic maps of the ten clusters, resulting from a detailed analysis of their 
\textit{Chandra} archival data, are being presented here for the first time.\footnote{\citet{Cavagnolo:2009} have earlier produced and 
analyzed the X-ray surface-brightness and entropy profiles for four of the ten clusters A193, A376, A2124 and A539.} 
Due to the small field of 
view of \textit{Chandra}, the observations did not cover the outer parts of the clusters and, therefore, for many of the clusters the optical 
subclusters reported by \citet{FlinKry:2006} and/or \citet{Ramella:2007} are found to be outside the field of view of \textit{Chandra}.

  The X-ray surface brightness profiles of all the clusters (except A1377) are found to be better fitted with the double-$\beta$ 
model than the single-$\beta$ model (see \S\ref{sec:xray_sb_prof}), suggesting the presence of a double-phase gas in their central regions, 
which is a characteristic of the cool-core clusters. The negligible improvement in the single-$\beta$ model fits, after removing the central 
galaxy regions of the clusters (see \S\ref{sec:xray_sb_prof}), implies that the X-ray emission from the ICM is not significantly affected by 
the emission from the central galaxies. The results from the surface-brightness profile fits, thus, seem to be in contrast with the central 
cooling time estimates, which suggest an absence of strong central cooling in all the ten clusters (see \S\ref{sec:Azimuth_spec_analys}). 
The central cooling time estimates of the clusters can be affected by mergers between the central galaxies. These 
mergers can lead to an increase in the central temperatures of the clusters which can result in an increase in 
the central cooling time estimates. Central galaxy mergers are evident in many of the clusters of our sample, 
that show multiple X-ray peaks and high temperature regions in their central regions (see \S\ref{sec:spec_discussion}).
Due to the large errors in the cooling time estimates of the clusters, the weak cool-core and non cool-core segregation is extremely difficult. 
Therefore, based on our results we argue for an absence of strong cooling in the clusters in our sample. Deeper \textit{Chandra} observations of the central parts are, therefore, required 
for a better estimation of the temperatures and densities in the central regions of the clusters, and to resolve the 
weak-cool-core/non-cool-core ambiguity.

  The average values of temperatures of all the clusters are found to be consistent with the typically observed values, and range between 
1.7 keV to 5.3 keV. The average elemental abundances of all clusters, except A1377, are also found to be similar to the normally observed 
values, and range between 0.2 to 0.5 times the solar value. 
The radial temperature profiles of all the clusters of the sample (except A193 and 
A1831B) seem to be consistent with the generally observed isothermal profiles 

in the centers of the non-cool core clusters. 
The temperature maps of many of these clusters 
show presence of significantly hotter/colder regions in their thermodynamic maps, specially in their central regions. This combined with the 
presence of multiple small peaks seen in the central bright regions of X-ray maps, indicate the presence of shocked hot gas or stripped off 
cold gas resulting from possible galaxy scale mergers in the centre of the clusters.

\subsection{L$_{\rm X}$-kT relation}
\label{sec:Lx_kT_rela}
\begin{figure}
\centering
\includegraphics[width=5.0in]{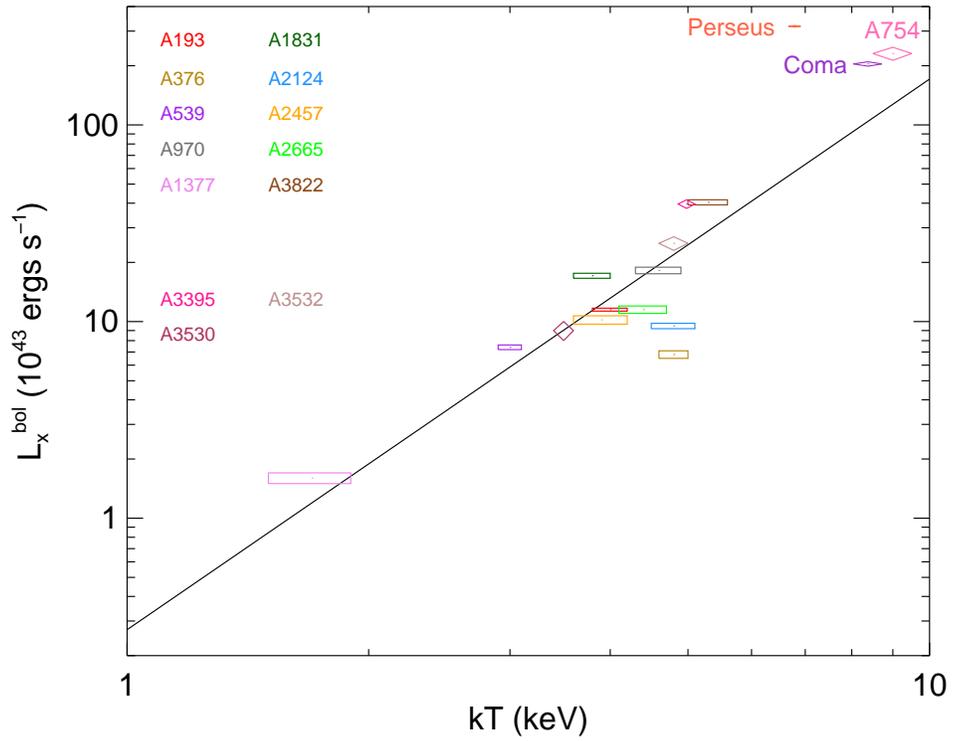}
\caption{L$_{X}$-kT relation of clusters of galaxies obtained by \citet{Takey:2013} (solid line). 
The values obtained for the ten clusters of the sample are shown as rectangles, 
and those of Coma, A754, Perseus, A3395, A3532 and A3530 are 
shown as diamonds.}
\label{fig:Lx_kT_relation}
\end{figure}

    The bolometric X-ray luminosities (L$_{\rm X}$) of clusters of galaxies are generally found to have a power-law dependence on their 
average X-ray temperatures (kT) i.e. L$_{\rm X}=\rm K\;(\rm kT)^{\alpha}$. Here \textquoteleft{}K\textquoteright{} is the constant of 
proportionality, and is equal to the 
expected bolometric X-ray luminosity for $\rm kT=1 keV$. 
Figure \ref{fig:Lx_kT_relation} shows the relationship between the L$_{X}$ and kT 
values obtained by 
\citet{Takey:2013}. The positions of the all the clusters of our sample are shown as rectangles and the well known merging clusters 
Coma and A754, the extremely relaxed cluster Perseus, and three of our previously studied clusters viz. A3395, A3532 and A3530 
\citep{Lakhchaura:2011,Lakhchaura:2013} are shown as diamonds. 
The results obtained by us for the ten clusters seem to be most consistent with 
the relation obtained by 
\citet{Takey:2013}. The only large outliers are the strong cool-core cluster Perseus which 
is shifted to the high luminosity side relative to all the L$_{\rm X}$-kT relations, and A376 and A2124, shifted to the low luminosity side.

\subsection{Individual Clusters}
\label{sec:spec_discussion}
  The X-ray morphologies of all the clusters of our sample show substructures and departures from spherical symmetry, especially in their 
central parts. Our 
analysis has revealed several interesting new features in the central regions of many clusters. Below we discuss the findings 
in each of the ten clusters.\\

\textbf{A193:} X-ray emission is detected (significance$>$3$\sigma$) up to a radius of about 400$^{\prime\prime}$ 
($\sim$0.4 Mpc $\sim$R$_{200}/5$) in this cluster 
(see Figure \ref{fig:xsb_prof_db_fit}). In the X-ray image of the central part of this cluster, three distinct X-ray peaks 
(significant at $>8\sigma$ above the immediate surroundings), 
marked as 1, 2 and 3 in Figure \ref{fig:cont2}(a), are seen. 
The peaks have very similar X-ray brightness and, therefore, none of them stands out as the single brightest X-ray peak of the cluster. 
The apparent spatial extents of the peaks 1, 2 and 3 are $19^{\prime\prime}\times30^{\prime\prime}$, 
$21^{\prime\prime}\times21^{\prime\prime}$ 
and $34^{\prime\prime}\times24^{\prime\prime}$, respectively, significantly larger than the width of the smoothing Gaussian kernel used in 
Figure \ref{fig:cont2}(a) ($=4^{\prime\prime}$).  
The peak separations 1-2, 2-3 and 1-3 are about 33$^{\prime\prime}$, 
24$^{\prime\prime}$ and 69$^{\prime\prime}$ ($\sim$32 kpc, 23 kpc and 66 kpc), respectively. 
In the optical image of the 
cluster, the brightest optical peak is 
found to be located near the X-ray peak 1, which is associated with the BCG of the cluster (see Figure \ref{fig:cont2}(b)). Another 
optical peak seen near the X-ray peak 2, is due to a foreground star located close to the line-of-sight to A193. The BCG in the optical 
image is separated from the X-ray peaks 1 and 2 by about 10$^{\prime\prime}$ and 30$^{\prime\prime}$ ($\sim$10 kpc and 
29 kpc), respectively . \citet{Seigar:2003} detected three infrared 
peaks in the I-band map of this cluster from the \textit{Hubble Space Telescope (HST)}, and in the K-band map from the \textit{United 
Kingdom InfraRed Telescope (UKIRT)}. Contours of infrared emission from the HST I-band map have been overlaid on the optical image of central 
part of the cluster, and are shown as an inset in Figure \ref{fig:cont2}(b). All the infrared peaks are found coinciding with the 
BCG of the cluster, and have been marked as a, b and c in Figure \ref{fig:cont2}(b), in a decreasing order of infrared surface brightness. 
Considering their very small spatial extents these peaks do not seem to be related with the X-ray peaks. The infrared peaks 
seem to be star forming \textquotedblleft{}hotspots\textquotedblright{} within the BCG. 
The irregular X-ray emission and multiple X-ray peaks in the central region of the cluster also point towards galaxy scale 
($\sim$20-60 kpc scale) mergers. 
Mergers are reinforced by the thermodynamic maps of the cluster. 
The temperature map (Figure \ref{fig:2D_temp_map}(a)) shows a high temperature 
in the central region of the cluster (marked as \textquoteleft{}a\textquoteright{}, average kT$=4.8^{+0.4}_{-0.3}$ keV). 
The temperature is significantly high compared to the average temperature of the immediate surrounding region 
(kT$=3.7\pm0.2$ keV). 
The density map (Figure \ref{fig:2D_dens_map}(a)) shows a high density for the three peaks in Figure \ref{fig:cont2}(a) 
(n$_{\rm e}=(9.0\pm0.2)\times10^{-3}$, $(10.2\pm0.5)\times10^{-3}$ and $(9.5\pm0.2)\times10^{-3}$ cm$^{-3}$ for peaks 1, 2 and 3 , 
respectively). 
The entropy map shows a low entropy at the location of the peaks 1 and 2 ($=87^{+13}_{11}$ and 77$^{+27}_{-17}$ keV cm$^{2}$, respectively) 
and a high entropy ($=115^{+28}_{-22}$ keV cm$^{2}$) at the location of peak 3 (see Figure \ref{fig:2D_entr_map}(a)).

\textbf{A376:} We have detected X-ray emission (significance$\geq$3$\sigma$) up to a radius of about 480$^{\prime\prime}$ ($\sim$0.45 Mpc 
$\sim$R$_{200}/5$) in this cluster (see Figure \ref{fig:xsb_prof_db_fit}). The X-ray image of the central region of A376 (Figure 
\ref{fig:cont2}(c)) shows a compression of inner contours towards the NE. Along with the brightest X-ray peak (marked as 
\textquoteleft{}1\textquoteright{} in Figure \ref{fig:cont2}(c)), the cluster shows a secondary X-ray peak (marked as 
\textquoteleft{}2\textquoteright{}), about 1.2$^{\prime}$ ($\sim$65 kpc) towards the SW of the brightest peak. 
Both the peaks are found to be 
significantly brighter than their immediate neighborhood (significance $>21\sigma$ for peak 1 and $>10\sigma$ for peak 2). The apparent spatial 
extent of peak 1 is $\sim130^{\prime\prime}\times103^{\prime\prime}$ and that of peak 2 is $\sim47^{\prime\prime}\times43^{\prime\prime}$, 
therefore both peaks are significantly larger than the width of the smoothing Gaussian kernel used in Figure \ref{fig:cont2}(c) 
($=8^{\prime\prime}$). 
The brightest X-ray peak (peak 1) is found to be coinciding with the brightest galaxy in the optical image of the cluster (Figure 
\ref{fig:cont2}(d)), although no optical counterpart is seen for the secondary X-ray peak. \citet{Proust:2003} studied this cluster using the 
position and velocity informations of 73 member galaxies and found a regular distribution of galaxies within the central 15$^{\prime}$ 
($\sim$0.84 Mpc). They also found a substructure bound to the main cluster, separated by about 20$^{\prime}$ ($\sim$1.1 Mpc) towards the 
southwest (outside the 
\textit{Chandra} FOV). The X-ray image of the cluster from the \textit{R\"{o}entgen Satellite (ROSAT)} High Resolution Imager 
(HRI) given in \citet{Proust:2003}, shows a contraction of contours towards the NE similar to the \textit{Chandra} image (see Figure 
\ref{fig:cont2}(c)), although the secondary X-ray peak detected by us is not seen in the HRI image. 
\citet{Proust:2010} again studied this cluster with 40 new galaxy velocity measurements, and found a bimodal 
distribution of galaxy velocities and a morphological segregation of late-type and early-type galaxies. They suggested a very 
complex dynamics for A376, and the possibility of an ongoing merger with the southwest subcluster. 
The temperature map of the cluster shown in Figure \ref{fig:2D_temp_map}(b) indicates a drop in the temperature at the location of the 
secondary peak (kT$=2.7^{+0.5}_{-0.4}$ keV vs. $3.6^{+1.5}_{-0.4}$ keV at peak 1 in Fig. \ref{fig:cont2}(c); $\Delta$kT$=0.9\pm0.6$ keV). 
Two higher temperature regions are seen to the north (marked as \textquoteleft{}a\textquoteright{}, 
kT$=4.7^{+0.9}_{-0.6}$ keV) and south (marked as \textquoteleft{}b\textquoteright{}, 
kT$=5.4^{+1.3}_{-1.0}$ keV) side of the central region in Fig. \ref{fig:2D_temp_map}(b). The entropy map of the cluster 
(Figure \ref{fig:2D_entr_map}(b)) also shows 
significantly higher values of entropies in the regions \textquoteleft{}a\textquoteright{} and \textquoteleft{}b\textquoteright{}
(S$_{a}=186^{+37}_{-26}$ keV cm$^{2}$ and S$_{b}=203^{+53}_{-41}$ keV cm$^{2}$). The entropy map also shows an entropy of 
$=83^{+19}_{-16}$ keV cm$^{2}$ near the secondary peak (peak 2 in Fig. \ref{fig:cont2}(c)) which is significantly lower 
than the value of entropy at the same distance from the centre (the brightest peak) on the diametrically opposite side 
(marked as \textquoteleft{}c\textquoteright{}, S$_{c}=178^{+45}_{-33}$ keV cm$^{2}$). 
The low values of temperature and entropy near the secondary X-ray peak suggest that the peak may be associated with the cool core of a 
small subcluster or a nearby galaxy that has already merged with the main cluster \citep[see][]{Kempner:2002,Ma:2012}. 
The distortions of the X-ray contours (compression along NE and elongation along SW) and the presence of high temperature and high entropy 
regions in various parts of the cluster, seem to be the possible results of an ongoing merger between A376 and the southwest subcluster 
detected by \citet{Proust:2003}, and in agreement with the results of \citet{Proust:2010}.\\

\textbf{A539:} This is the nearest cluster in our sample with a redshift of 0.0284 (SIMBAD). In this cluster, we have detected X-ray 
emission (significance$\geq$3$\sigma$) up to a radius of about 700$^{\prime\prime}$ ($\sim$0.4 Mpc $\sim$R$_{200}/5$; see Figure 
\ref{fig:xsb_prof_db_fit}). We find a double 
peak in the X-ray image of the central regions of this cluster (Figure \ref{fig:cont2}(e), the brighter peak is marked as 
\textquoteleft{}1\textquoteright{} and the other peak is marked as \textquoteleft{}2\textquoteright{}). 
The peaks are found to be 
brighter than their immediate surroundings at a significance level of about 14$\sigma$. The spatial extents of peak 1 and 2 
are $69^{\prime\prime}\times44^{\prime\prime}$ and $57^{\prime\prime}\times47^{\prime\prime}$, respectively, which are significantly 
larger than the width of the smoothing Gaussian kernel used in Figure \ref{fig:cont2}(e) ($=8^{\prime\prime}$). 
The projected separation between the two peaks is about 45$^{\prime\prime}$ ($\sim$26 kpc). In the optical image the cluster 
(Figure \ref{fig:cont2}(f)), many (at least five) bright galaxies are seen near the centre. In Figure \ref{fig:cont2}(f), 
the X-ray peak 1 is found close to three of these galaxies (including the brightest optical galaxy of the cluster), although 
shifted from all of them by about 10$^{\prime\prime}$ ($\sim$6 kpc). The X-ray peak 2 is close to the 
other two galaxies, and is at a projected separation of about 7$^{\prime\prime}$ ($\sim$4 kpc) from them. 
\citet{Ostriker:1988} detected the presence of two major 
spatially overlapping structures in A539, separated in velocity space by more than 4000 km s$^{-1}$ (see also Girardi et al. 1997). Using 
wavelet analysis, \citet{FlinKry:2006} also detected two groups of galaxies in the central parts of this cluster, at scales of 258 kpc 
($\sim$7.5$^{\prime}$) and 188 kpc ($\sim$5.5$^{\prime}$). The X-ray images produced by us, however, do not show substructuring at these scales 
in this cluster. The double X-ray peaks at the centre of the cluster seem to be formed by the overlapping X-ray halos of the nearby five 
galaxies. All these galaxies have redshifts very close to the mean redshift of the cluster, and therefore are members 
of the cluster. The close proximity of these galaxies with the BCG of the cluster indicates possible interactions between them. However, the 
absence of high temperature and high entropy regions in between or at the location of these peaks rules out an ongoing merger between the 
galaxies (see Figures \ref{fig:2D_temp_map}(c) and \ref{fig:2D_entr_map}(c)). It appears that the galaxies have started coming together to 
merge with the BCG in due course of time. 
In the temperature map, Fig. \ref{fig:2D_temp_map}(c), and entropy map, Fig. \ref{fig:2D_entr_map}(c), the two bright peaks in Fig. 
\ref{fig:cont2}(e) seem to have significantly lower temperature ($2.7^{+0.3}_{-0.2}$ keV at peak 1 and 
$2.8\pm0.1$ keV at peak 2) and entropy ($58^{+7}_{-5}$ keV cm$^{2}$ at peak 1 and $71^{+6}_{-5}$ keV cm$^{2}$ at peak 2) than their 
surrounding medium on the northern, western and southern side (marked as \textquoteleft{}a\textquoteright{}, with average kT$\sim3.1\pm0.1$ keV 
and average S$\sim106\pm3$ keV cm$^{2}$). The outermost regions and the regions on the eastern side have a low temperature similar to that of the two 
peaks, although a small region on the westernmost edge of the map (marked as \textquoteleft{}b\textquoteright{}) seems to have a slightly 
higher temperature of $3.0\pm0.3$ keV.\\

\textbf{A970:} In A970, the X-ray emission extends to a radius of about 360$^{\prime\prime}$ ($\sim$0.41 Mpc $\sim$R$_{200}/4.5$; 
see Figure \ref{fig:xsb_prof_db_fit}). The cluster shows a highly asymmetric X-ray morphology (see Figure \ref{fig:cont1}(g)). The X-ray 
contours seem to be highly compressed along the south-east and elongated along the north-west. 
A similar compression of contours towards the south-east was also observed in the \textit{Einstein} Imaging Proportional Counter (IPC) image 
of the cluster \citep[see][]{Sodre:2001}. The X-ray peak seems to be significantly shifted (by about 2$^{\prime}\sim$140 kpc) 
towards the south-east from the centre of the outer X-ray contours. The cluster therefore, seems to be moving along the 
south-east. The brightest central galaxy, in the optical image of the cluster (Figure \ref{fig:cont1}(h)), seems coinciding with the location 
of the compression of the X-ray contours. 
The temperature map of the cluster in Figure \ref{fig:2D_temp_map}(d) seems to be bimodal with the 
eastern parts colder than the western parts (kT$_{\rm east}=4.1\pm0.4$ keV and kT$_{\rm west}=5.1\pm0.5$ keV$\implies
\Delta$kT$\sim1.0\pm0.6$ keV). 
The results obtained by us, thus, suggest that the 
cluster is out of dynamical equilibrium, in agreement with the findings of \citet{Sodre:2001}. \citeauthor{Sodre:2001} obtained a very high 
value of the velocity dispersion ($\sim$845 km s$^{-1}$) and large scale velocity gradients in the cluster, and, therefore, suggested 
that the cluster is out of equilibrium. \citeauthor{Sodre:2001} also found a substructure towards the north-west of the cluster with a very 
different galaxy dispersion ($\sim$381 km s$^{-1}$) and without any X-ray emission. The subcluster was also detected by \citet{Ramella:2007} 
in the galaxy velocity distribution maps. In the X-ray image of the cluster (Figure \ref{fig:cont1}(g)), we do not see any X-ray emission 
from this subcluster. It seems that the cluster has had a past merger with the optical group towards the northwest and, as a 
result of it, is now out of dynamical equilibrium.\\

\textbf{A1377:} X-ray emission is detected (significance$\geq$3$\sigma$) up to a radius of about 470$^{\prime\prime}$ 
($\sim$0.47 Mpc $\sim$R$_{200}/3$) in this cluster 
(see Figure \ref{fig:xsb_prof_db_fit}). In its X-ray image (Figure \ref{fig:cont1}(i)), 
the cluster shows a highly irregular X-ray emission. A zoomed-in image of the central part of the cluster 
shows several weak peaks (see Figure \ref{fig:cont2}(g)). Three of the brightest peaks are 
marked as 1, 2 and 3 in decreasing order of brightness. 
Peak 1 is brighter than its immediate surrounding at a significance level of greater than 
$\sim5\sigma$, whereas peaks 2 and 3 are brighter than their immediate surroundings at a significance level of greater than 
$\sim3\sigma$. The apparent spatial extents of peaks 1, 2 and 3 are 
$\sim54^{\prime\prime}\times48^{\prime\prime}$, $\sim66^{\prime\prime}\times44^{\prime\prime}$ 
and $\sim54^{\prime\prime}\times48^{\prime\prime}$, respectively. Thus, all the peaks are significantly larger than the 
width of smoothing Gaussian kernel used in Figure \ref{fig:cont1}(i) ($=8^{\prime\prime}$). 
The 1-2, 2-3 and 1-3 peak separations are about 
43$^{\prime\prime}$, 67$^{\prime\prime}$ and 90$^{\prime\prime}$ ($\sim$43 kpc, 67 kpc and 90 kpc), respectively. In its optical 
image (Figure \ref{fig:cont2}(h)), the BCG is found coinciding with the second brightest X-ray peak (peak 2), although shifted towards the 
south by about 10$^{\prime\prime}$ ($\sim$10 kpc). Quite intriguingly, the brightest peak 1 does not seem to have 
an optical counterpart. The cluster has the lowest X-ray 
temperature (kT$=1.7\pm0.2$ keV) and the lowest bolometric X-ray luminosity (L$_{X}^{bol}=(1.05\pm0.03)\times10^{43}$ ergs s$^{-1}$) in the 
entire sample (Tables \ref{tab:total_regions_spectral_results} and \ref{tab:cluster_properties}). The cluster also has the lowest value of 
the average elemental abundance A1377 ($=0.09\pm0.04$ times the solar value), which is significantly lower than the 
typical values observed for clusters ($\sim$0.3 times the solar value). This indicates that the ICM of the cluster has not been processed 
sufficiently. A1377, therefore, seems to be a young unrelaxed cluster, which is consistent with its extremely irregular morphology. 
The temperature map (Figure \ref{fig:2D_temp_map}(e)) of the cluster shows an irregular distribution and the average temperature of the 
western parts (kT$_{\rm west}=2.0\pm0.2$ keV) seems to be slightly lower than the average temperature of the eastern parts 
(kT$_{\rm east}=2.4\pm0.4$ keV), although due to the large errors the difference is not significant. 
Unlike all the other clusters of the 
sample, A1377 shows anisotropic and irregular density, entropy and pressure distributions (Figures \ref{fig:2D_dens_map}(e), 
\ref{fig:2D_entr_map}(e) and \ref{fig:2D_pres_map}(e), respectively).\\

\textbf{A1831B:} The cluster A1831, is a visual superposition of the two clusters, A1831A and A1831B, located along very 
close line-of-sights, but at very different average redshifts (cz $\sim$ 18870 km s$^{-1}$ and 22629 km s$^{-1}$, respectively) 
\citep{Kopylov:2010}. 
\citet{Kopylov:2010} found that A1831B is a much richer cluster than A1831A, the latter being a poor foreground cluster. They also found 
that the X-ray emission from this region is associated with A1831B. We detected significant X-ray emission ($\geq$3$\sigma$) from 
A1831B up to a radius of about 360$^{\prime\prime}$ ($\sim$0.52 Mpc $\sim$R$_{200}/4$; see Figure \ref{fig:xsb_prof_db_fit}). In the X-ray 
image (Figure \ref{fig:cont1}(k)), the emission from the cluster seems to be smooth, although with a 
significant departure from spherical symmetry. In the optical image 
(Figure \ref{fig:cont1}(l)), the brightest galaxy in the field is found very close to the peak of the X-ray emission. The galaxy has a 
redshift of about $\sim$0.076 \citep{smith:2004} (equal to the mean redshift of A1831B) and, therefore, is the BCG of A1831B. 
No distinguishable X-ray emission is detected from the cluster A1831A, which is located to the southwest at about a projected separation of 
1$^{\prime}$ ($\sim$86 kpc) from A1831B \citep{Kopylov:2010}. The outermost X-ray contours in Figure \ref{fig:cont1}(k) show an extension along 
the north-west. Interestingly, \citet{Ramella:2007} found an optical subcluster towards the northwest of the A1831B BCG, at a projected distance 
of $\sim$10$^{\prime}$ ($\sim$860 kpc) and the same redshift. The subcluster lies just outside the FOV of \textit{Chandra}. It seems that the 
extension of X-ray 
emission in A1831B towards the northwest is due to a possible tidal interaction of the cluster with the NW subcluster. A shift of about 
2$^{\prime\prime}$ ($\sim$3 kpc) between the position of the BCG and the peak/centroid of the X-ray emission seen in Figure \ref{fig:cont1}(l), 
suggests that the cluster is out of dynamical equilibrium. In the temperature profile of the cluster, a gradual decrease is seen from the centre 
outwards. 
However, in the temperature map of the cluster (Figure \ref{fig:2D_temp_map}(f)), a higher temperature can be seen 
in the central region (marked as \textquoteleft{}a\textquoteright{}, average kT$=4.3\pm0.3$ keV) and towards the east (marked as 
\textquoteleft{}b\textquoteright{}, average kT$=3.9\pm0.4$ keV) as 
compared to the average temperature of $2.9\pm0.3$ keV seen in the outer parts of the cluster.\\

\textbf{A2124:} We have detected X-ray emission (significance$\geq$3$\sigma$) extending to a radius of about 380$^{\prime\prime}$ 
($\sim$0.48 Mpc $\sim$R$_{200}/4$) in this cluster 
(see Figure \ref{fig:xsb_prof_db_fit}). In the X-ray image of the cluster, the emission seems to be regular on large scales, but is aligned 
along the NW-SE direction (see Figure \ref{fig:cont1}(m)). However, a zoomed-in image of the central region of the cluster shows four peaks 
(see Figure \ref{fig:cont2}(i), the peaks are numbered from 1 to 4, in decreasing order of brightness). 
Peaks 1, 2, 3 and 4 are brighter than 
their immediate surroundings at significance levels of about 22$\sigma$, 15$\sigma$, 15$\sigma$ and 11$\sigma$, respectively. 
The apparent spatial extents of peaks 1, 2, 3 and 4 in Figure \ref{fig:cont1}(m) are $\sim38^{\prime\prime}\times38^{\prime\prime}$, 
$\sim28^{\prime\prime}\times28^{\prime\prime}$, $\sim20^{\prime\prime}\times18^{\prime\prime}$ and 
$\sim13^{\prime\prime}\times28^{\prime\prime}$, 
respectively. Therefore, all the peaks seem to be significantly larger than the width of the smoothing Gaussian kernel used in 
Figure \ref{fig:cont1}(m) (=4$^{\prime\prime}$). 
The 1-2, 1-3, 1-4, 2-3, 2-4 and 3-4 
peak separations are about 23$^{\prime\prime}$, 26$^{\prime\prime}$, 48$^{\prime\prime}$, 48$^{\prime\prime}$, 26$^{\prime\prime}$ and 
75$^{\prime\prime}$ ($\sim$29, 33, 60, 60, 33 and 94 kpc), respectively. The BCG in the optical image of the 
cluster (Figure \ref{fig:cont2}(j)) is found nearest to the brightest of the X-ray peaks (peak 1), although, shifted towards the NE by about 
6$^{\prime\prime}$ ($\sim$7 kpc). No optical emission is seen from the other three X-ray peaks. \citet{FlinKry:2006} had detected two small 
optical subclusters in the central part of the cluster at a scale of 129 kpc ($\sim$1.7$^{\prime}$). The X-ray image of the central region of 
the cluster (Figure \ref{fig:cont2}(i)) does not show substructures at this scale, although there seems to be a 
connection between the optical subclusters and the multiple central X-ray peaks. 
The temperature map of the cluster (Figure \ref{fig:2D_temp_map}(g)) shows a high temperature in the central region 
(marked as \textquoteleft{}a\textquoteright{}, average kT$=4.7^{+0.4}_{-0.3}$ keV). In general, a significantly lower temperature 
of $\sim3.5\pm0.3$ keV is observed in the outermost parts of the map. However, a significantly higher temperature region is seen 
in the easternmost parts (marked as \textquoteleft{}b\textquoteright{}, kT$=4.6^{+0.6}_{-0.5}$ keV). A small region with a higher 
temperature is indicated in the southeastern parts (marked as \textquoteleft{}c\textquoteright{}, kT$=4.4^{+0.8}_{-0.6}$ keV), and a very low 
temperature is seen in the southwestern corner of the map (marked as \textquoteleft{}d\textquoteright{}, kT$=1.7^{+0.6}_{-0.4}$ keV). 
The region, marked as \textquoteleft{}e\textquoteright{}, lying between the high temperature regions \textquoteleft{}a\textquoteright{} 
and \textquoteleft{}b\textquoteright{}, seems to have a lower temperature of kT$=3.9\pm0.3$ keV. 
The mismatch between the optical and X-ray peaks, and the high temperature regions in the centre of the cluster suggest ongoing mergers 
between the central galaxies. The X-ray peaks can, therefore, be related to the displaced X-ray halos of the central galaxies of the cluster.\\

\textbf{A2457:} X-ray emission is detected (significance$\geq$3$\sigma$) up to a radius of about 360$^{\prime\prime}$ 
($\sim$0.41 Mpc $\sim$R$_{200}/3$) in this cluster
(see Figure \ref{fig:xsb_prof_db_fit}). The large scale X-ray emission of the cluster seems to be regular and aligned along the 
E-W direction (see Figure \ref{fig:cont1}(o)). However, X-ray image of the central region of the cluster 
(Figure \ref{fig:cont2}(k)) shows the central X-ray morphology of the cluster to be highly irregular. Multiple peaks are seen in the 
central region of the cluster, four of which have been numbered as 1 to 4 in Figure \ref{fig:cont2}(k), in a decreasing order of brightness. 
The peaks 1, 2, 3 and 4 are brighter than their immediate surroundings at significance levels of more than 17$\sigma$, 17$\sigma$, 
17$\sigma$ and 12$\sigma$, respectively. The spatial extents of peaks 1, 2, 3 and 4 are $\sim13^{\prime\prime}\times19^{\prime\prime}$, 
$\sim13^{\prime\prime}\times13^{\prime\prime}$, $\sim45^{\prime\prime}\times26^{\prime\prime}$ and 
$\sim19^{\prime\prime}\times21^{\prime\prime}$, respectively. Therefore, all the peaks are found to 
be significantly larger than the width of smoothing Gaussian kernel used in 
Figure \ref{fig:cont1}(o) (=4$^{\prime\prime}$). However, due to their close proximity peaks 1 and 2 are not well separated 
and the boundaries are not well defined. The two peaks could either be separate peaks or a single broader peanut shaped bright region. 
The peak separations 1-2, 1-3, 1-4, 2-3, 2-4 and 3-4 are about 12$^{\prime\prime}$, 20$^{\prime\prime}$, 28$^{\prime\prime}$, 
21$^{\prime\prime}$, 28$^{\prime\prime}$ and 41$^{\prime\prime}$ (14, 23, 32, 24, 32 and 47 kpc), respectively. The BCG in the optical image of 
the cluster (Figure \ref{fig:cont2}(l)), is located in between the two brightest X-ray peaks 1 and 2, and is separated from them by about 
6$^{\prime\prime}$ and 8$^{\prime\prime}$ (7 kpc and 9 kpc), respectively. The other two X-ray peaks do not seem to have an optical counterpart. 
\citet{FlinKry:2006} found a group of galaxies in the centre of the cluster at a scale of 129 kpc ($\sim$1.9$^{\prime}$). 
Therefore, the multiple X-ray peaks seen in A2457 seem to be connected with the hot gas of the galaxy group. In its temperature profile, 
A2457 shows a drop in its innermost annulus followed by an isothermal profile outwards. The cluster seems to host a (weak) cool-core. 
The temperature map of the cluster (Figure \ref{fig:2D_temp_map}(h)), however, shows a higher temperature region in the 
central and eastern parts (marked as  \textquoteleft{}a\textquoteright{}, average kT$=4.0\pm0.2$ keV) relative to the regions seen in 
the northern, western and southern parts of the cluster (average  kT$=2.8\pm0.2$ keV). The high temperature region has a large spatial extent 
and seems to be an indication of the central group of galaxies having merged with the rest of the cluster.\\

\textbf{A2665:} Significant ($\geq$3$\sigma$) X-ray emission extends to a radius of about 400$^{\prime\prime}$ 
($\sim$0.43 Mpc) in this cluster (see 
Figure \ref{fig:xsb_prof_db_fit}). The X-ray image of the cluster (Figure \ref{fig:cont2}(q)) shows that the emission in its central 
region is aligned 
along the NE-SW direction and the central X-ray contours are compressed along the SE. The BCG in the optical image of the 
cluster (Figure \ref{fig:cont2}(r)) is slightly shifted (by about 7$^{\prime\prime}\sim$8 kpc) towards the NW with respect to centroid of 
the innermost X-ray 
contours. The cluster has an average X-ray temperature of 4.4$\pm$0.2 keV and a high elemental abundance of $=0.5\pm0.1$ times the solar 
value. 
The thermodynamic maps of the cluster show only slightly lower temperature and entropy in the central region 
(marked as \textquoteleft{}a\textquoteright{}, average kT$_{1}=4.1\pm0.2$ keV, average S$=106\pm6$ keV cm$^{2}$), as compared to 
the somewhat higher temperature and higher entropy regions seen in the outer parts of the cluster, towards 
the south (marked as \textquoteleft{}b\textquoteright{}, kT$=4.6^{+0.8}_{-0.7}$ keV, S$=158^{+28}_{-25}$ keV cm$^{2}$) and east 
(marked as \textquoteleft{}c\textquoteright{}, kT$=4.6^{+0.8}_{-0.6}$ keV, S$=155^{+28}_{-22}$ 
keV cm$^{2}$; see Figures \ref{fig:2D_temp_map}(i) and \ref{fig:2D_entr_map}(i)). The temperature difference, however, does not seem to be 
significant due to large measurement errors. 
The cluster is not well studied, therefore, little information is available in the literature.\\

\textbf{A3822:} In this cluster, we detected significant X-ray emission (significance$\geq$3$\sigma$) up to a radius of 430$^{\prime\prime}$ 
($\sim$0.62 Mpc, see Figure \ref{fig:xsb_prof_db_fit}), which is about one-fourth of the virial radius (R$_{200}$) of the cluster. The 
innermost contours in the X-ray image of A3822 (Figure \ref{fig:cont1}(s)) show an elongation towards the North 
and a contraction along the South. Also, the BCG in the optical image of the cluster (Figure \ref{fig:cont1}(t)) seems to be slightly shifted 
($\sim$10$^{\prime\prime}\sim$14 kpc) 
from the X-ray peak southwards. The cluster is found to have 
the highest average X-ray temperature of $5.3\pm0.3$ keV and the highest bolometric X-ray luminosity of $(40.4\pm1.2)\times10^{43}$ ergs 
s$^{-1}$, in the entire sample (see Tables 
\ref{tab:total_regions_spectral_results} and \ref{tab:cluster_properties}). 
The thermodynamic maps of the cluster show an average temperature of $4.7\pm0.2$ keV and entropy of $150\pm10$ keV cm$^{2}$ 
in the central region (marked as 
\textquoteleft{}a\textquoteright{}). Although higher temperature (average kT$=5.4^{+1.9}_{-1.2}$ keV) and higher 
entropy (average S$=456^{+168}_{-106}$ keV cm$^{2}$) values are measured  towards the northern part 
(marked as \textquoteleft{}b\textquoteright{} in Figures \ref{fig:2D_temp_map}(j) and \ref{fig:2D_entr_map}(j)), the difference is 
not statistically significant. A higher temperature 
region (marked as \textquoteleft{}c\textquoteright{}, kT$=6.0^{+1.3}_{-0.9}$ keV) is also indicated at a distance of about 
2$^{\prime}$ ($\sim$170 kpc) towards the southwest from the centre of the cluster. 
Very little information is available on A3822 in the literature.\\
\section{Summary}

 We have presented a detailed study of a sample of ten low-redshift clusters of galaxies with optical substructures, using their 
\textit{Chandra} X-ray observations. X-ray images of all the clusters of the sample (except A1831, A2665 and A3822) exhibit significant 
sub-structures, especially in their central regions. Multiple central X-ray intensity peaks are detected in six clusters viz. A193, A376, 
A539, A1377, A2124 and A2457. A total of 22 peaks are detected, of which eleven are found to have optical galaxy counterparts (10 
coincident with the BCGs of the clusters and one coincident with the second brightest galaxy of A539). 
All the peaks 
are found to be brighter (significance$>$3$\sigma$) than their immediate surroundings and are well resolved 
even after taking into account the smoothing kernels used. In the thermodynamic maps, three peaks 
(one in A376 and two in A539) are found to be significantly colder than their immediate surroundings. For A539 for which both the cold 
peaks are found to have optical galaxy counterparts, the peaks seem to be due to the cold gas stripped off from the two galaxies, as a result 
of interactions between them. In the absence of high temperature regions between the two peaks, ongoing 
mergers between the two galaxies cannot be established. However, to rule out the mergers completely, higher resolution temperature maps 
are required. 
For the secondary peak in A376, with significantly lower temperature and lower entropy than its surroundings, no counterpart 
is seen in the optical image. Therefore, the peak seems to be due 
to a remnant of a cool core of a subcluster that has already merged with the main cluster. 
The X-ray morphology of A970 is 
found to be azimuthally asymmetric and its temperature map seems to be bimodal. The cluster seems to be out of dynamical equilibrium which 
is probably due to a past merger with the optical subcluster in the northwest. The cluster A1831B seems to be tidally interacting with a 
subcluster towards the northwest. The values of average X-ray temperature and bolometric X-ray luminosity are found to be maximum for A3822 
and minimum for A1377. A1377 is found to have the most irregular X-ray morphology and 
thermodynamic maps, and has the smallest value of average elemental abundance in the 
entire sample. The cluster, therefore, seems to be the youngest cluster of the sample. 
The asymmetric thermodynamic maps and presence of small significantly higher temperature regions in various parts of at least five of the 
ten clusters (A193, A376, A1831, A2124 and A3822) suggest the possibility of galaxy scale mergers in them. 
The non-detection of substructures in the X-ray intensity maps coinciding with the hotter regions (that would have confirmed the 
presence of a shocked gas), however, raises some doubts regarding the reality of these features. 
A better and more significant detection of the substructures and variations in thermodynamic maps will require much deeper \textit{Chandra} 
observations of the clusters. 
The profiles of X-ray surface-brightness show very good fits with the double-$\beta$ model for all 
the clusters (except A1377), thereby indicating the presence of a (weak) cool core in them. From their cooling time estimates none of the 
clusters seems to have a strong cool core in it, although presence of weak cool cores in some of the clusters cannot be ruled out. The 
bolometric X-ray luminosities and the average X-ray temperature of all the clusters of our sample are found to be consistent with the 
L$_{X}$-kT relation of clusters obtained by \citet{Takey:2013}.

\section{Acknowledgements}

  The X-ray data used in this research have been obtained from the High Energy Astrophysics Science Archive Research Center (HEASARC), provided 
by NASA's Goddard Space Flight Center. We have used observations done with the \textit{Chandra} X-ray Observatory, managed by NASA's Marshall 
Center.


\bibliographystyle{apj}
\bibliography{ref1}

\begin{thebibliography}{53}
\expandafter\ifx\csname natexlab\endcsname\relax\def\natexlab#1{#1}\fi

\bibitem[{{Anders} \& {Ebihara}(1982)}]{AndEbi:1982}
{Anders}, E., \& {Ebihara}, M. 1982, \gca, 46, 2363

\bibitem[{{Anders} \& {Grevesse}(1989)}]{AndGre:1989}
{Anders}, E., \& {Grevesse}, N. 1989, \gca, 53, 197

\bibitem[{{Burgett} {et~al.}(2004){Burgett}, {Vick}, {Davis}, {Colless}, {De
  Propris}, {Baldry}, {Baugh}, {Bland-Hawthorn}, {Bridges}, {Cannon}, {Cole},
  {Collins}, {Couch}, {Cross}, {Dalton}, {Driver}, {Efstathiou}, {Ellis},
  {Frenk}, {Glazebrook}, {Hawkins}, {Jackson}, {Lahav}, {Lewis}, {Lumsden},
  {Maddox}, {Madgwick}, {Norberg}, {Peacock}, {Percival}, {Peterson},
  {Sutherland}, \& {Taylor}}]{Burgett:2004}
{Burgett}, W.~S., {Vick}, M.~M., {Davis}, D.~S., {et~al.} 2004, \mnras, 352,
  605

\bibitem[{{Cavagnolo} {et~al.}(2009){Cavagnolo}, {Donahue}, {Voit}, \&
  {Sun}}]{Cavagnolo:2009}
{Cavagnolo}, K.~W., {Donahue}, M., {Voit}, G.~M., \& {Sun}, M. 2009, \apjs,
  182, 12

\bibitem[{{Domainko} {et~al.}(2005){Domainko}, {Kapferer}, {Schindler}, {van
  Kampen}, {Kimeswenger}, {Mair}, {Kronberger}, {Ruffert}, \&
  {Mangete}}]{Domainko:2005}
{Domainko}, W., {Kapferer}, W., {Schindler}, S., {et~al.} 2005, Advances in
  Space Research, 36, 685

\bibitem[{{Donnelly} {et~al.}(2001){Donnelly}, {Forman}, {Jones}, {Quintana},
  {Ramirez}, {Churazov}, \& {Gilfanov}}]{Donnelly:2001}
{Donnelly}, R.~H., {Forman}, W., {Jones}, C., {et~al.} 2001, \apj, 562, 254

\bibitem[{{Dressler} \& {Shectman}(1988)}]{DressShect:1988}
{Dressler}, A., \& {Shectman}, S.~A. 1988, \aj, 95, 985

\bibitem[{{Ehlert} {et~al.}(2011){Ehlert}, {Allen}, {von der Linden},
  {Simionescu}, {Werner}, {Taylor}, {Gentile}, {Ebeling}, {Allen}, {Applegate},
  {Dunn}, {Fabian}, {Kelly}, {Million}, {Morris}, {Sanders}, \&
  {Schmidt}}]{Ehlert:2011}
{Ehlert}, S., {Allen}, S.~W., {von der Linden}, A., {et~al.} 2011, \mnras, 411,
  1641

\bibitem[{{Fabian}(1994)}]{Fabian:1994}
{Fabian}, A.~C. 1994, \araa, 32, 277

\bibitem[{{Flin} \& {Krywult}(2006)}]{FlinKry:2006}
{Flin}, P., \& {Krywult}, J. 2006, \aap, 450, 9

\bibitem[{{Forman} \& {Jones}(1982)}]{ForJon:1982}
{Forman}, W., \& {Jones}, C. 1982, \araa, 20, 547

\bibitem[{{Forman} \& {Jones}(1990)}]{ForJon:1990}
{Forman}, W., \& {Jones}, C. 1990, in Clusters of Galaxies, ed. W.~R.
  {Oegerle}, J.~{Fitchett}, \& L.~{Danly}, 257

\bibitem[{{Fukazawa} {et~al.}(1994){Fukazawa}, {Ohashi}, {Fabian}, {Canizares},
  {Ikebe}, {Makishima}, {Mushotzky}, \& {Yamashita}}]{Fukazawa:1994}
{Fukazawa}, Y., {Ohashi}, T., {Fabian}, A.~C., {et~al.} 1994, \pasj, 46, L55

\bibitem[{{Geller} \& {Beers}(1982)}]{GeBe:1982}
{Geller}, M.~J., \& {Beers}, T.~C. 1982, \pasp, 94, 421

\bibitem[{{Girardi} {et~al.}(1997){Girardi}, {Escalera}, {Fadda}, {Giuricin},
  {Mardirossian}, \& {Mezzetti}}]{Girardi:1997}
{Girardi}, M., {Escalera}, E., {Fadda}, D., {et~al.} 1997, \apj, 482, 41

\bibitem[{{Gitti} {et~al.}(2010){Gitti}, {O'Sullivan}, {Giacintucci}, {David},
  {Vrtilek}, {Raychaudhury}, \& {Nulsen}}]{Gitti:2010}
{Gitti}, M., {O'Sullivan}, E., {Giacintucci}, S., {et~al.} 2010, \apj, 714, 758

\bibitem[{{Gu} {et~al.}(2012){Gu}, {Xu}, {Gu}, {Kawaharada}, {Nakazawa}, {Qin},
  {Wang}, {Wang}, {Zhang}, \& {Makishima}}]{Gu:2012}
{Gu}, L., {Xu}, H., {Gu}, J., {et~al.} 2012, \apj, 749, 186

\bibitem[{{Gu} {et~al.}(2010){Gu}, {Wang}, {Gu}, {Wang}, {Qin}, {Yao}, {Yang},
  \& {Xu}}]{Gu:2010}
{Gu}, L.-Y., {Wang}, Y., {Gu}, J.-H., {et~al.} 2010, Research in Astronomy and
  Astrophysics, 10, 1005

\bibitem[{{Henry} {et~al.}(2004){Henry}, {Finoguenov}, \& {Briel}}]{Henry:2004}
{Henry}, J.~P., {Finoguenov}, A., \& {Briel}, U.~G. 2004, \apj, 615, 181

\bibitem[{{Jones} \& {Forman}(1992)}]{JonFor:1992}
{Jones}, C., \& {Forman}, W. 1992, in NATO ASIC Proc. 366: Clusters and
  Superclusters of Galaxies, ed. A.~C. {Fabian}, 49

\bibitem[{{Jones} \& {Forman}(1999)}]{JonFor:1999}
{Jones}, C., \& {Forman}, W. 1999, \apj, 511, 65

\bibitem[{{Kalberla} {et~al.}(2005){Kalberla}, {Burton}, {Hartmann}, {Arnal},
  {Bajaja}, {Morras}, \& {P{\"o}ppel}}]{Kalberla:2005}
{Kalberla}, P.~M.~W., {Burton}, W.~B., {Hartmann}, D., {et~al.} 2005, \aap,
  440, 775

\bibitem[{{Kellogg} {et~al.}(1972){Kellogg}, {Gursky}, {Tananbaum}, {Giacconi},
  \& {Pounds}}]{Kellogg:1972}
{Kellogg}, E., {Gursky}, H., {Tananbaum}, H., {Giacconi}, R., \& {Pounds}, K.
  1972, \apjl, 174, L65

\bibitem[{{Kempner} {et~al.}(2002){Kempner}, {Sarazin}, \&
  {Ricker}}]{Kempner:2002}
{Kempner}, J.~C., {Sarazin}, C.~L., \& {Ricker}, P.~M. 2002, \apj, 579, 236

\bibitem[{{Kopylov} \& {Kopylova}(2010)}]{Kopylov:2010}
{Kopylov}, A.~I., \& {Kopylova}, F.~G. 2010, Astrophysical Bulletin, 65, 205

\bibitem[{{Kopylova} \& {Kopylov}(2006)}]{Kopylova:2006}
{Kopylova}, F.~G., \& {Kopylov}, A.~I. 2006, Astronomy Letters, 32, 84

\bibitem[{{Kriessler} \& {Beers}(1997)}]{KriessBe:1997}
{Kriessler}, J.~R., \& {Beers}, T.~C. 1997, \aj, 113, 80

\bibitem[{{Lakhchaura} {et~al.}(2011){Lakhchaura}, {Singh}, {Saikia}, \&
  {Hunstead}}]{Lakhchaura:2011}
{Lakhchaura}, K., {Singh}, K.~P., {Saikia}, D.~J., \& {Hunstead}, R.~W. 2011,
  \apj, 743, 78

\bibitem[{{Lakhchaura} {et~al.}(2013){Lakhchaura}, {Singh}, {Saikia}, \&
  {Hunstead}}]{Lakhchaura:2013}
---. 2013, \apj, 767, 91

\bibitem[{{Ma} {et~al.}(2012){Ma}, {Owers}, {Nulsen}, {McNamara}, {Murray}, \&
  {Couch}}]{Ma:2012}
{Ma}, C.-J., {Owers}, M., {Nulsen}, P.~E.~J., {et~al.} 2012, \apj, 752, 139

\bibitem[{{Makishima} {et~al.}(2001){Makishima}, {Ezawa}, {Fukuzawa}, {Honda},
  {Ikebe}, {Kamae}, {Kikuchi}, {Matsushita}, {Nakazawa}, {Ohashi}, {Takahashi},
  {Tamura}, \& {Xu}}]{Makishima:2001}
{Makishima}, K., {Ezawa}, H., {Fukuzawa}, Y., {et~al.} 2001, \pasj, 53, 401

\bibitem[{{Markevitch} \& {Vikhlinin}(2007)}]{MarkeVikh:2007}
{Markevitch}, M., \& {Vikhlinin}, A. 2007, \physrep, 443, 1

\bibitem[{{Maurogordato} {et~al.}(2011){Maurogordato}, {Sauvageot}, {Bourdin},
  {Cappi}, {Benoist}, {Ferrari}, {Mars}, \& {Houairi}}]{Maurogordato:2011}
{Maurogordato}, S., {Sauvageot}, J.~L., {Bourdin}, H., {et~al.} 2011, \aap,
  525, A79

\bibitem[{{McNamara} \& {Nulsen}(2007)}]{McNamaNul2007}
{McNamara}, B.~R., \& {Nulsen}, P.~E.~J. 2007, \araa, 45, 117

\bibitem[{{Mitchell} {et~al.}(1979){Mitchell}, {Dickens}, {Burnell}, \&
  {Culhane}}]{Mitchell:1979}
{Mitchell}, R.~J., {Dickens}, R.~J., {Burnell}, S.~J.~B., \& {Culhane}, J.~L.
  1979, \mnras, 189, 329

\bibitem[{{Morrison} \& {McCammon}(1983)}]{MorrMcCam:1983}
{Morrison}, R., \& {McCammon}, D. 1983, \apj, 270, 119

\bibitem[{{Ostriker} {et~al.}(1988){Ostriker}, {Huchra}, {Geller}, \&
  {Kurtz}}]{Ostriker:1988}
{Ostriker}, E.~C., {Huchra}, J.~P., {Geller}, M.~J., \& {Kurtz}, M.~J. 1988,
  \aj, 96, 1775

\bibitem[{{Owers} {et~al.}(2011){Owers}, {Randall}, {Nulsen}, {Couch}, {David},
  \& {Kempner}}]{Owers:2011}
{Owers}, M.~S., {Randall}, S.~W., {Nulsen}, P.~E.~J., {et~al.} 2011, \apj, 728,
  27

\bibitem[{{Proust} {et~al.}(2003){Proust}, {Capelato}, {Hickel}, {Sodr{\'e}},
  {Lima Neto}, \& {Cuevas}}]{Proust:2003}
{Proust}, D., {Capelato}, H.~V., {Hickel}, G., {et~al.} 2003, \aap, 407, 31

\bibitem[{{Proust} {et~al.}(2010){Proust}, {Capelato}, {Lima Neto}, \&
  {Sodr{\'e}}}]{Proust:2010}
{Proust}, D., {Capelato}, H.~V., {Lima Neto}, G.~B., \& {Sodr{\'e}}, L. 2010,
  \aap, 515, A57

\bibitem[{{Ramella} {et~al.}(2007){Ramella}, {Biviano}, {Pisani}, {Varela},
  {Bettoni}, {Couch}, {D'Onofrio}, {Dressler}, {Fasano}, {Kj{\o}rgaard},
  {Moles}, {Pignatelli}, \& {Poggianti}}]{Ramella:2007}
{Ramella}, M., {Biviano}, A., {Pisani}, A., {et~al.} 2007, \aap, 470, 39

\bibitem[{{Reiprich}(2001)}]{Reiprich:2001}
{Reiprich}, T.~H. 2001, PhD thesis, Max-Planck-Institut f{\"u}r
  extraterrestrische Physik, P.O.~Box 1312, Garching bei M{\"u}nchen, Germany

\bibitem[{{Roettiger} {et~al.}(1996){Roettiger}, {Burns}, \&
  {Loken}}]{Roettiger:1996}
{Roettiger}, K., {Burns}, J.~O., \& {Loken}, C. 1996, \apj, 473, 651

\bibitem[{{Rosati} {et~al.}(2002){Rosati}, {Borgani}, \&
  {Norman}}]{Rosati:2002}
{Rosati}, P., {Borgani}, S., \& {Norman}, C. 2002, \araa, 40, 539

\bibitem[{{Sarazin}(1988)}]{Sarazin:1988}
{Sarazin}, C.~L. 1988, {X-ray emission from clusters of galaxies}

\bibitem[{{Schuecker} {et~al.}(2001){Schuecker}, {B{\"o}hringer}, {Reiprich},
  \& {Feretti}}]{Schuecker:2001}
{Schuecker}, P., {B{\"o}hringer}, H., {Reiprich}, T.~H., \& {Feretti}, L. 2001,
  \aap, 378, 408

\bibitem[{{Seigar} {et~al.}(2003){Seigar}, {Lynam}, \& {Chorney}}]{Seigar:2003}
{Seigar}, M.~S., {Lynam}, P.~D., \& {Chorney}, N.~E. 2003, \mnras, 344, 110

\bibitem[{{Smith} {et~al.}(2004){Smith}, {Hudson}, {Nelan}, {Moore}, {Quinney},
  {Wegner}, {Lucey}, {Davies}, {Malecki}, {Schade}, \& {Suntzeff}}]{smith:2004}
{Smith}, R.~J., {Hudson}, M.~J., {Nelan}, J.~E., {et~al.} 2004, \aj, 128, 1558

\bibitem[{{Smith} {et~al.}(2001){Smith}, {Brickhouse}, {Liedahl}, \&
  {Raymond}}]{Smith:2001}
{Smith}, R.~K., {Brickhouse}, N.~S., {Liedahl}, D.~A., \& {Raymond}, J.~C.
  2001, \apjl, 556, L91

\bibitem[{{Sodr{\'e}} {et~al.}(2001){Sodr{\'e}}, {Proust}, {Capelato}, {Lima
  Neto}, {Cuevas}, {Quintana}, \& {Fouqu{\'e}}}]{Sodre:2001}
{Sodr{\'e}}, Jr., L., {Proust}, D., {Capelato}, H.~V., {et~al.} 2001, \aap,
  377, 428

\bibitem[{{Takahashi} {et~al.}(2009){Takahashi}, {Kawaharada}, {Makishima},
  {Matsushita}, {Fukazawa}, {Ikebe}, {Kitaguchi}, {Kokubun}, {Nakazawa},
  {Okuyama}, {Ota}, \& {Tamura}}]{Takahashi:2009}
{Takahashi}, I., {Kawaharada}, M., {Makishima}, K., {et~al.} 2009, \apj, 701,
  377

\bibitem[{{Takey} {et~al.}(2013){Takey}, {Schwope}, \& {Lamer}}]{Takey:2013}
{Takey}, A., {Schwope}, A., \& {Lamer}, G. 2013, \aap, 558, A75

\bibitem[{{Vulcani} {et~al.}(2011){Vulcani}, {Poggianti}, {Dressler}, {Fasano},
  {Valentinuzzi}, {Couch}, {Moretti}, {Simard}, {Desai}, {Bettoni},
  {D'Onofrio}, {Cava}, \& {Varela}}]{Vulcani:2011}
{Vulcani}, B., {Poggianti}, B.~M., {Dressler}, A., {et~al.} 2011, \mnras, 413,
  921

\end{thebibliography}

\clearpage


\begin{sidewaystable}
 \caption{Redshifts and positions ($\alpha$ (J2000), $\delta$ (J2000)) of the ten clusters of the sample along with the details 
(observation ID, date of observations ad exposure time) of their \textit{Chandra} observations.}
\label{tab:observation_table}
\vskip 0.5cm
\centering
{\small
\begin{tabular}{c c c c c c c}
\hline
Cluster & Redshift &$\alpha$ (J2000)&$\delta$ (J2000)& Observation ID & Date of & Exposure \\
 & & & & & Observation & Time (ks) \\
\hline
\hline
A193   & 0.0490 & 01 25 07.3 & 08 41 36.0 &  6931 & 2005 Nov 10 & 18.2 \\
A376   & 0.0475 & 02 45 48.0 & 36 51 36.0 & 12277 & 2010 Nov 07 & 10.6 \\
A539   & 0.0284 & 05 16 35.1 & 06 27 14.0 &  5808 & 2005 Nov 18 & 24.6 \\
       &        &            &            &  7209 & 2005 Nov 20 & 18.8 \\
A970   & 0.0588 & 10 17 34.3 & -10 42 01.0 & 12285 & 2011 Feb 04 & 10.1 \\
A1377  & 0.0514 & 11 47 15.7 &	55 43 01.6 &  6943 & 2006 Sep 08 & 44.1 \\
A1831B & 0.0755 & 13 59 15.1 & 27 58 34.5 & 12283 & 2011 Apr 19 & 10.1 \\
A2124  & 0.0654 & 15 44 59.3 &	36 03 40.0 &  3238 & 2002 May 20 & 19.6 \\
A2457  & 0.0591 & 22 35 40.3 &	01 31 33.6 & 12276 & 2011 Jan 12 & 10.1 \\
A2665  & 0.0556 & 23 50 50.6 &	06 09 00.0 & 12280 & 2011 Jan 17 & 10.1 \\
A3822  & 0.0760 & 21 54 06.2 & -57 50 49  &  8269 & 2007 Mar 14 &  8.2 \\
\hline
\end{tabular}}
\end{sidewaystable}

\begin{sidewaystable}
 \caption{Results obtained from the double-$\beta$ model fitting of the X-ray surface brightness profiles of the clusters, 
as described in \S\ref{sec:xray_sb_prof}. Here, S$_{01}$ and S$_{02}$ are the peak surface brightness, 
and r$_{c1}$ and r$_{c2}$ are the corresponding core radii of the two components of the double-$\beta$ model. Values of the index 
$\beta$ (assumed same for both the components) are also given. 
For A1377, results of the single-$\beta$ model fitting are given. The values 
of the reduced chi-square minimum $(\chi^{2}_{\nu})_{min}$ and the degrees of freedom of each fit are also given.}
\label{tab:xsb_db_fit_results}
\vskip 0.3cm
\centering
  \begin{threeparttable}
\begin{tabular}{c c c c c c c}
 \hline
\hline
Cluster & S$_{01}$ & S$_{02}$ & r$_{c1}$ & r$_{c2}$ & $\beta\; (=\beta 1 = \beta 2$) & $(\chi^{2}_{\nu})_{min}$ (DOF)\\\
 & (10$^{-5}$ counts s$^{-1}$ arcsec$^{-2}$) & (10$^{-5}$ counts s$^{-1}$ arcsec$^{-2}$) & (kpc) & (kpc) & & \\
\hline
 A193 & 3.3$\pm$0.4 & 4.0$\pm$0.5 & 64$\pm$13 & 192$\pm$28 & 0.63$\pm$0.05 & 1.38 (45) \\
 A376 & 2.3$\pm$0.3 & 0.8$\pm$0.1 &  27$\pm$4 & 193$\pm$19 & 0.67$\pm$0.04 & 0.84 (55) \\
 A539 & 5.9$\pm$0.6 & 2.3$\pm$0.1 &  18$\pm$2 & 169$\pm$8 & 0.48$\pm$0.01 & 1.28 (81) \\
 A970 & 1.5$\pm$0.2 & 1.4$\pm$0.2 & 37$\pm$11 & 142$\pm$21 & 0.65$\pm$0.05 & 0.93 (40) \\
A1377 & 0.43$\pm$0.01 & -- & 252$\pm$46 & -- & 0.75$\pm$0.15 & 2.42 (19) \\
A1831B & 2.8$\pm$0.2 & 0.8$\pm$0.1 & 118$\pm$28 & 374$\pm$89 &  2.20$\pm$0.70 & 0.86 (40) \\
A2124 & 4.7$\pm$0.3 & 2.5$\pm$0.2 & 40$\pm$5 & 171$\pm$16 & 0.70$\pm$0.04 & 1.59 (43) \\
A2457 & 1.9$\pm$0.3 & 0.7$\pm$0.1 & 37$\pm$8 & 232$\pm$39 & 1.10$\pm$0.20 & 0.48 (39) \\
A2665 & 3.1$\pm$0.3 & 1.5$\pm$0.1 & 47$\pm$7 & 190$\pm$22 & 1.10$\pm$0.10 & 1.45 (24) \\
A3822 & 0.9$\pm$0.3 & 1.6$\pm$0.1 & 31$\pm$12 & 154$\pm$14 & 0.76$\pm$0.05 & 0.95 (48) \\
\hline
\end{tabular}
     \begin{tablenotes}
\footnotesize 
       \item[] All errors are quoted at 90\% confidence level based on $\chi^{2}_{min}$+2.71.
     \end{tablenotes}
  \end{threeparttable}
\end{sidewaystable}
\clearpage

\begin{sidewaystable}
 \caption{Average spectral properties of all the clusters of the sample resulting from the analysis described in \S\ref{sec:Total-Spec-Analy}. 
the N$_{\rm H}$ values obtained from the LAB survey for all the clusters, are given in the second column.}
\label{tab:total_regions_spectral_results}
\vskip 0.3cm
\centering
  \begin{threeparttable}
\begin{tabular}{c c c c c c}
 \hline
\hline
Cluster & N$_{\rm H}$ & kT & Abundance & \textit{apec} norm.& $(\chi^{2}_{\nu})_{min}$ (DOF)\\
 & & & & & \\
 & (10$^{20}$ cm$^{-2}$) & (keV)&(Rel. to solar)& ($10^{-3}$ cm$^{-5}$) & \\
\hline
A193  & 4.29 & $4.0\pm0.1$           & $0.2\pm0.1$          & $10.7\pm0.2$   & 1.4 (40)\\
A376  & 5.49 & $4.8\pm0.4$           & $0.3^{+0.2}_{-0.1}$  & $6.1\pm0.3$    & 0.68 (40)\\
A539  & 10.5 & $3.0\pm0.1$           & $0.19\pm0.03$        & $14.1\pm0.2$   & 1.0 (77)\\
A970  & 4.39 & $4.6\pm0.3$           & $0.2\pm0.1$          & $12.8\pm0.5$   & 1.1 (37)\\
A1377 & 0.75 & $1.7\pm0.2$           & $0.09\pm0.04$        & $2.1\pm0.2$    & 0.9 (25)\\
A1831B & 1.38 & $3.9\pm0.2$           & $0.3\pm0.1$          & $8.3\pm0.4$    & 0.93 (40)\\
A2124 & 1.86 & $4.8\pm0.3$           & $0.3\pm0.1$          & $5.4\pm0.2$    & 1.1 (40)\\
A2457 & 5.62 & $3.9\pm0.3$           & $0.2\pm0.1$          & $7.7\pm0.4$    & 1.5 (40)\\
A2665 & 6.02 & $4.4^{+0.3}_{-0.2}$   & $0.5\pm0.1$          & $8.9\pm0.4$    & 0.93 (40)\\
A3822 & 2.18 & $5.3\pm0.3$           & $0.2\pm0.1$          & $16.0\pm0.5$   & 1.3 (39)\\
\hline
\end{tabular}
     \begin{tablenotes}
\footnotesize 
       \item[] All errors are quoted at 90\% confidence level based on $\chi^{2}_{min}$+2.71.
     \end{tablenotes}
  \end{threeparttable}
\end{sidewaystable}
\clearpage

\begin{sidewaystable}
 \caption{Estimates of 0.5-7.0 keV X-ray luminosity (L$_{X}$), bolometric X-ray luminosity (L$_{bol}$) calculated within R$_{200}$, cooling 
 times (t$_{cool}$) and gas mass (M$_{\rm gas}(\rm r)$) (for both r= 1 Mpc and r=R$_{\rm 200}$) for the cluster sample. The second column 
 gives the R$_{200}$ values used here and their references.}
\label{tab:cluster_properties}
\vskip 0.3cm
\centering
  \begin{threeparttable}
\begin{tabular}{c c c c c c c}
 \hline
\hline
Cluster & R$_{200}$ & L$_{X}$ & L$_{X}^{bol}$ & t$_{cool}$ & M$_{\rm gas}(\rm r = 1 Mpc)$ & M$_{\rm gas}(\rm r = R_{200})$\\
 & & (0.5-7.0 keV) & (0.1-100 keV) & & & \\
 &(Mpc) & (10$^{43}$ erg s$^{-1}$) & (10$^{43}$ erg s$^{-1}$) & ($10^{10}$ yr) & ($10^{13}$ $\rm M_{\odot}$) & ($10^{13}$ $\rm M_{\odot}$) \\
\hline
 & & & \\
A193  & 1.83$^{1}$ & $6.64\pm0.04$         & $11.5\pm0.2$        & $1.2\pm0.6$ & $6.7\pm0.7$ & $18.5\pm1.9$\\
A376  & 2.06$^{1}$ & $3.91\pm0.05$         & $6.8\pm0.3$        & $1.1\pm0.4$ & $6.2\pm0.9$ & $10.4\pm1.5$\\
A539  & 1.87$^{2}$ & $3.67\pm0.03$         & $7.4\pm0.2$        & $0.8\pm0.2$ & $12.4\pm1.0$ & $40.9\pm3.2$\\
A970  & 1.84$^{1}$ & $11.9\pm0.1$          & $18.2\pm0.7$         & --  & -- & -- \\
A1377 & 1.40$^{3}$ & $0.88\pm0.03$         & $1.6\pm0.1$        & --  & -- & -- \\
A1831B & 2.27$^{4}$ & $12.3\pm0.3$           & $17.1\pm0.5$          & $1.0\pm0.2$ & $4.7\pm0.7$ & $21.9\pm3.3$\\
A2124 & 1.92$^{1}$ & $6.4\pm0.1$           & $9.5\pm0.3$          & $1.1\pm0.7$ & $2.7\pm0.3$ & $7.6\pm1.0$\\
A2457 & 1.40$^{1}$ & $6.8\pm0.1$           & $10.2\pm0.5$          & $1.5\pm0.7$ & $5.3\pm1.0$ & $12.7\pm2.4$\\
A2665 & 1.50$^{5}$ & $8.0\pm0.1$           & $11.5\pm0.5$          & $1.0\pm0.3$ & $2.5\pm0.3$ & $4.1\pm0.6$\\
A3822 & 2.43$^{2}$ & $25.9\pm0.2$          & $40.4\pm1.2$         & $1.2\pm0.4$ & $6.0\pm0.8$ & $30.6\pm4.1$\\
\hline
\end{tabular}
     \begin{tablenotes}
\footnotesize 
       \item[] All errors are quoted at 90\% confidence level based on $\chi^{2}_{min}$+2.71.
       \item[] $^{1}$ \citet{Vulcani:2011}, $^{2}$ \citet{Reiprich:2001}, $^{3}$ \citet{Kopylova:2006}, $^{4}$ \citet{Kopylov:2010}, \\$^{5}$ assumed.
     \end{tablenotes}
  \end{threeparttable}
\end{sidewaystable}
\clearpage


\begin{center}
  \begin{longtable}{cc}
\includegraphics[height=2.2in]{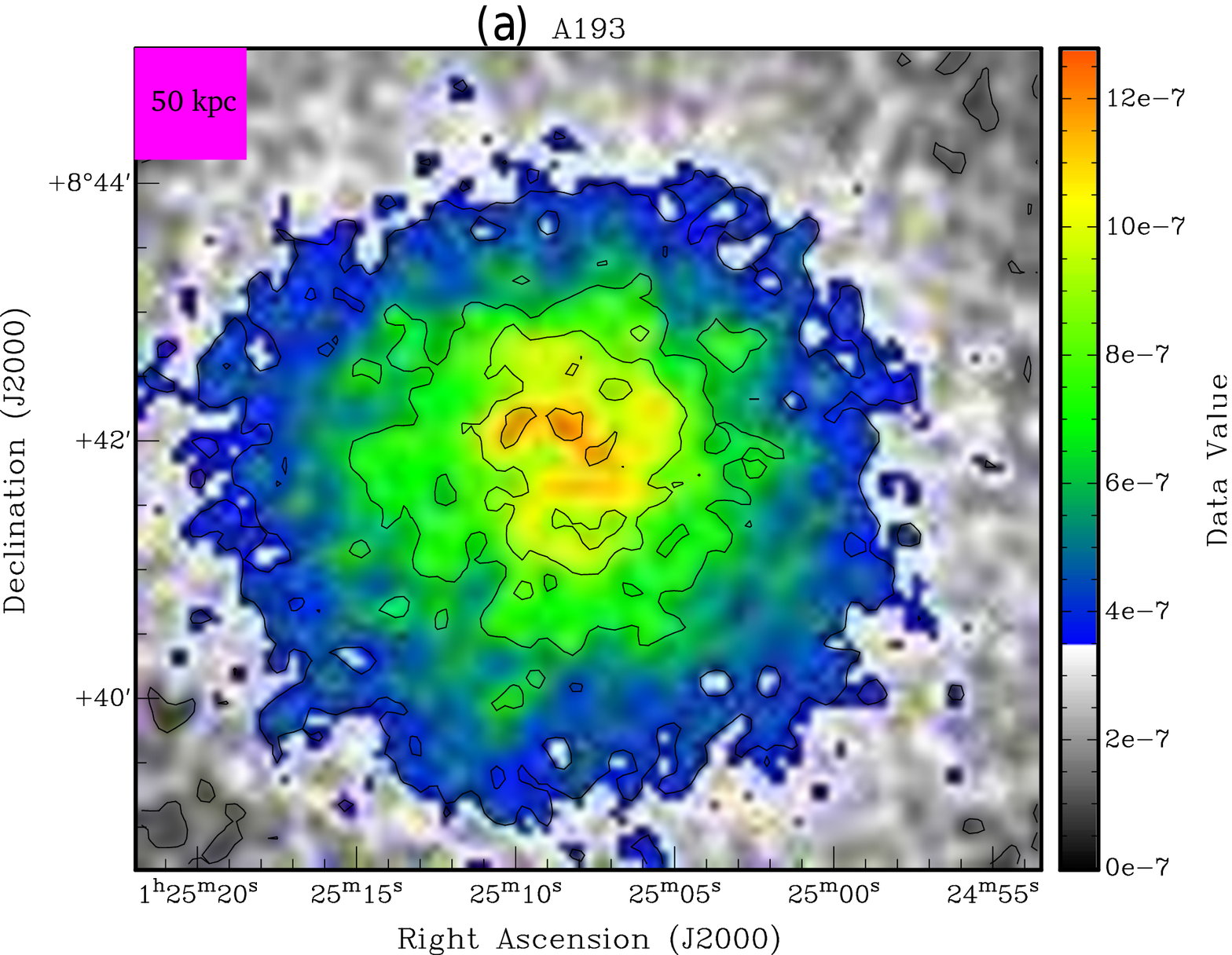}
\includegraphics[height=2.2in]{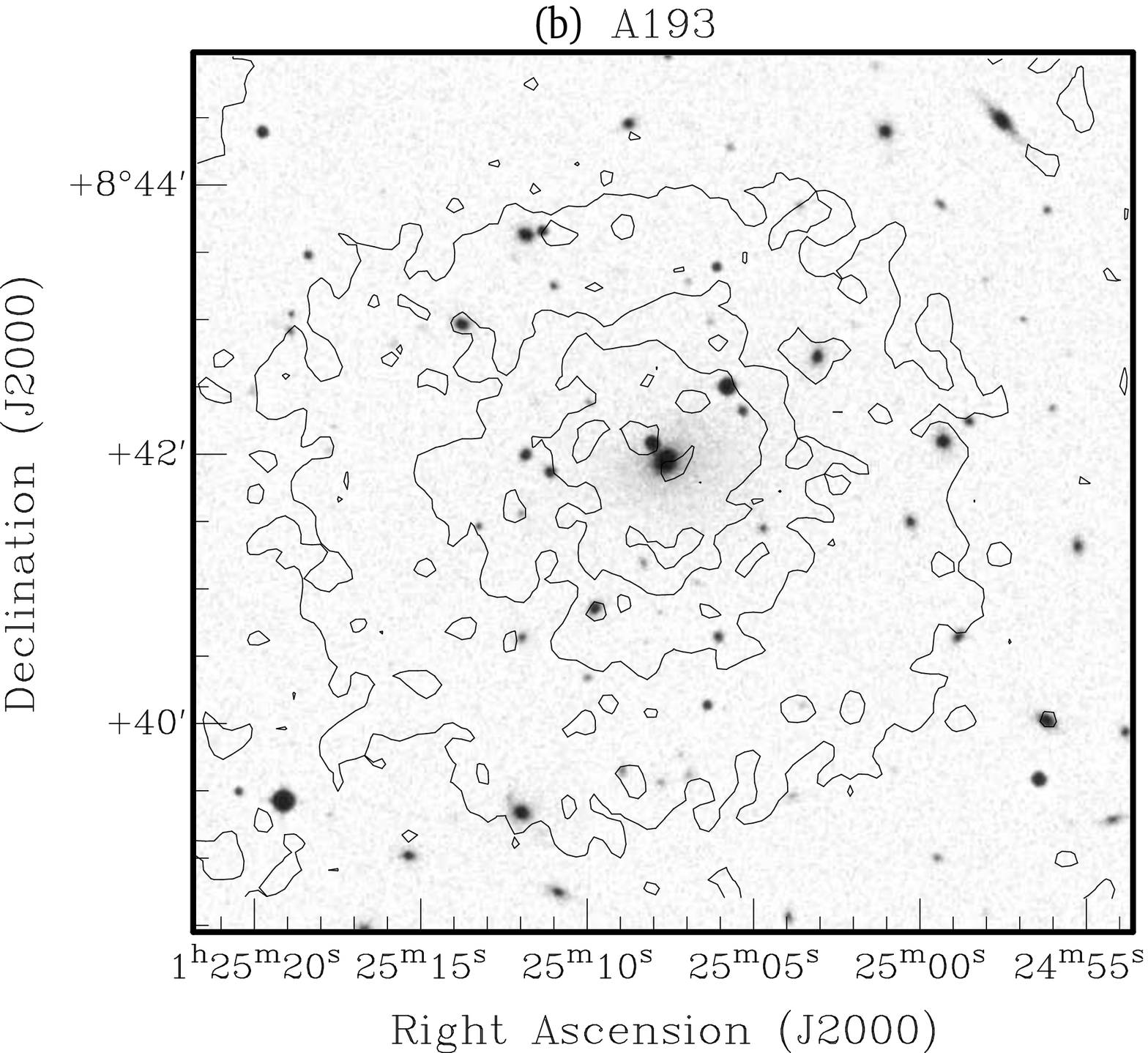}\\
\includegraphics[height=2.3in]{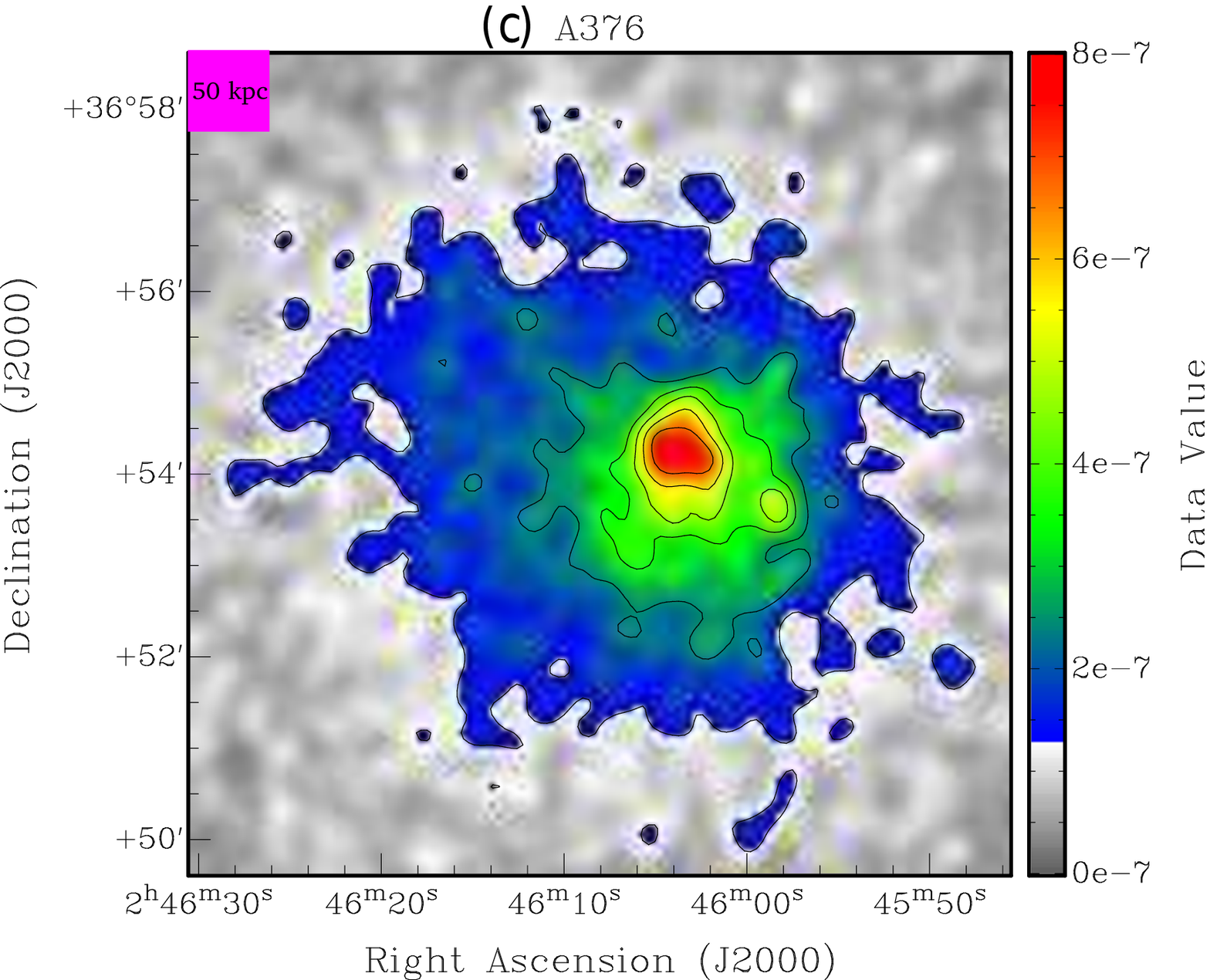}
\includegraphics[height=2.3in]{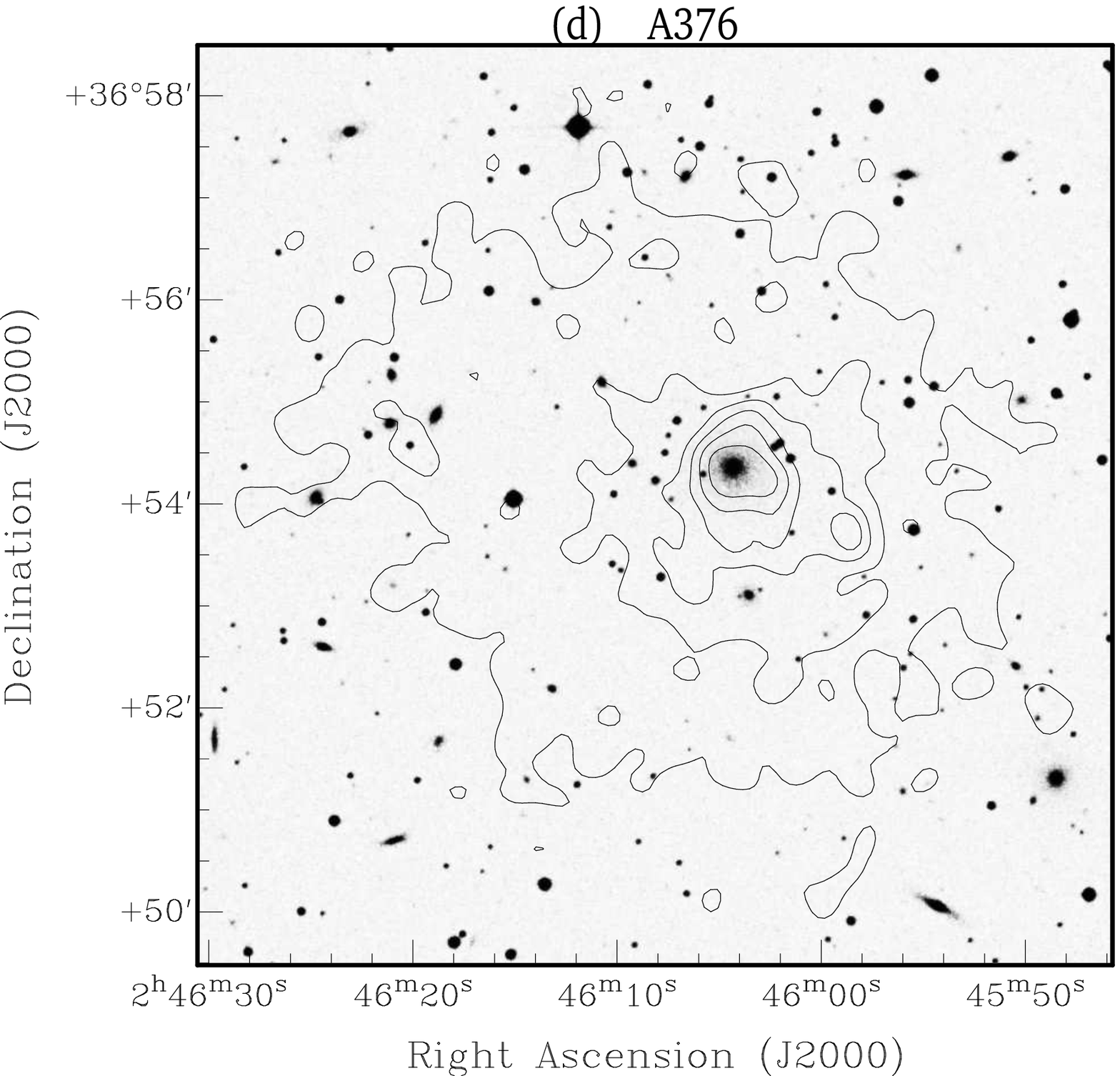}\\
\includegraphics[height=2.4in]{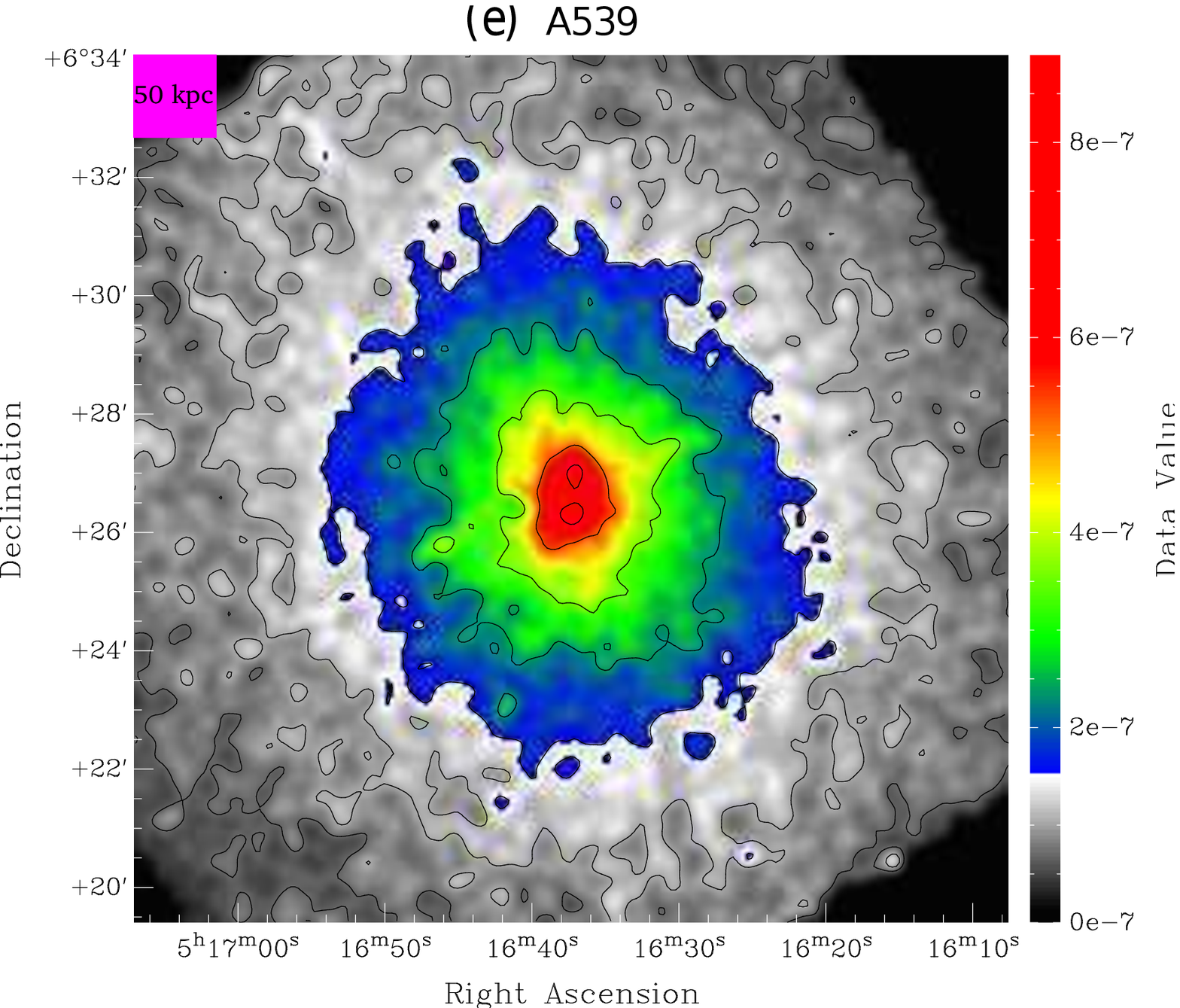}
\includegraphics[height=2.4in]{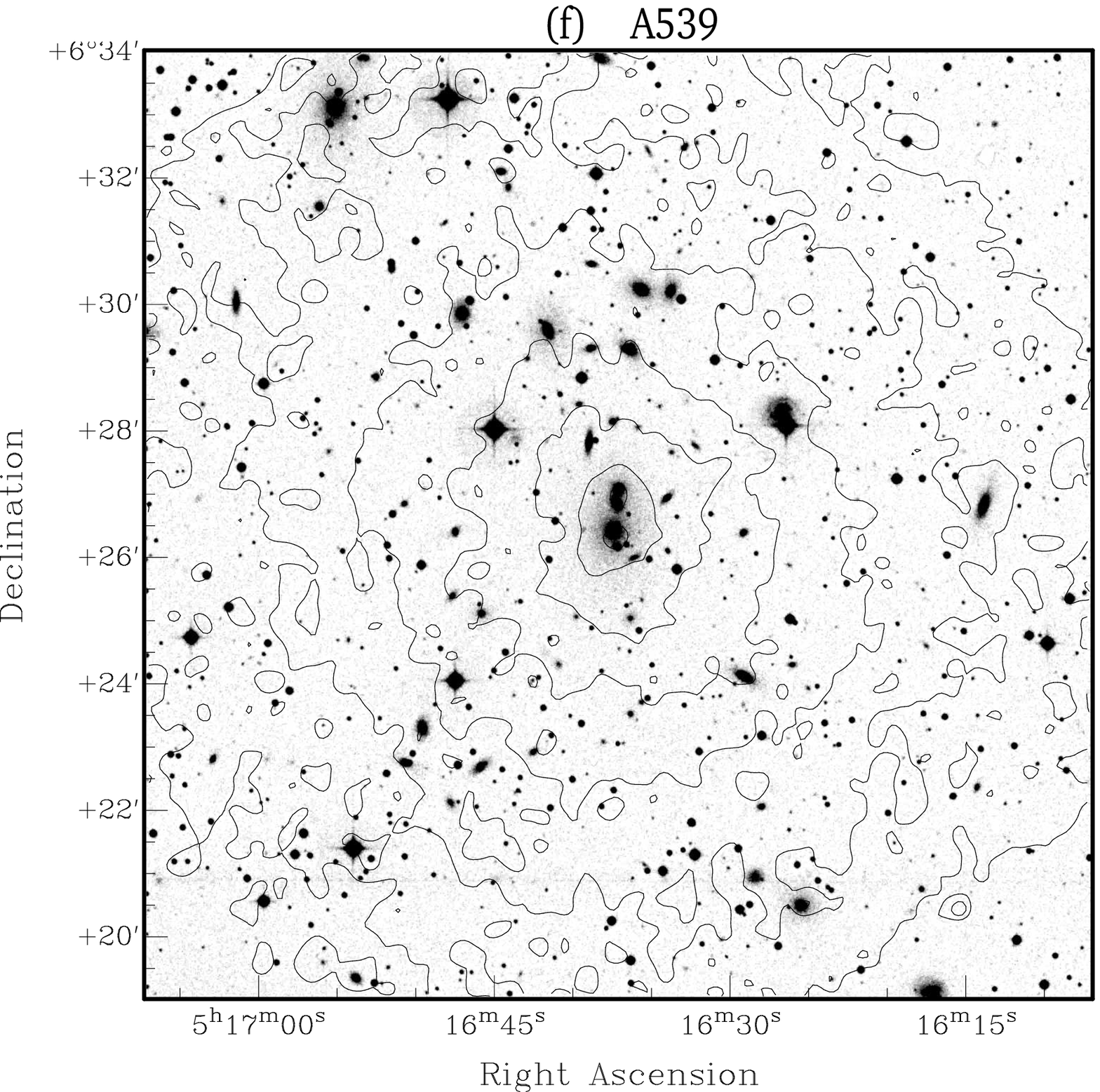}\\
\includegraphics[height=2.4in]{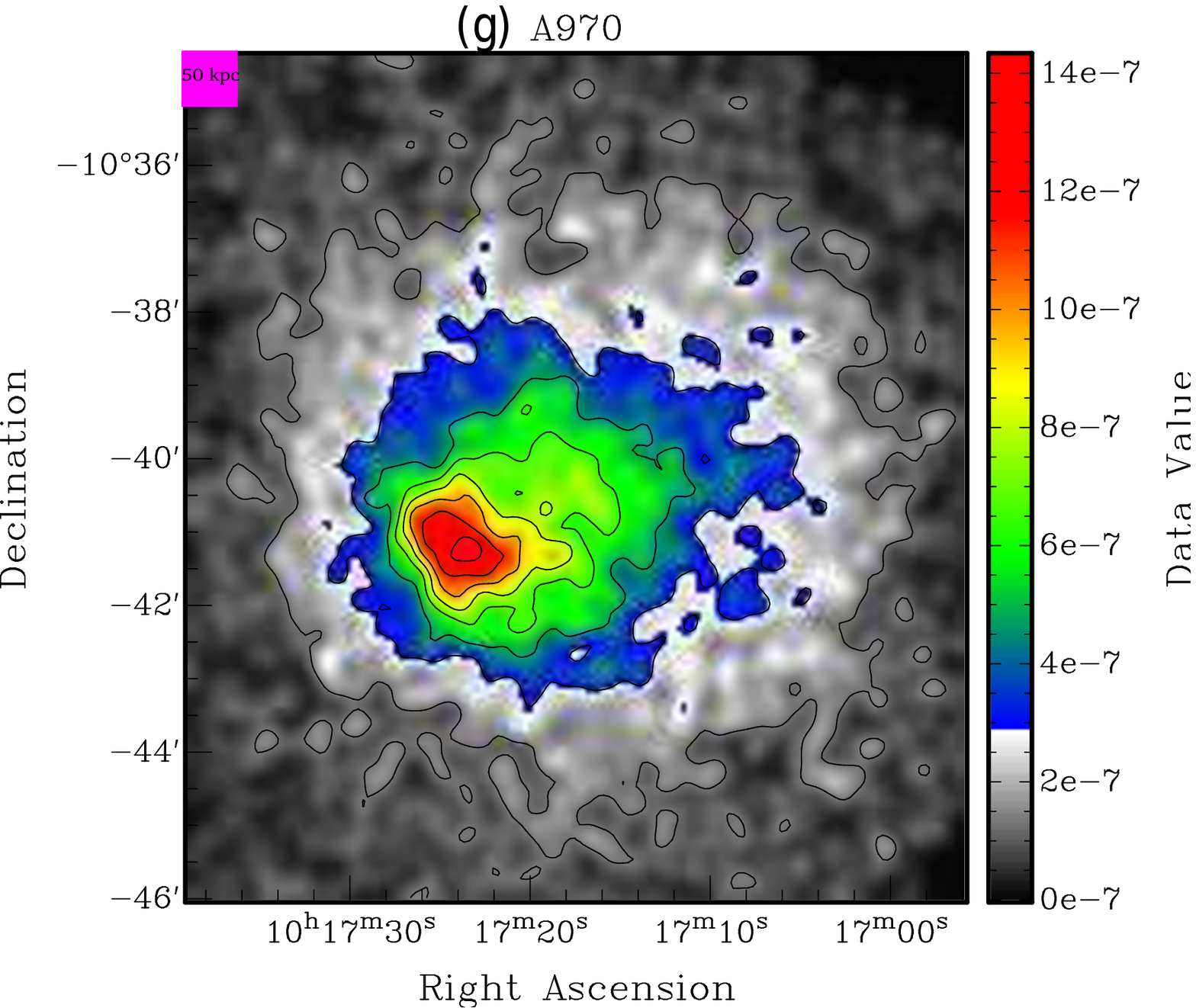}
\includegraphics[height=2.4in]{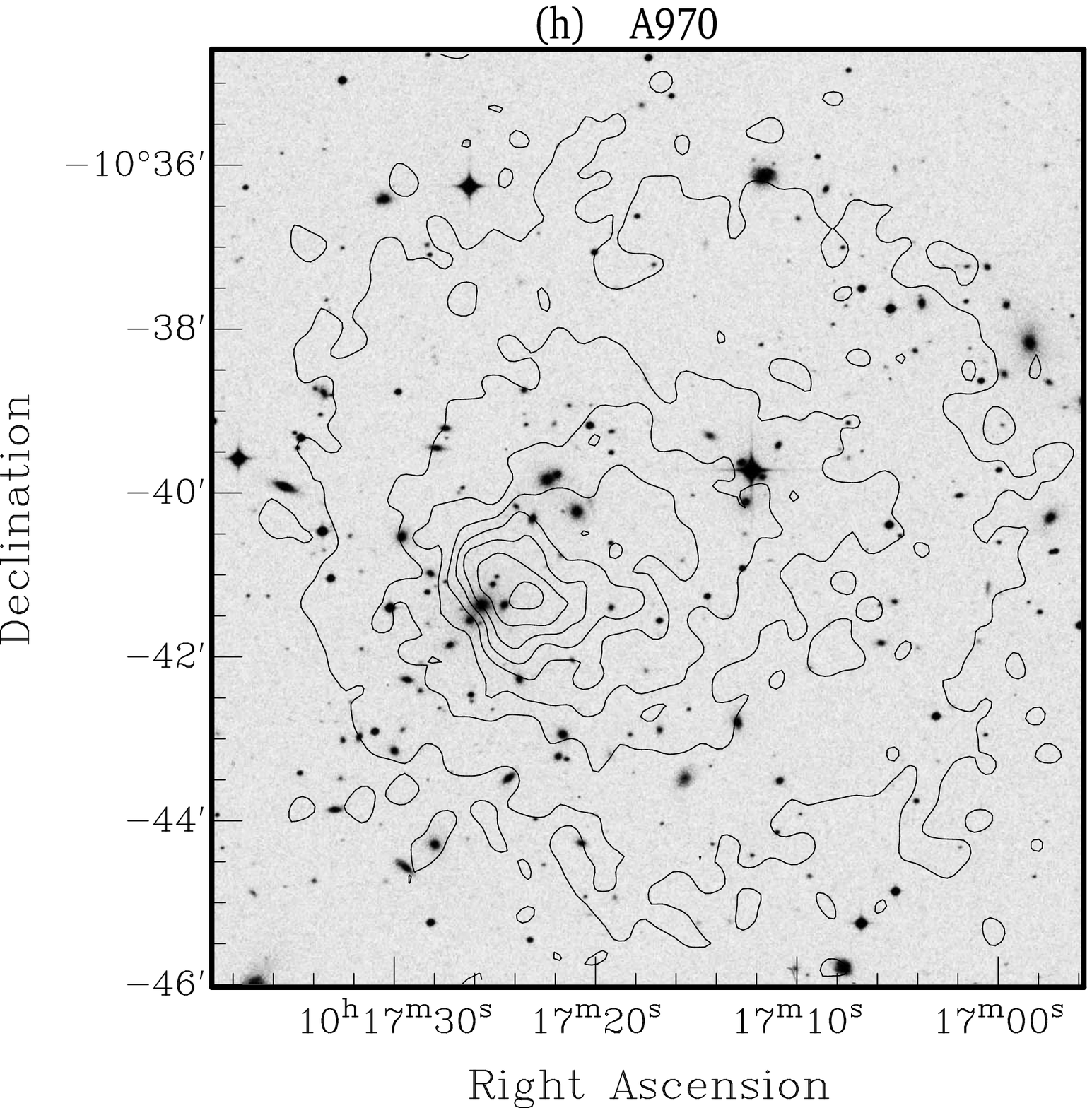}\\
\includegraphics[height=2.3in]{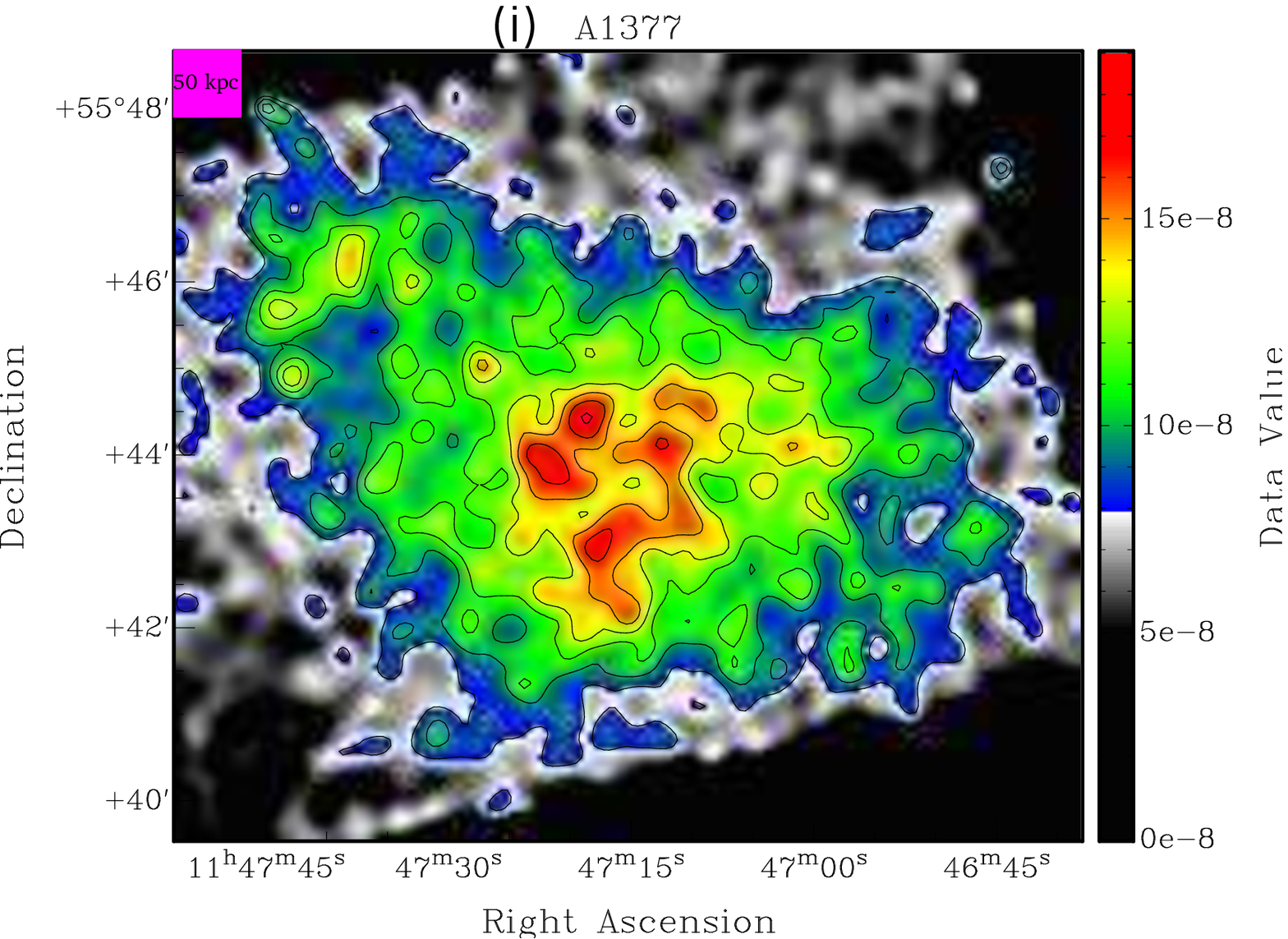}
\includegraphics[height=2.3in]{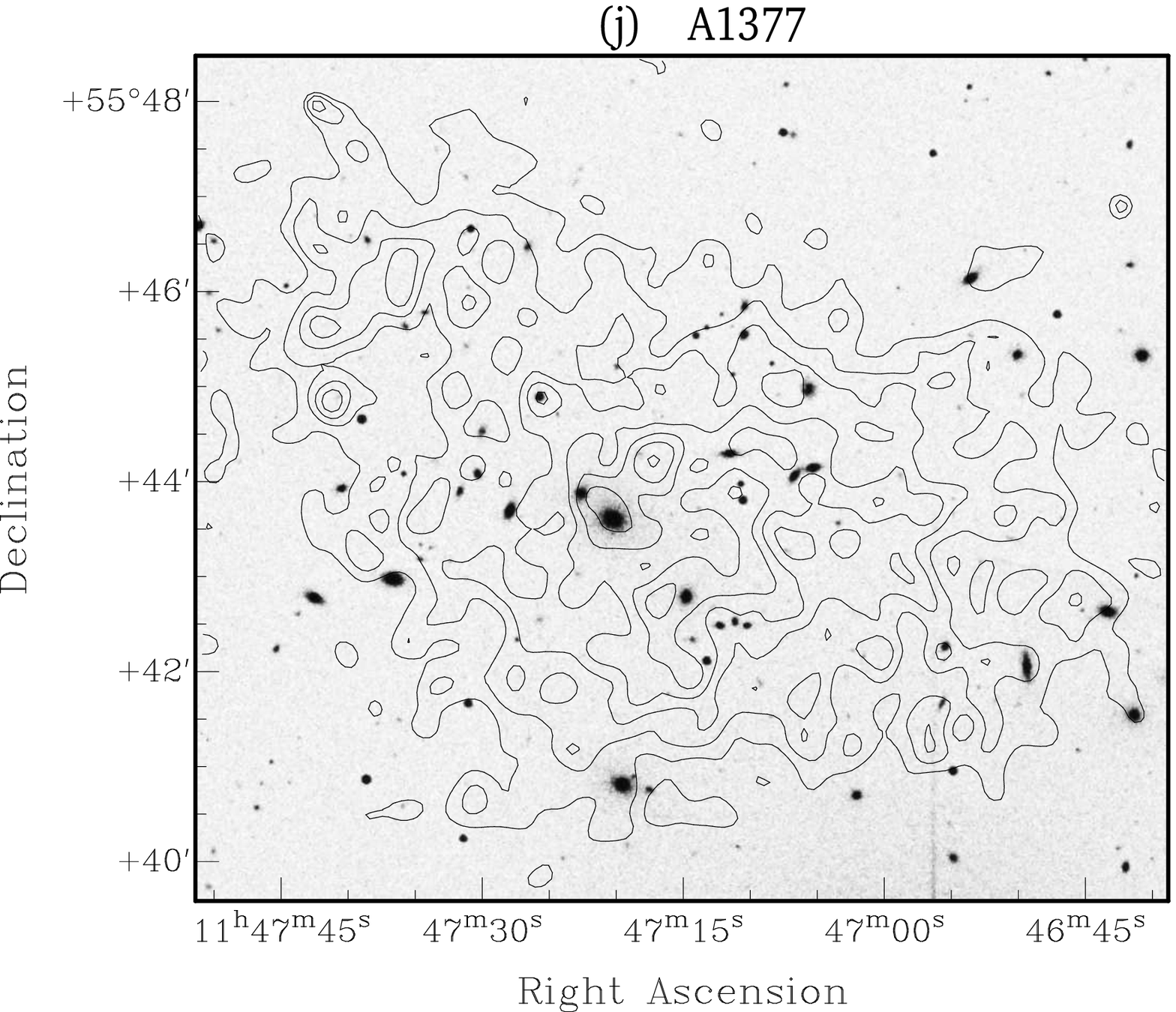}\\
\includegraphics[height=2.45in]{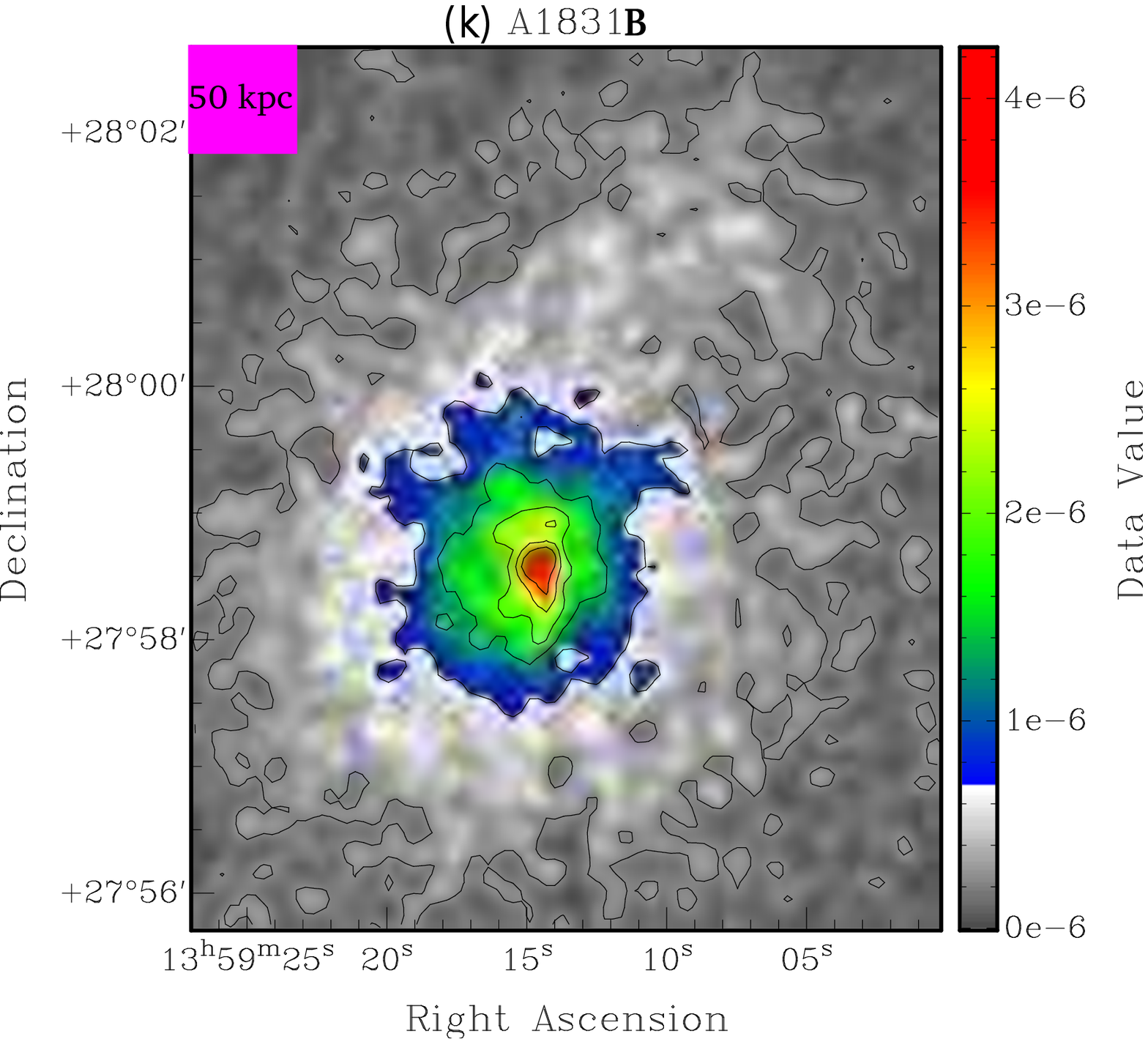}
\includegraphics[height=2.45in]{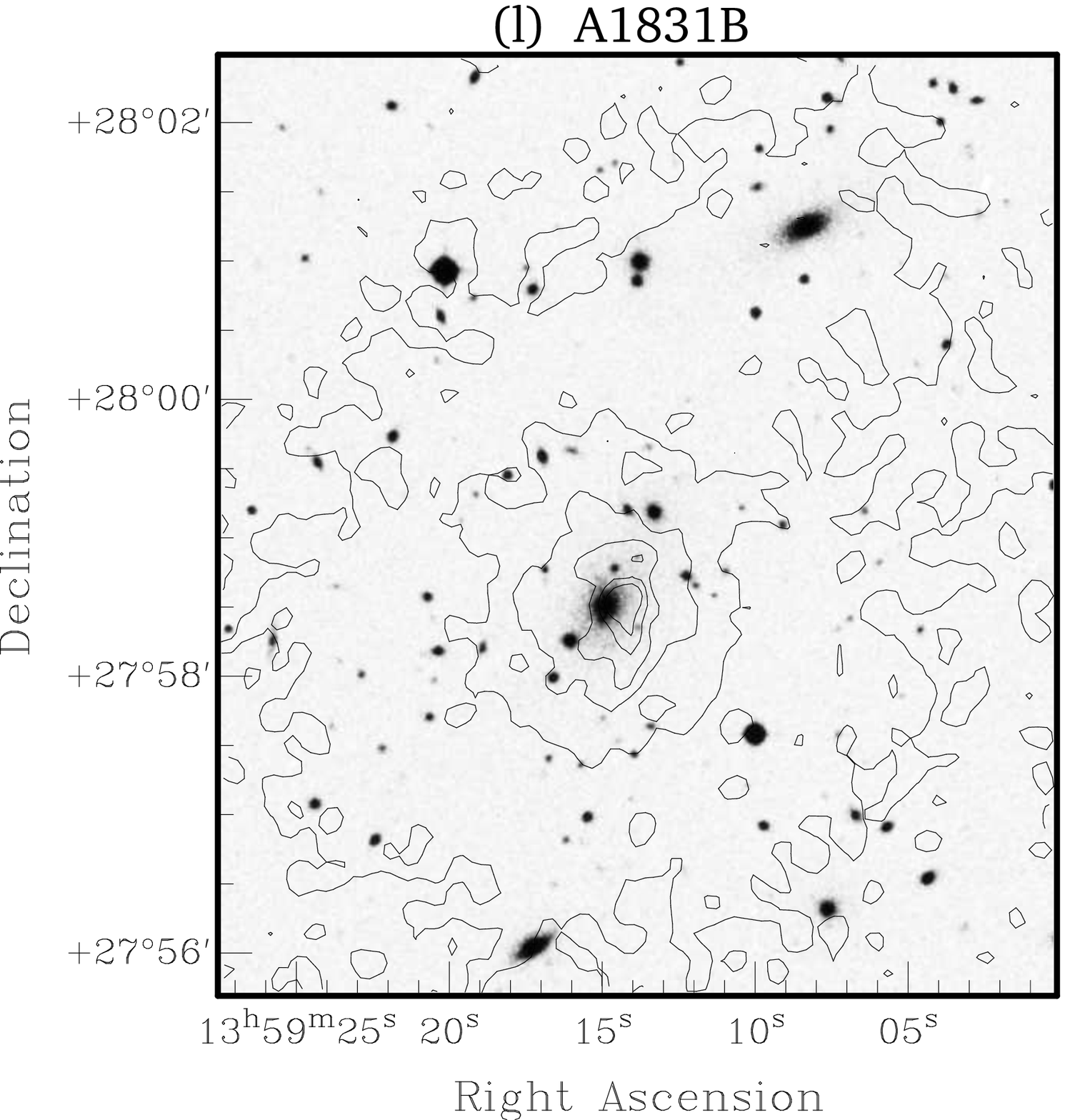}\\
\includegraphics[height=2.3in]{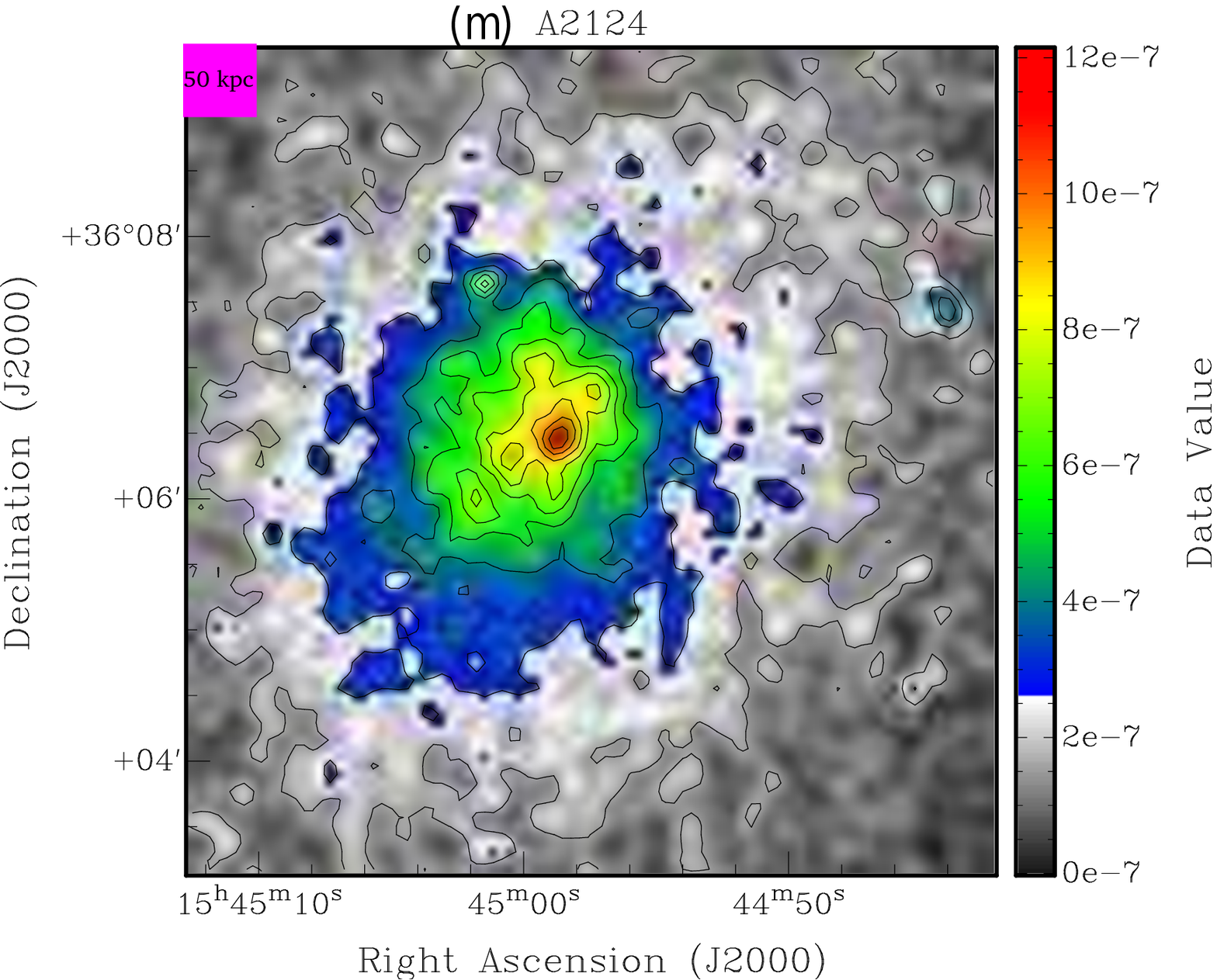}
\includegraphics[height=2.3in]{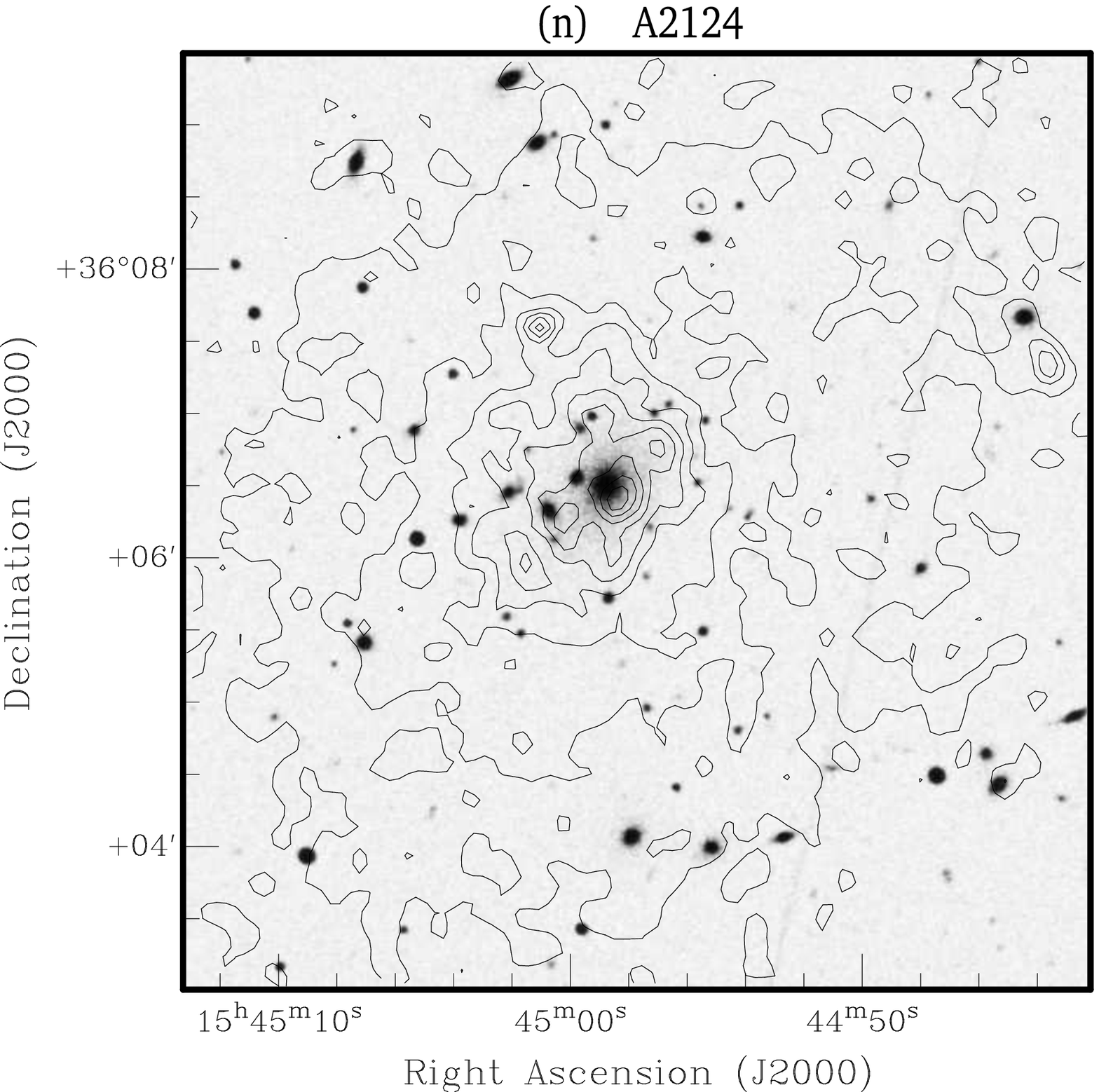}\\
\includegraphics[height=2.05in]{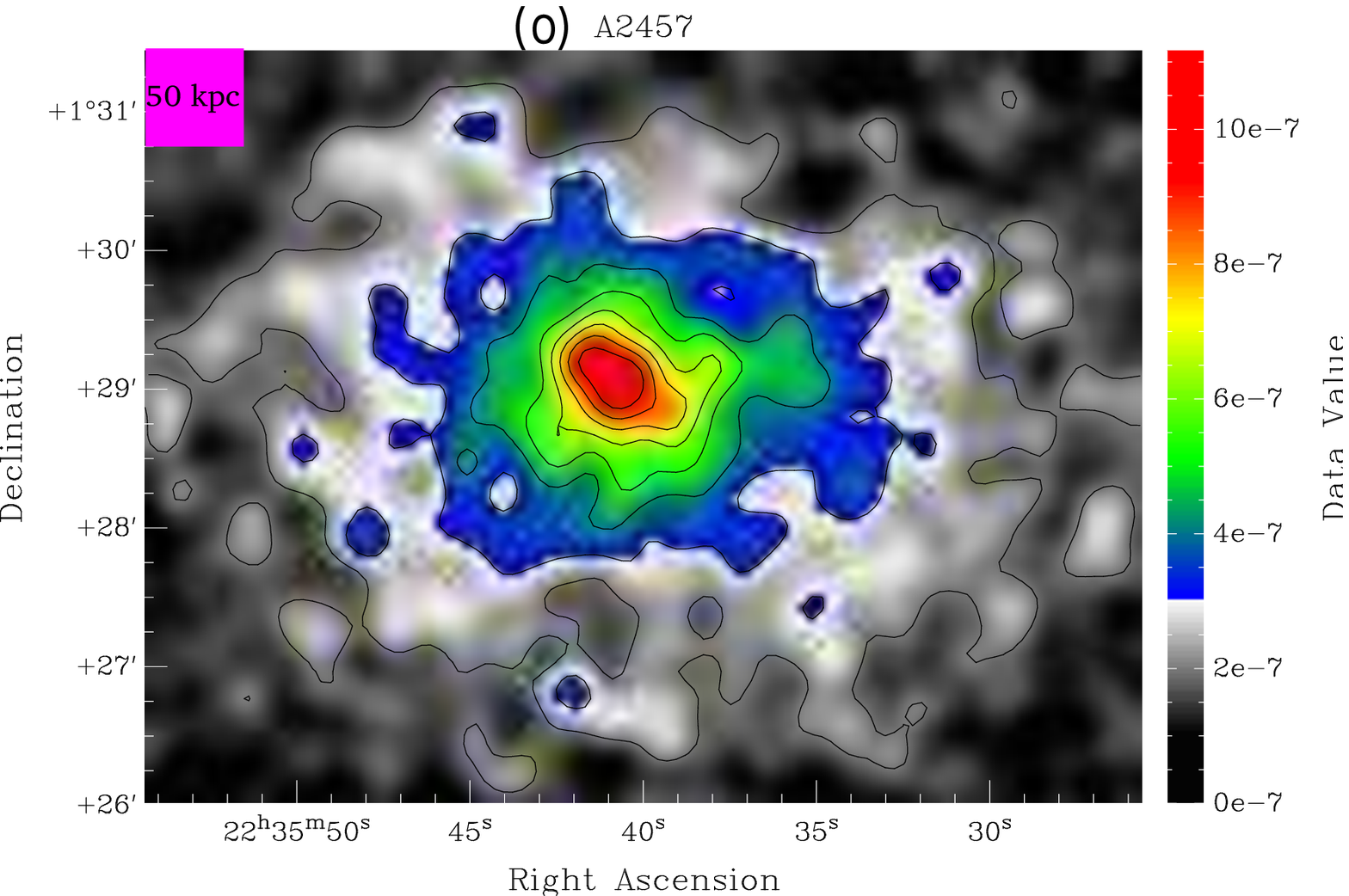}
\includegraphics[height=2.05in]{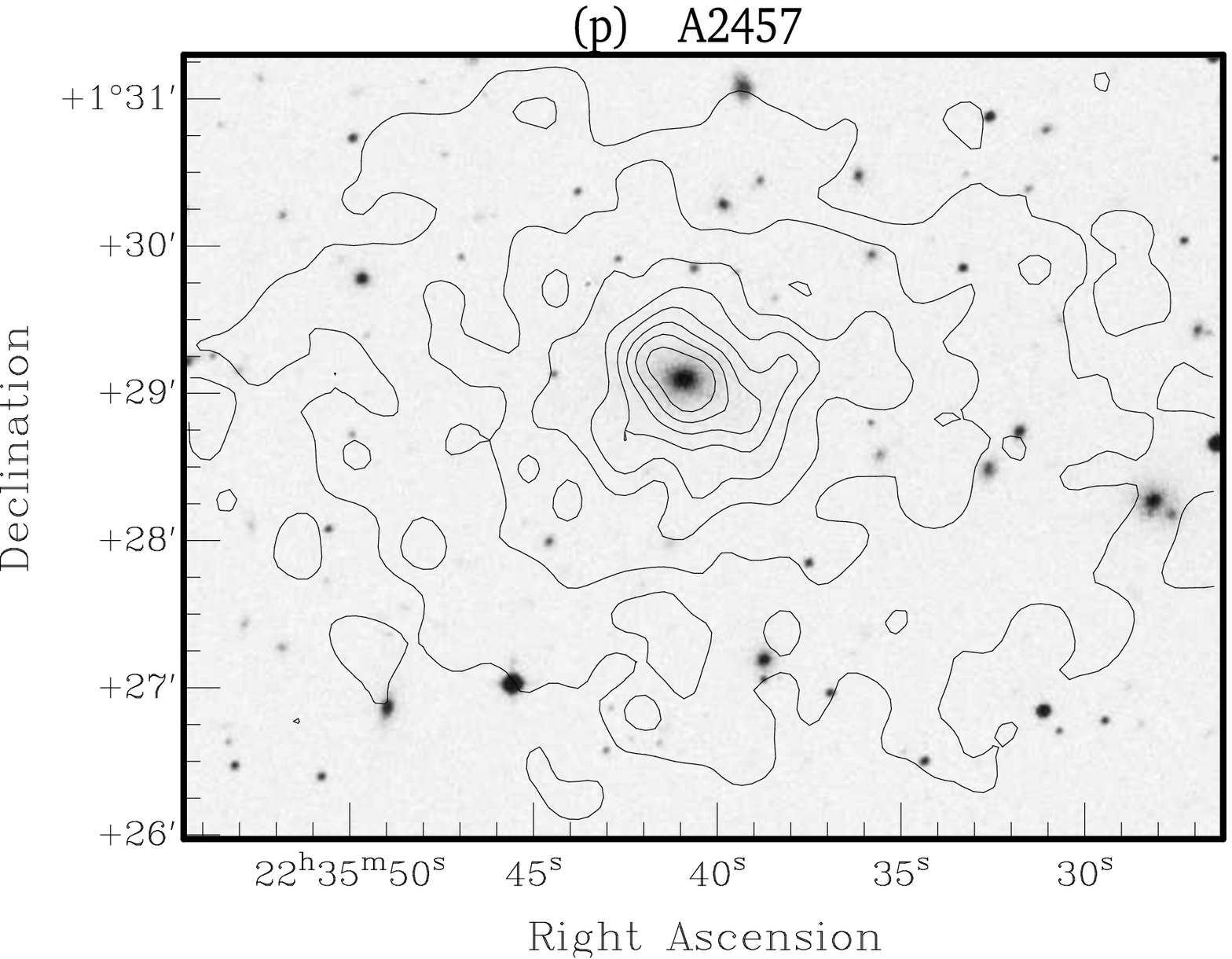}\\
\includegraphics[height=2.25in]{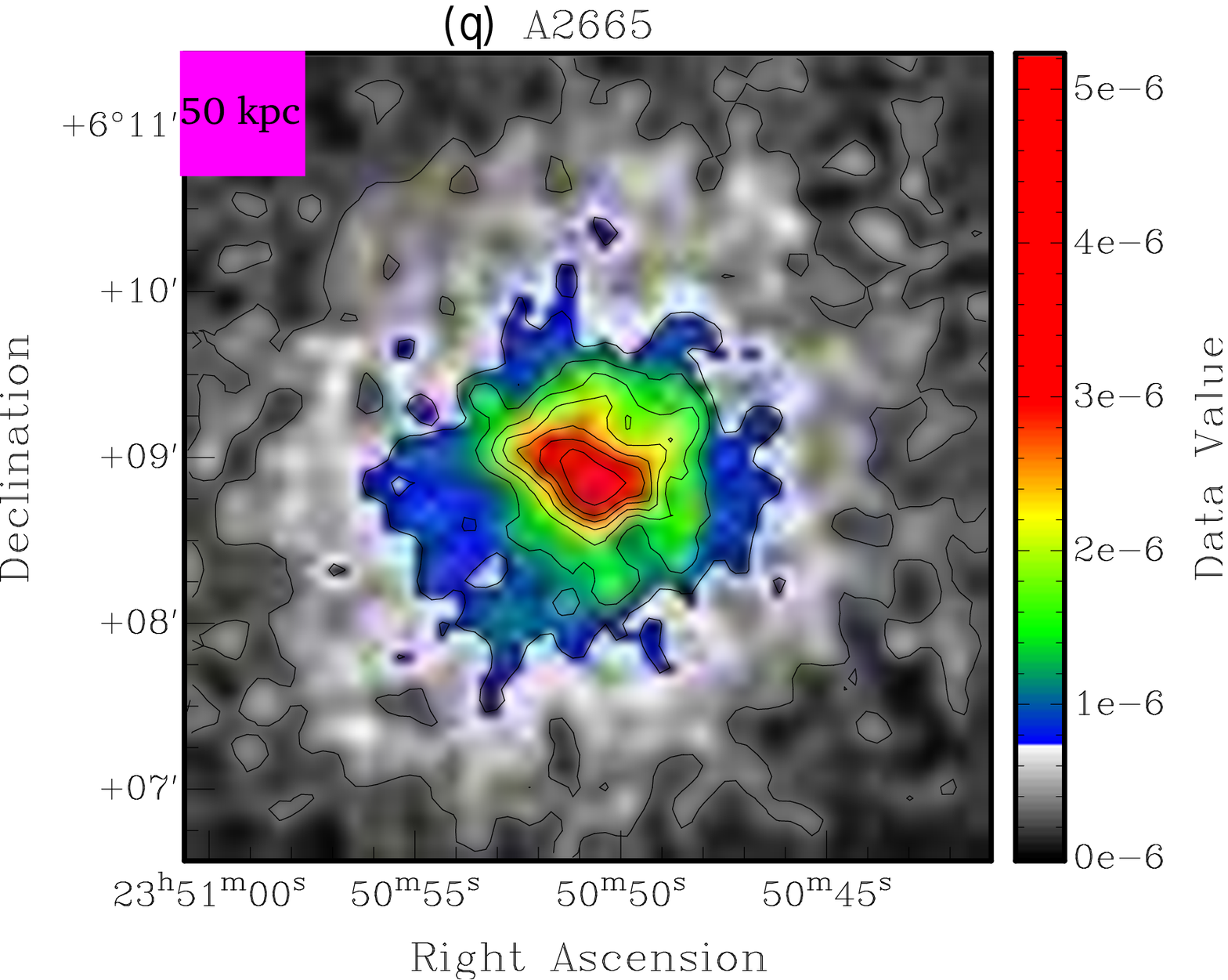}
\includegraphics[height=2.25in]{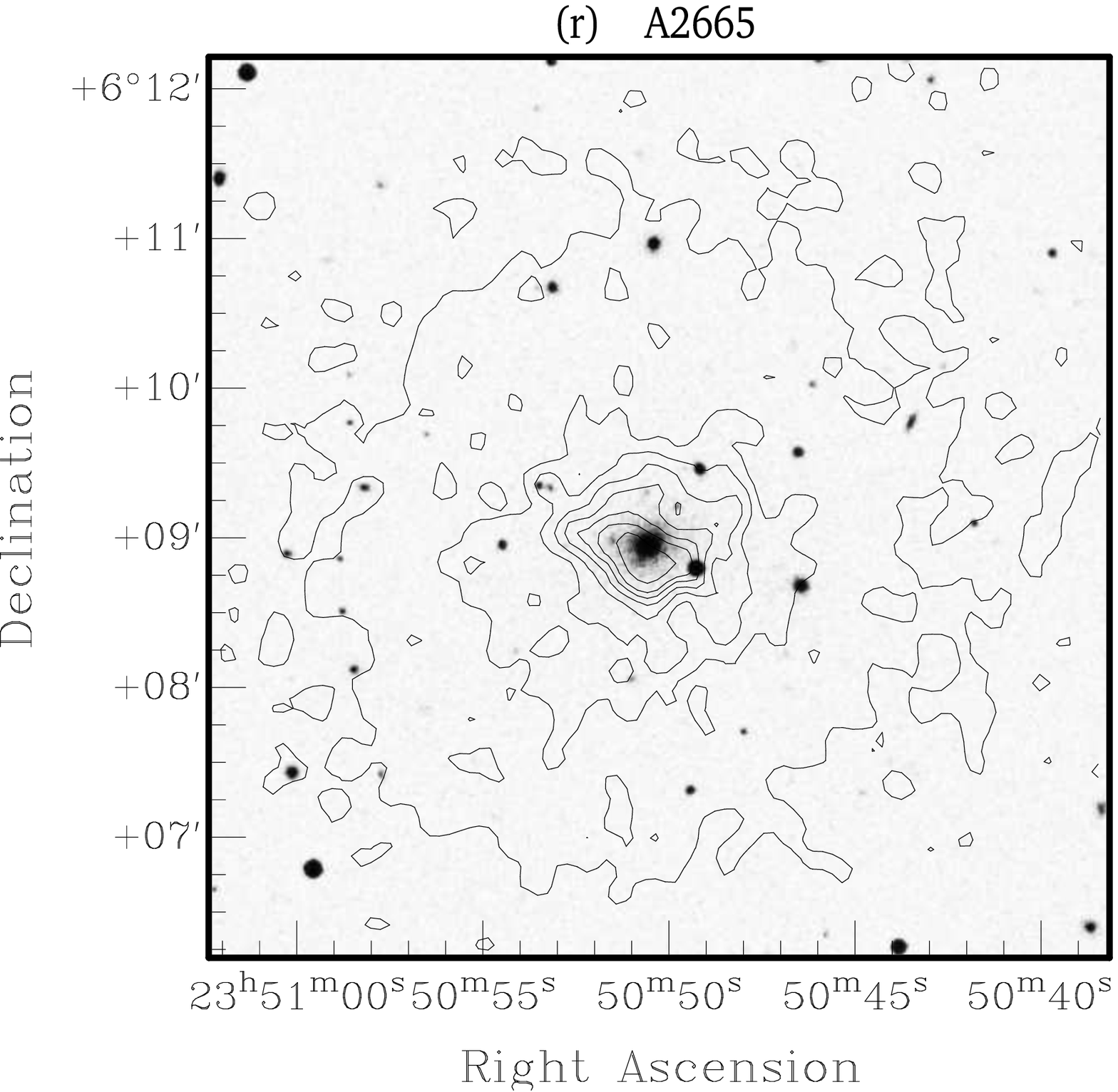}\\
\includegraphics[height=2.15in]{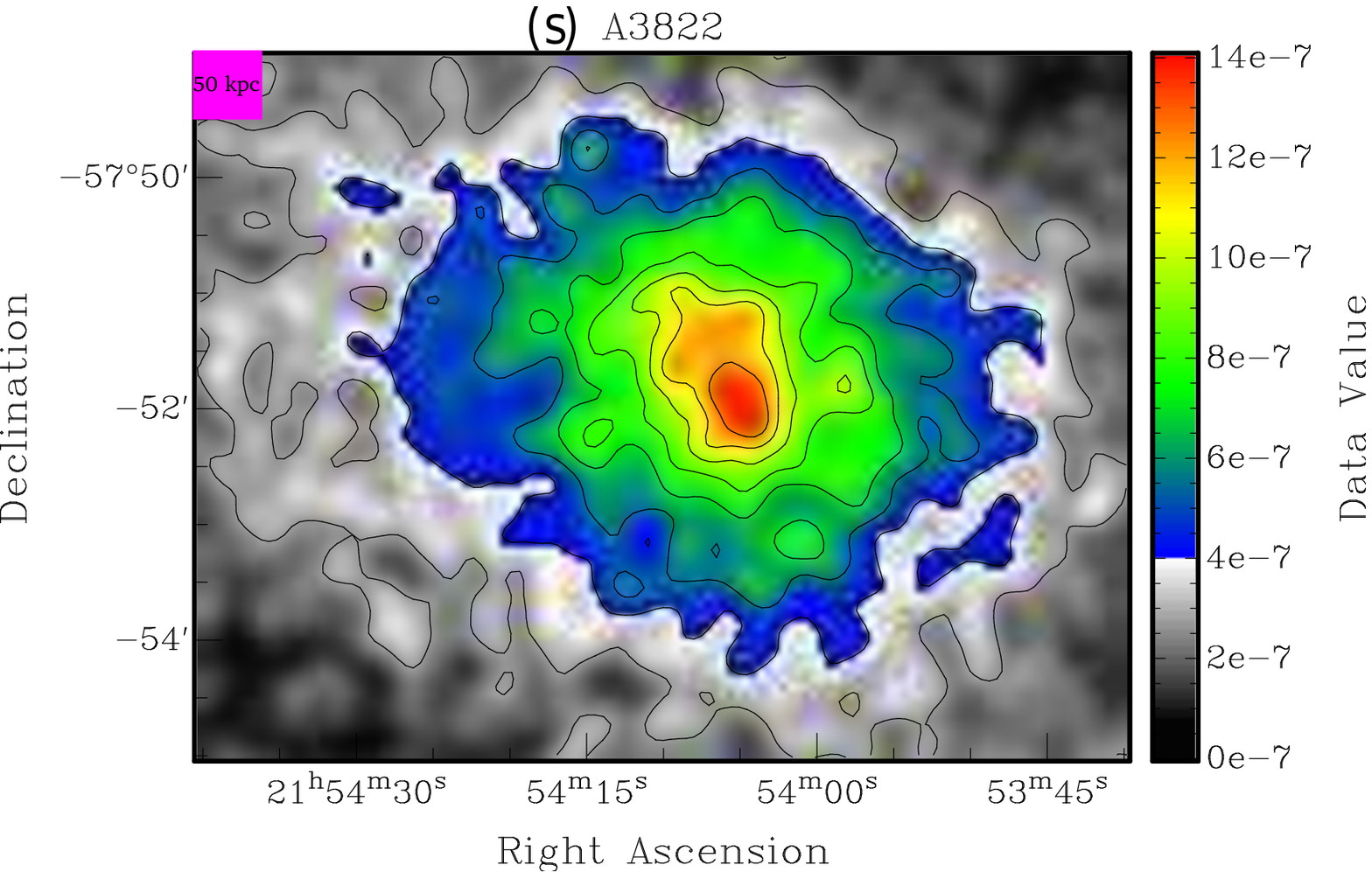}
\includegraphics[height=2.15in]{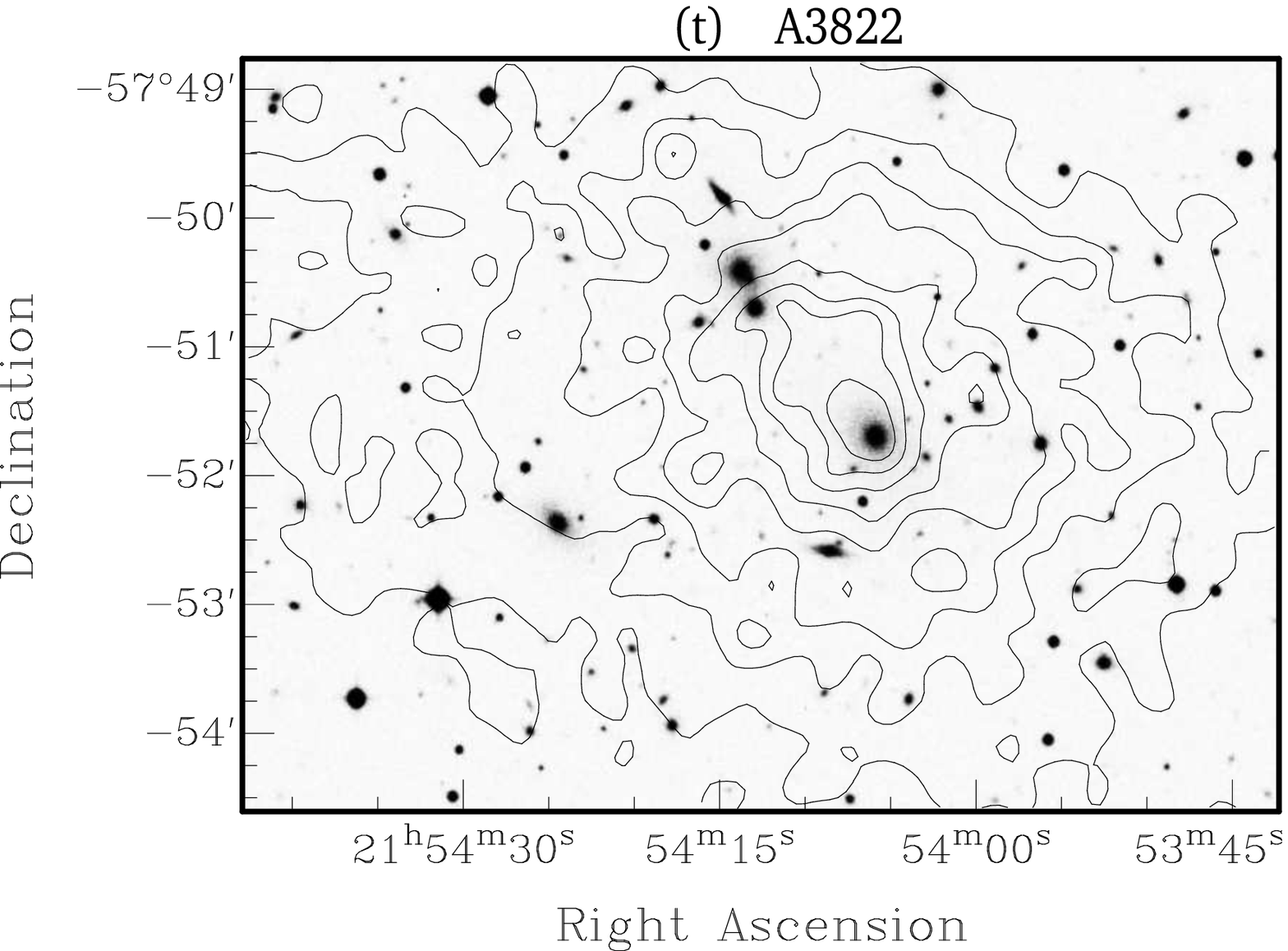}\\
 \end{longtable}%
  \captionof{figure}{\small{(a), (c), (e), (g), (i), (k), (m), (o), (q), and (s): \textit{Chandra} X-ray image of the clusters A193, A376, 
A539, A970, A1377, A1831B, A2124, A2457, A2665, and A3822, smoothed with Gaussian kernels of width 4$^{\prime\prime}$, 8$^{\prime\prime}$, 
8$^{\prime\prime}$, 8$^{\prime\prime}$, 8$^{\prime\prime}$, 4$^{\prime\prime}$, 4$^{\prime\prime}$, 8$^{\prime\prime}$, 4$^{\prime\prime}$, 
and 8$^{\prime\prime}$, respectively. The contour levels are distributed from 4$\sigma$ to 25$\sigma$, 
3$\sigma$ to 28$\sigma$, 3$\sigma$ to 59$\sigma$, 3$\sigma$ to 60$\sigma$, 3$\sigma$ to 13$\sigma$, 3$\sigma$ to 78$\sigma$, 
3$\sigma$ to 32$\sigma$, 3$\sigma$ to 42$\sigma$, 5$\sigma$ to 105$\sigma$, and 3$\sigma$ to 19$\sigma$ 
above the mean background, for the clusters A193, A376, A539, A970, A1377, A1831B, A2124, A2457, A2665, and A3822, 
respectively. The scale is expressed in units of counts s$^{-1}$ cm$^{-2}$ pixel$^{-1}$. (b), (d), (f), (h), (j), (l), (n), (p), (r), and (t): 
Optical images of the clusters A193, A376, A539, A970, A1377, A1831B, A2124, A2457, A2665, and A3822, from the SuperCOSMOS 
survey overlaid with the X-ray contours from the LHS. The pink boxes on the top left corner of the images on the left side 
mark a 50 kpc$\times$50 kpc region.}\label{fig:cont1}}%
  \addtocounter{table}{-1}%
\end{center}

\clearpage

\begin{center}
  \begin{longtable}{cc}
\includegraphics[totalheight=2.2in]{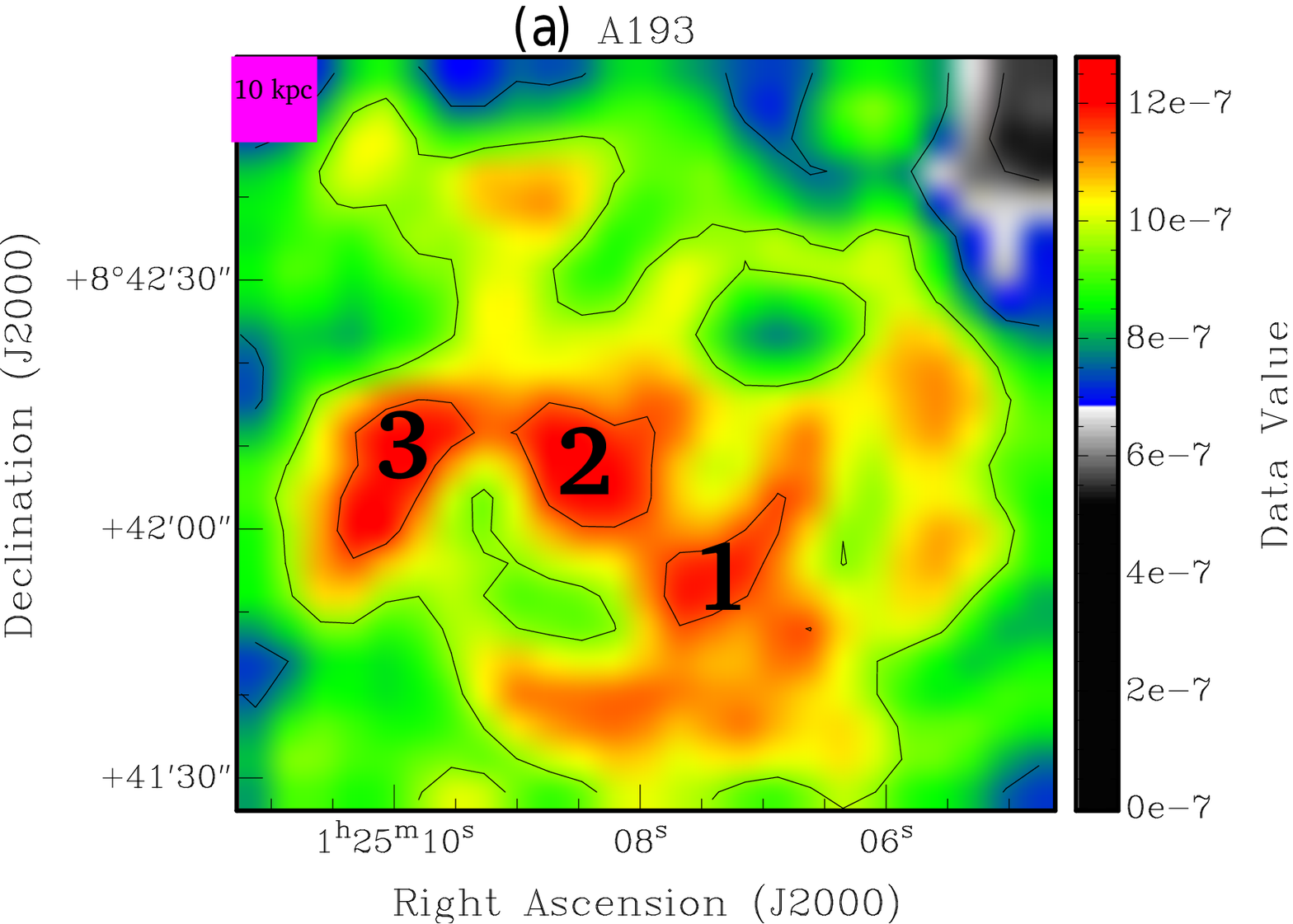}
\includegraphics[totalheight=2.2in]{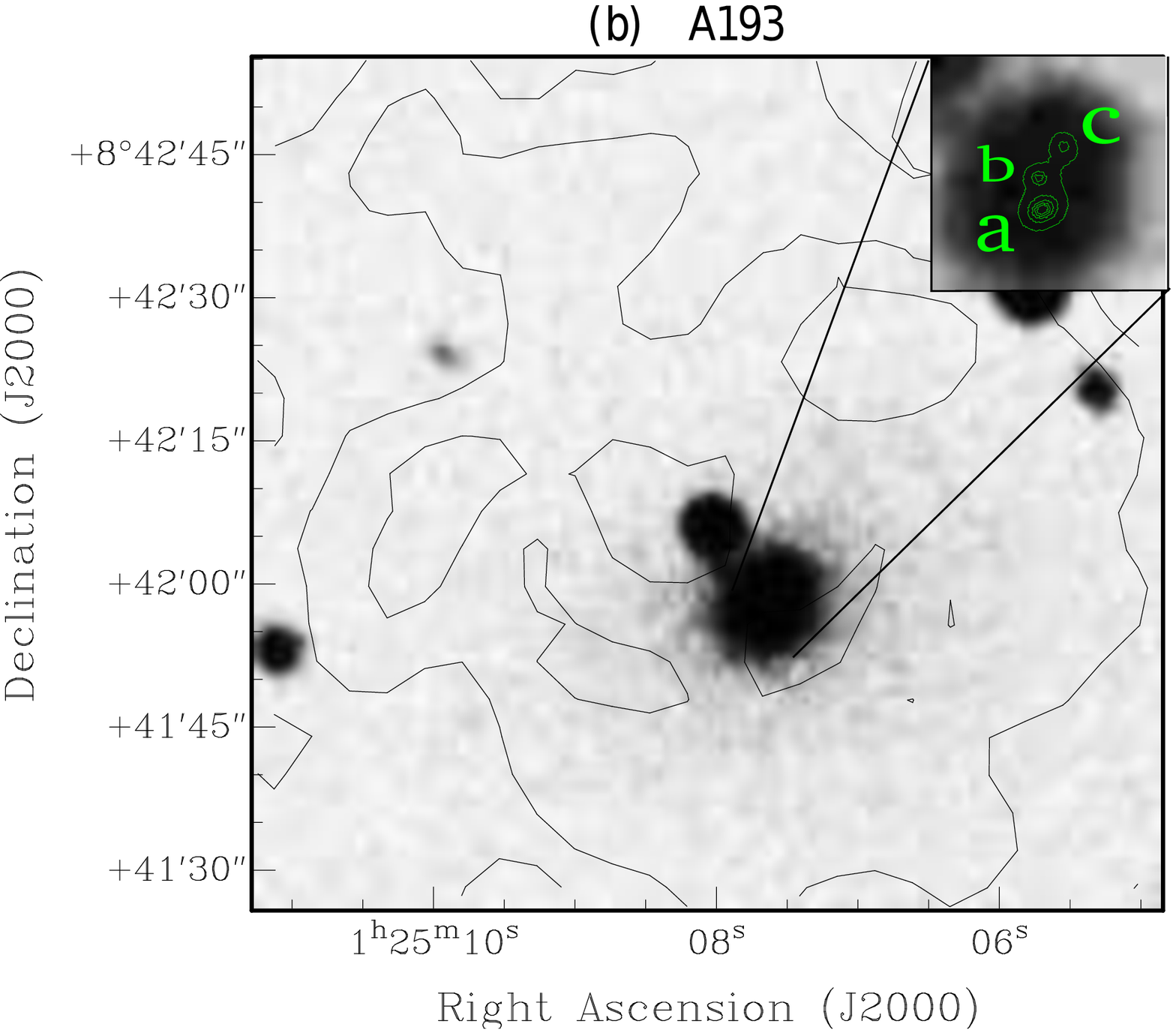}\\
\includegraphics[totalheight=2.1in]{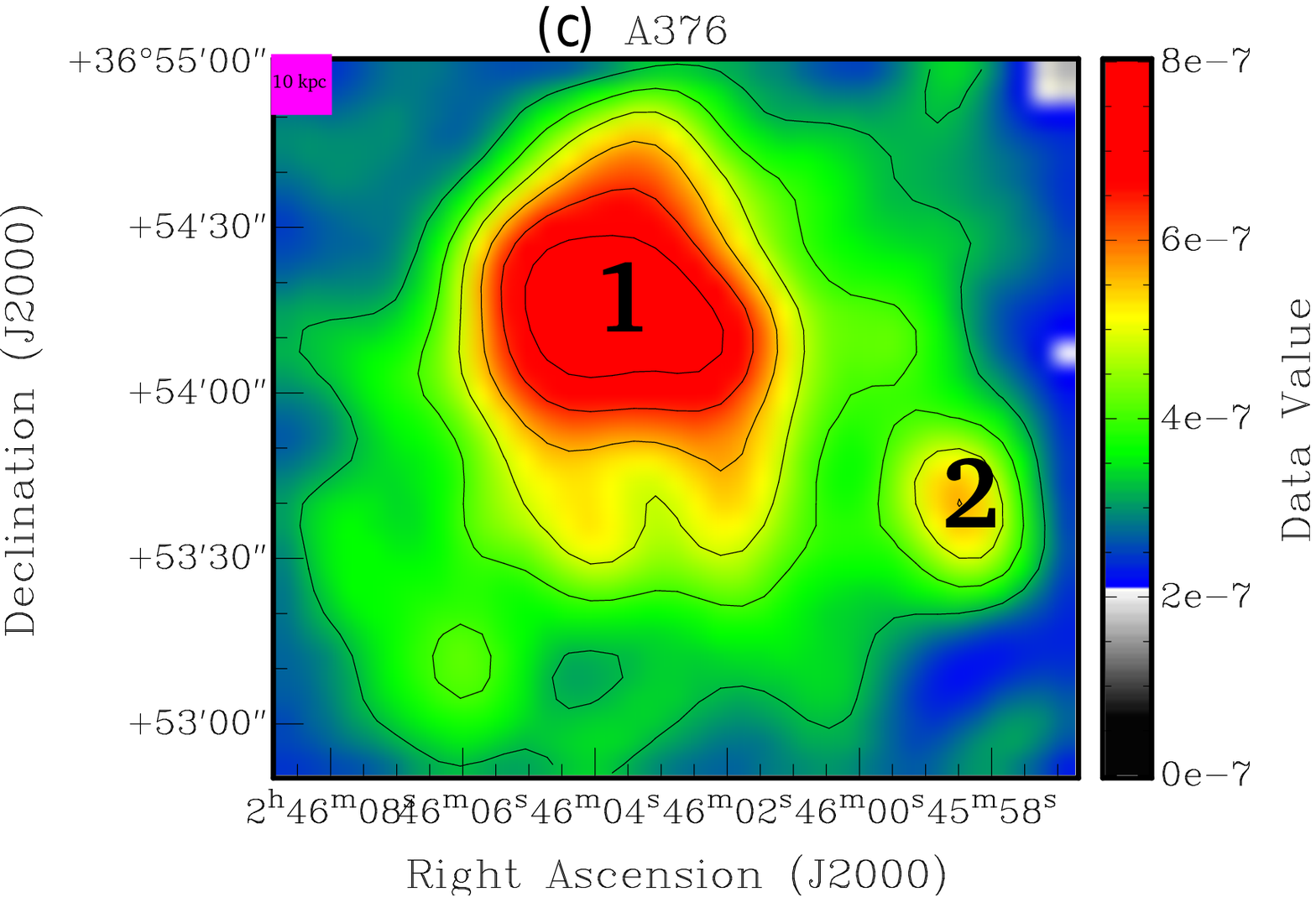}
\includegraphics[totalheight=2.1in]{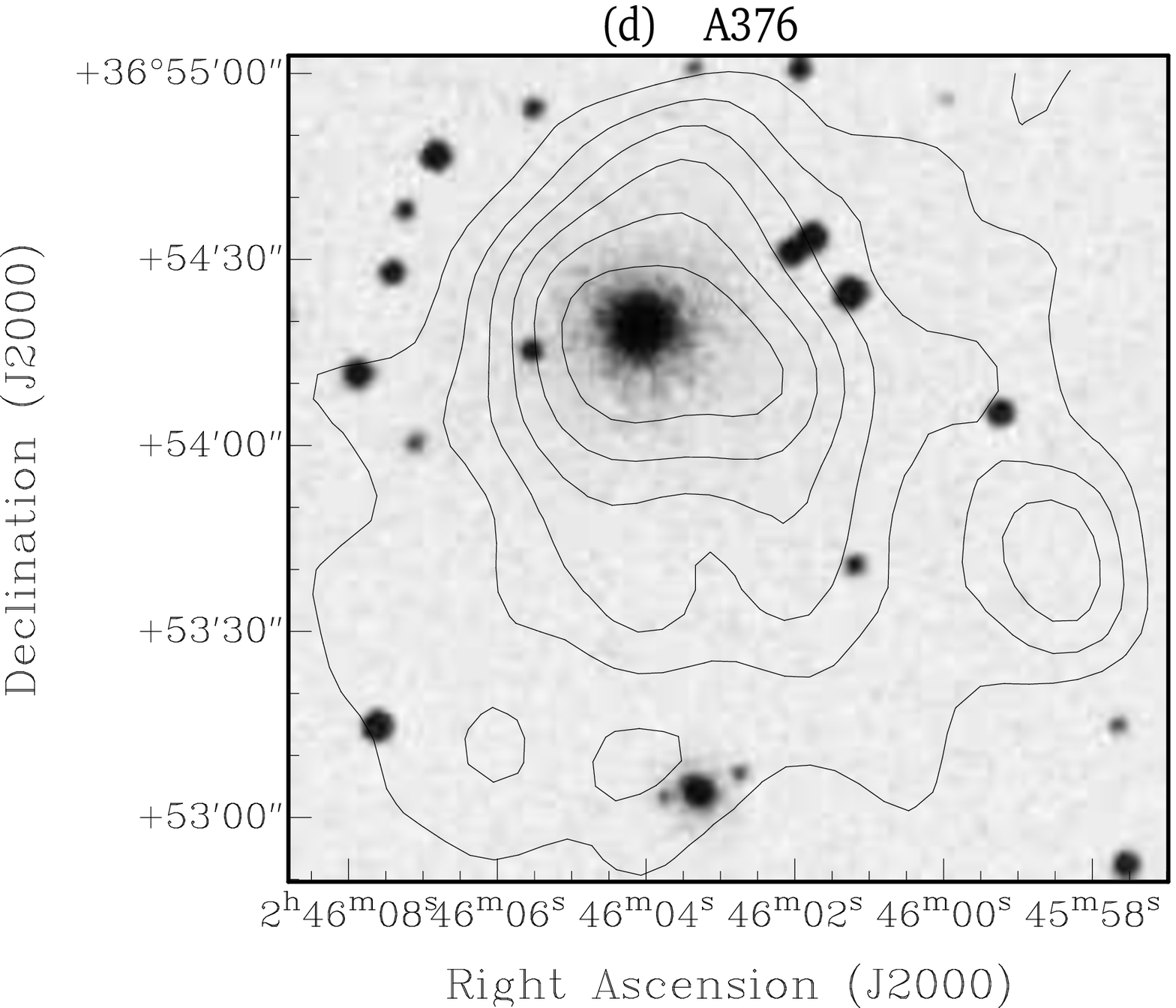}\\
\includegraphics[totalheight=2.5in]{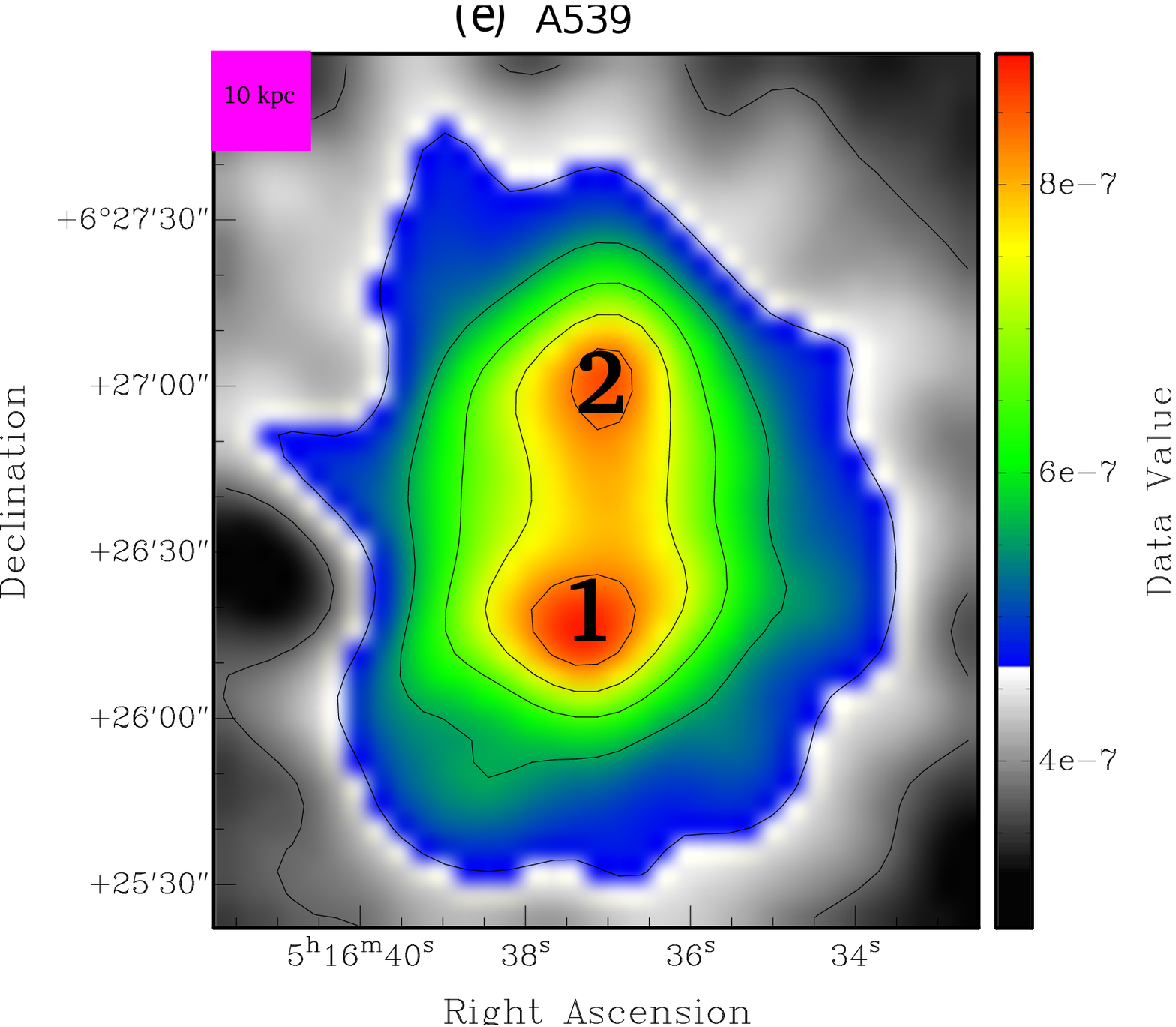}
\includegraphics[totalheight=2.5in]{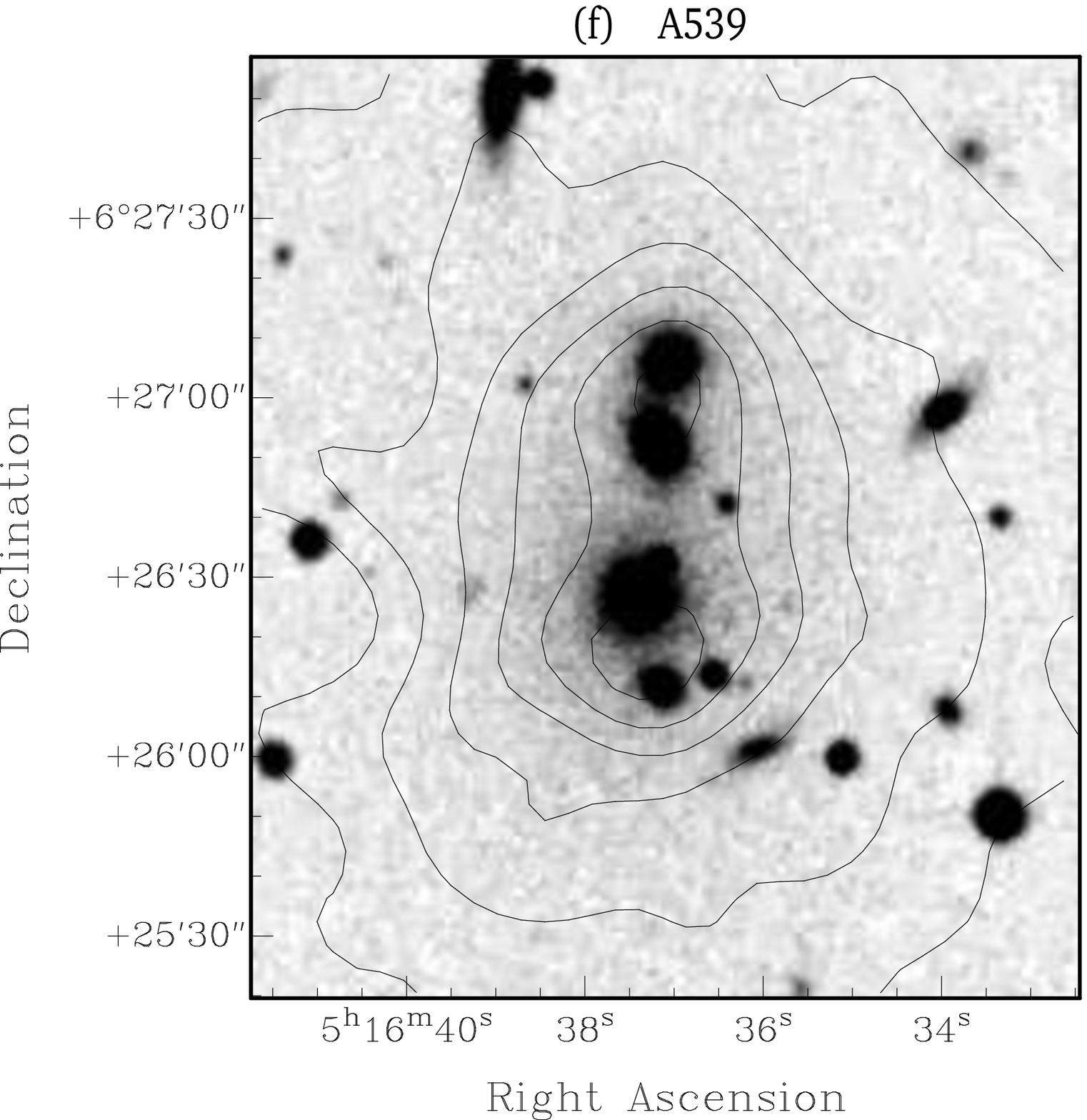}\\\\
\includegraphics[totalheight=1.9in]{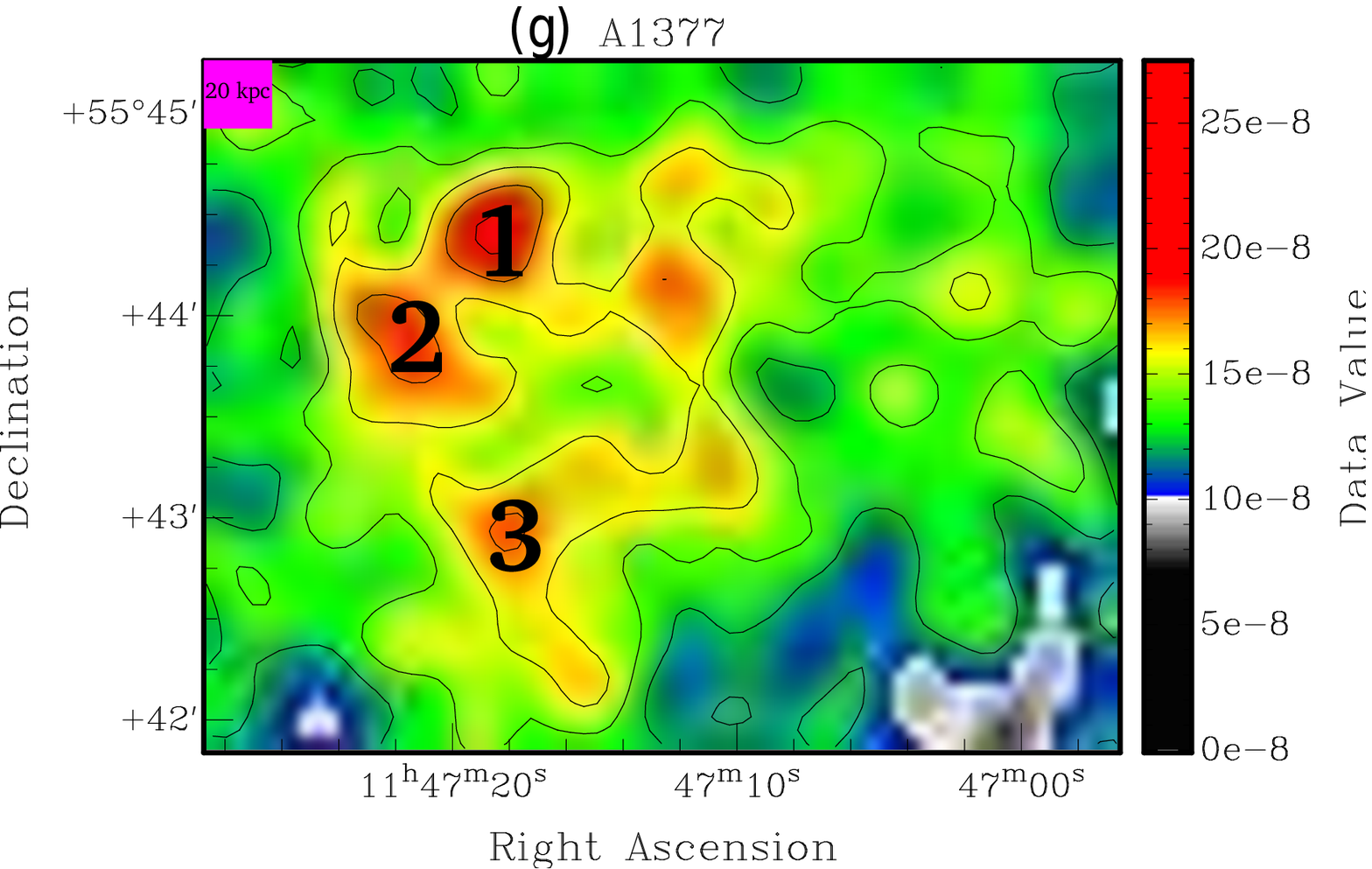}
\includegraphics[totalheight=1.9in]{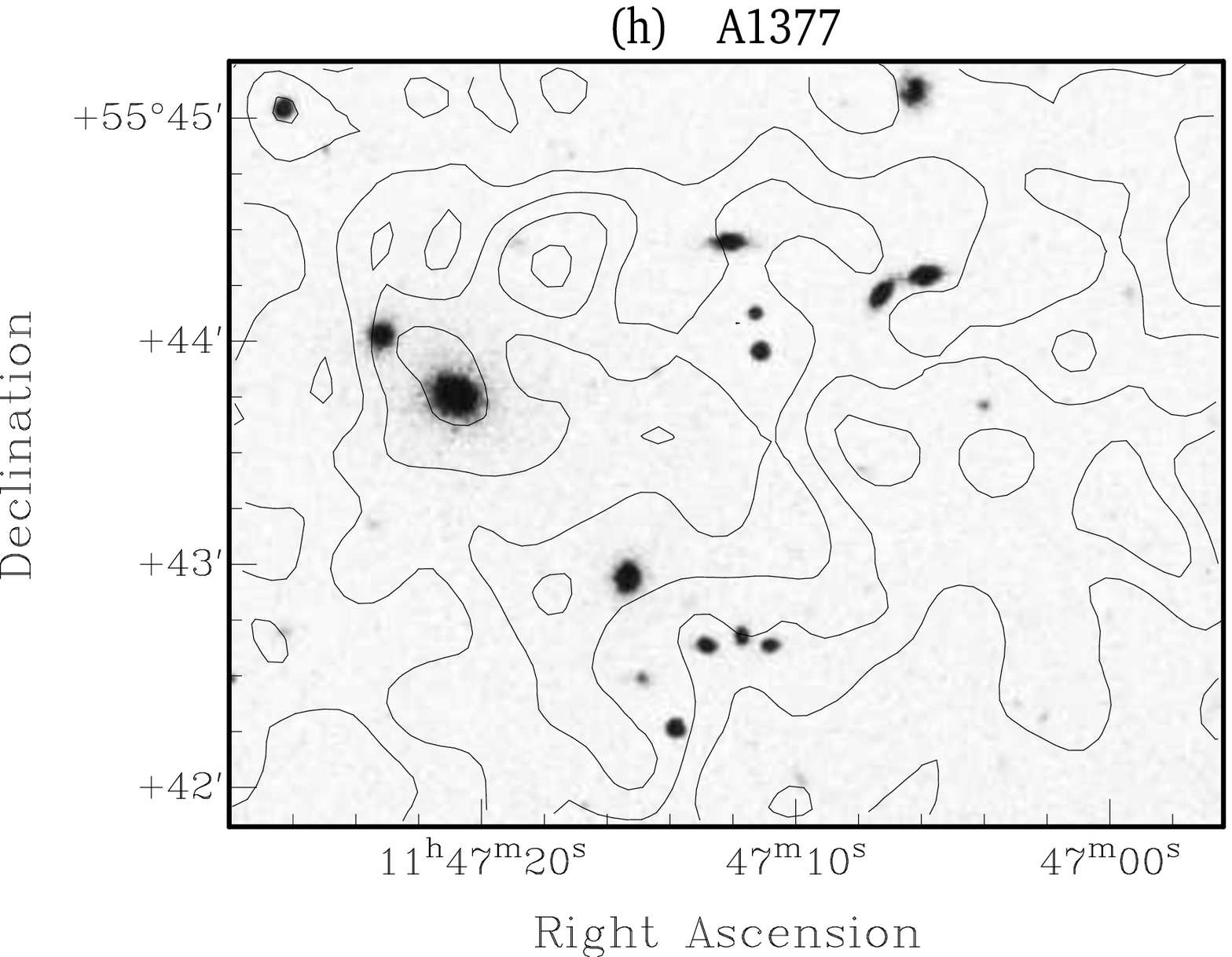}\\
\includegraphics[totalheight=2.4in]{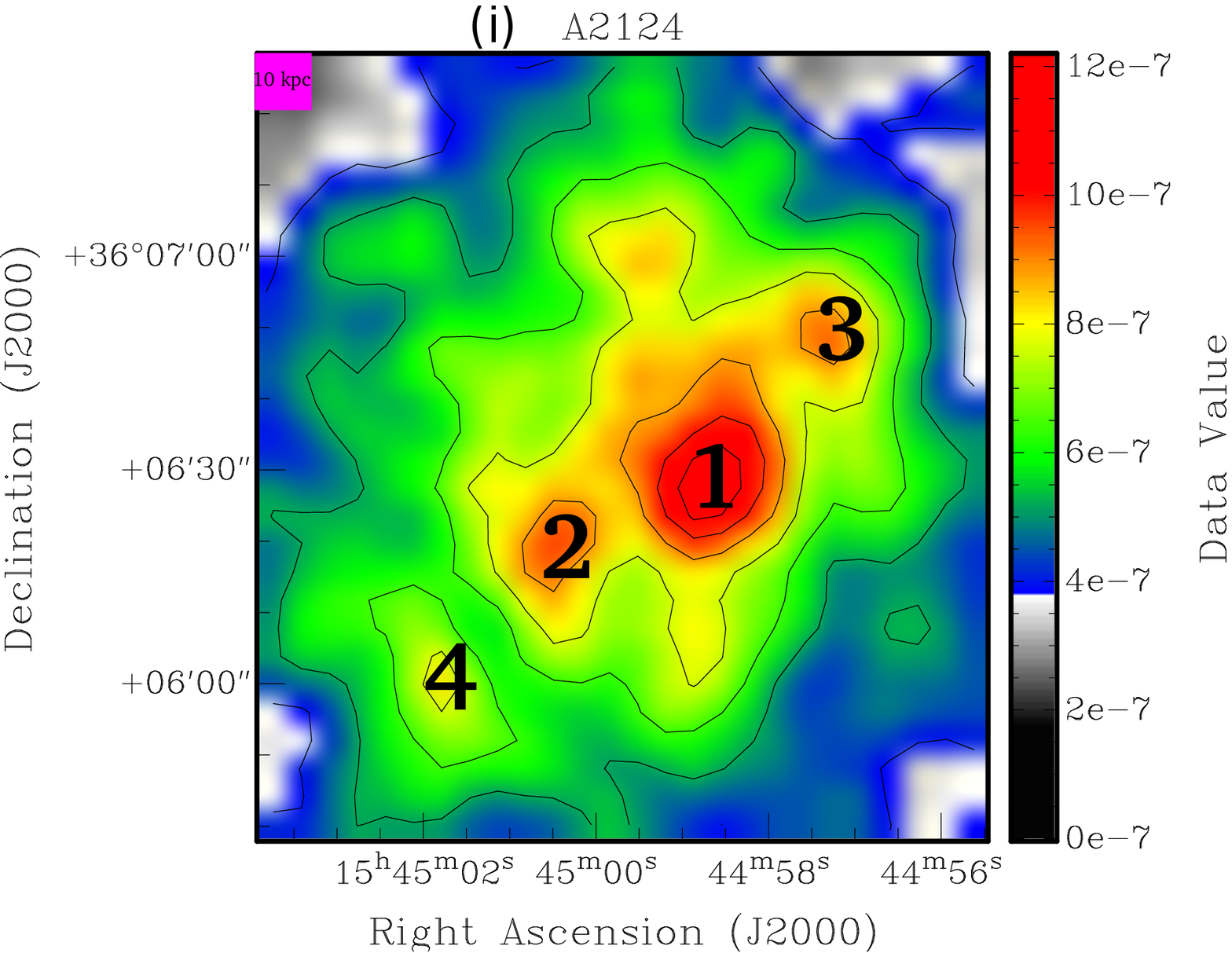}
\includegraphics[totalheight=2.4in]{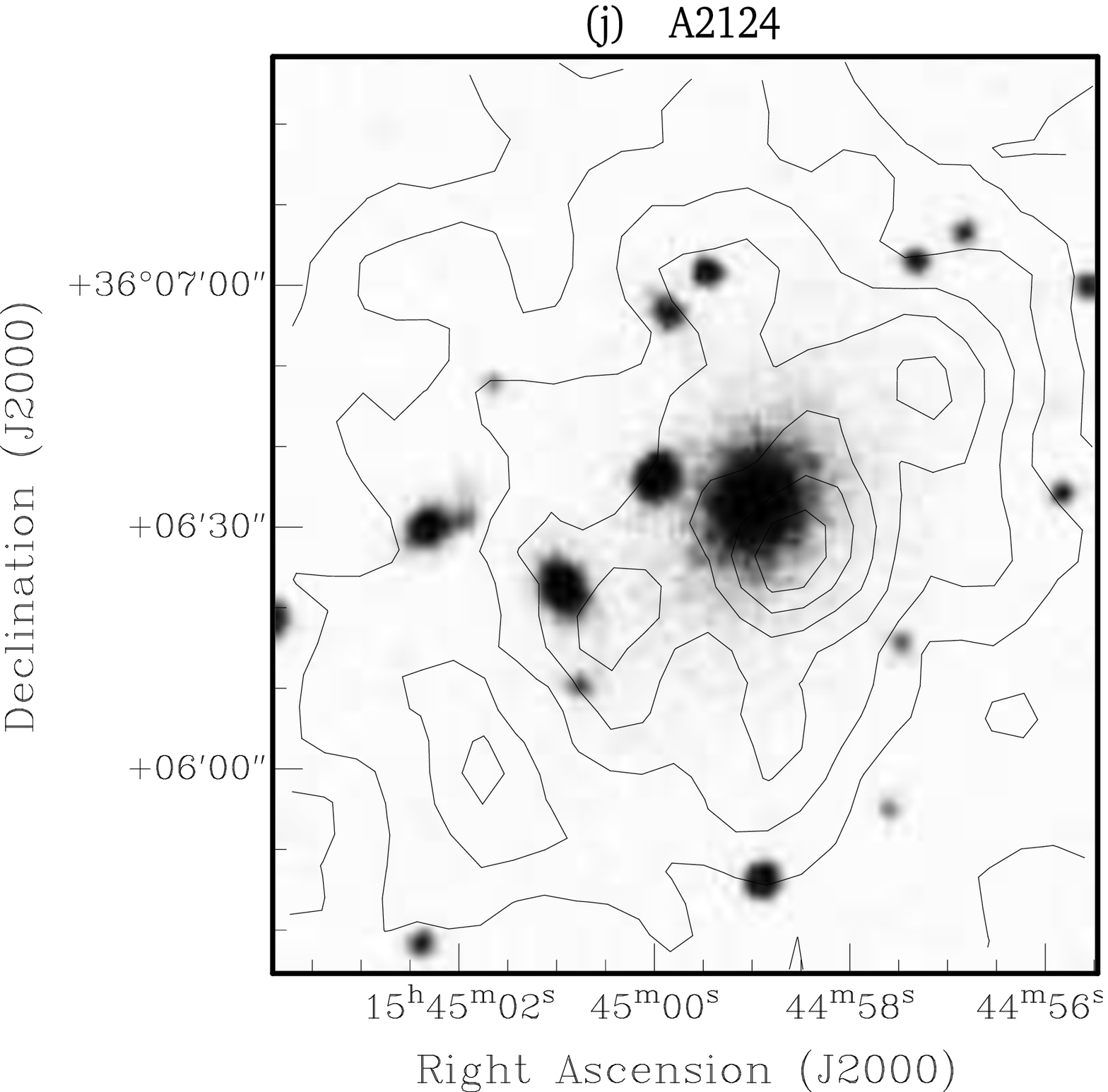}\\
\includegraphics[totalheight=2.3in]{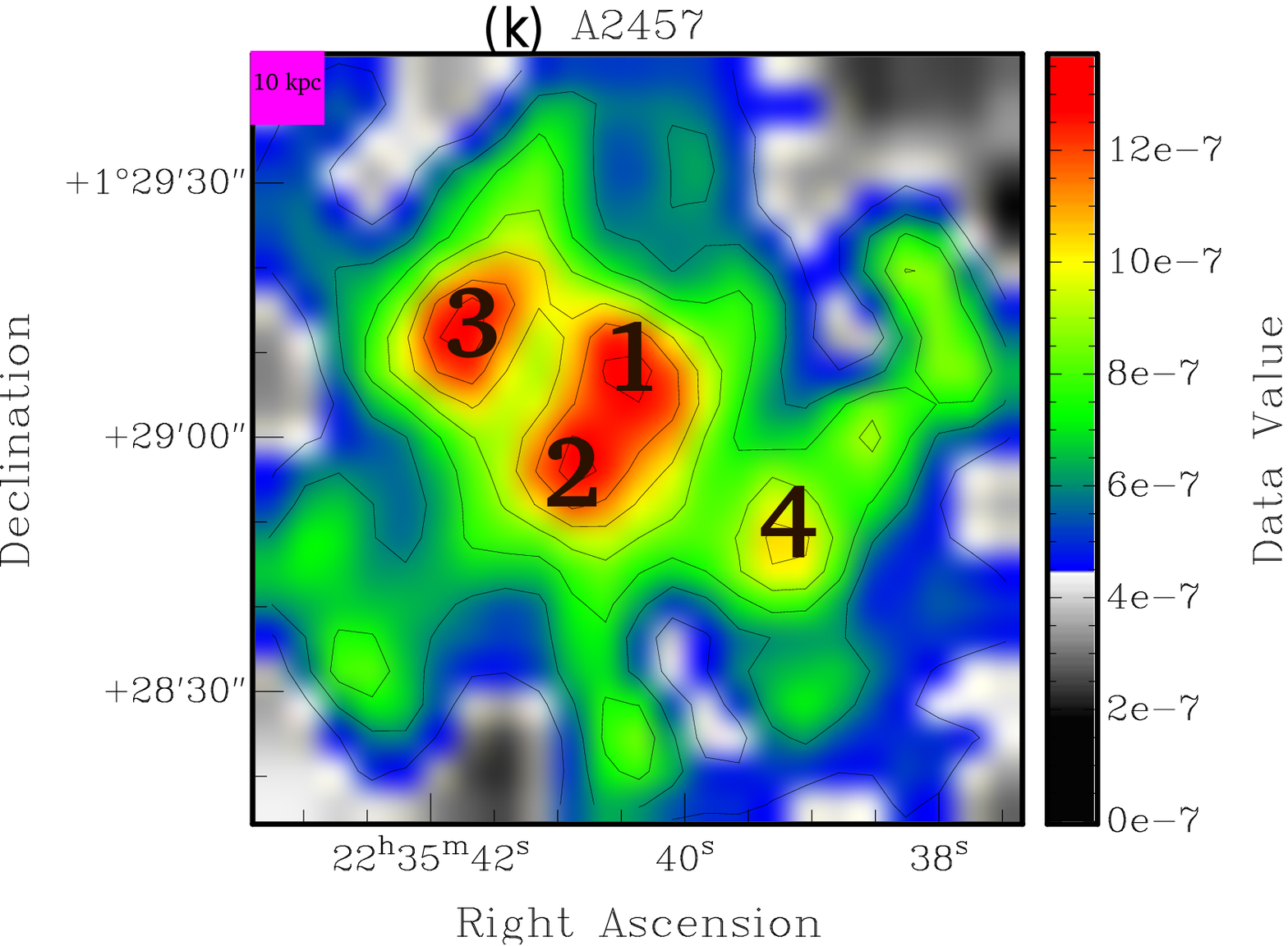}
\includegraphics[totalheight=2.3in]{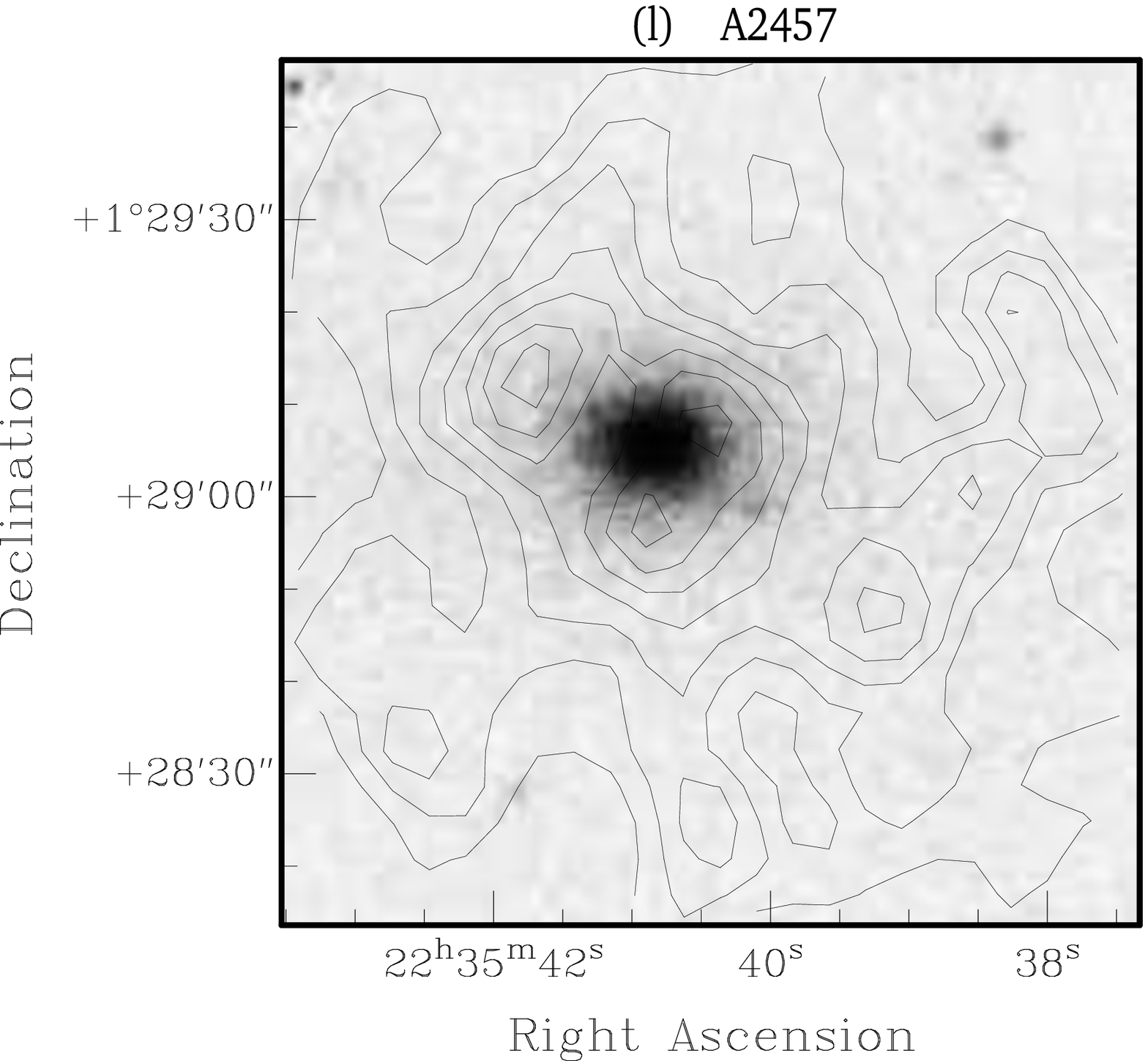}
 \end{longtable}%
  \captionof{figure}{\small{(a), (c), (e), (g), (i) and (k) : \textit{Chandra} X-ray images of the central parts of the clusters A193, A376, 
A539, A1377 and A2124 and A2457, smoothed with Gaussian kernels of width 4$^{\prime\prime}$, 8$^{\prime\prime}$, 
8$^{\prime\prime}$, 8$^{\prime\prime}$, 4$^{\prime\prime}$ and 4$^{\prime\prime}$ respectively.
The contour levels are distributed from 12$\sigma$ to 25$\sigma$, 12$\sigma$ to 
30$\sigma$, 11$\sigma$ to 89$\sigma$, 7$\sigma$ to 16$\sigma$, 10$\sigma$ to 32$\sigma$ and 7$\sigma$ to 
22$\sigma$ above the mean background, for the clusters A193, A376, A539, A1377, A2124 and A2457, 
respectively. The scales are expressed in units of counts s$^{-1}$ cm$^{-2}$ pixel$^{-1}$. (b), (d), (f), (h), (j) and (l) : Optical images 
of the central parts of the clusters A193, A376, A539, A1377, A2124 and A2457, respectively, from the SuperCOSMOS survey overlaid 
with the X-ray contours from the LHS. The pink boxes on the top left corner of the images on the left side 
mark a 10 kpc$\times$10 kpc region (20 kpc$\times$20 kpc for A1377).}\label{fig:cont2}}%
  \addtocounter{table}{-1}%
\end{center}
  
\clearpage  
  
\begin{center}
\begin{longtable}{cc}
\includegraphics[width=2.8in]{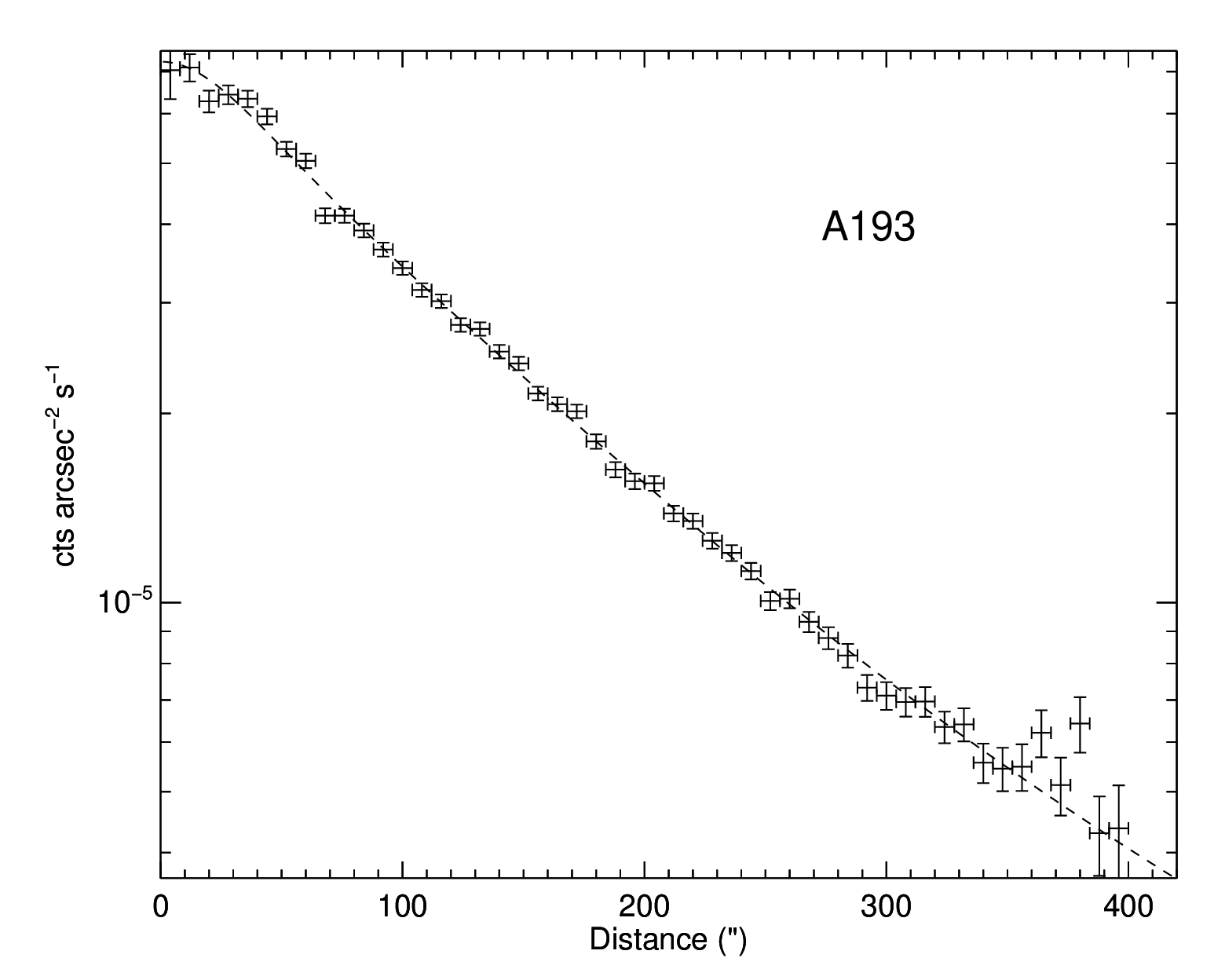}
\includegraphics[width=2.8in]{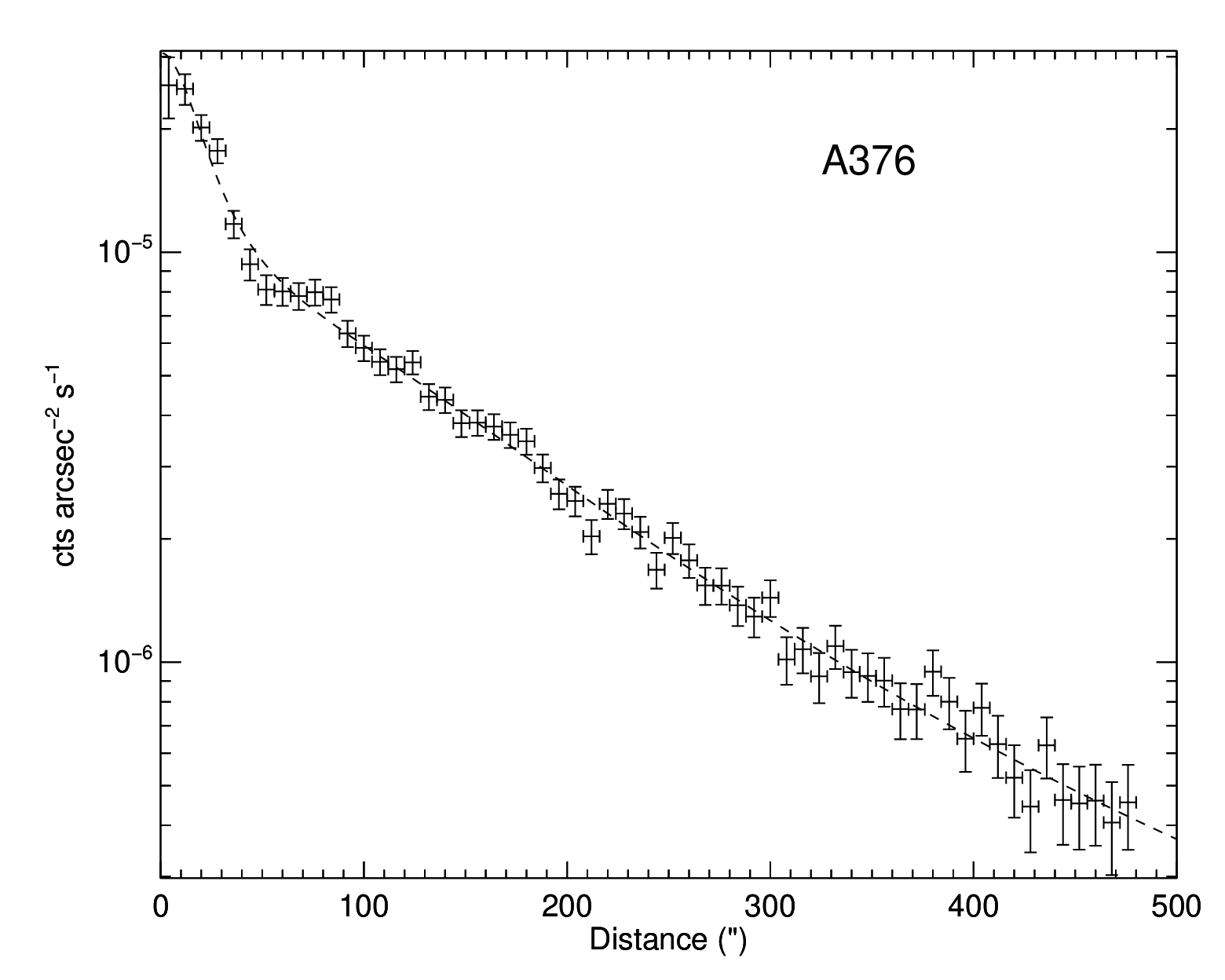}\\
\includegraphics[width=2.8in]{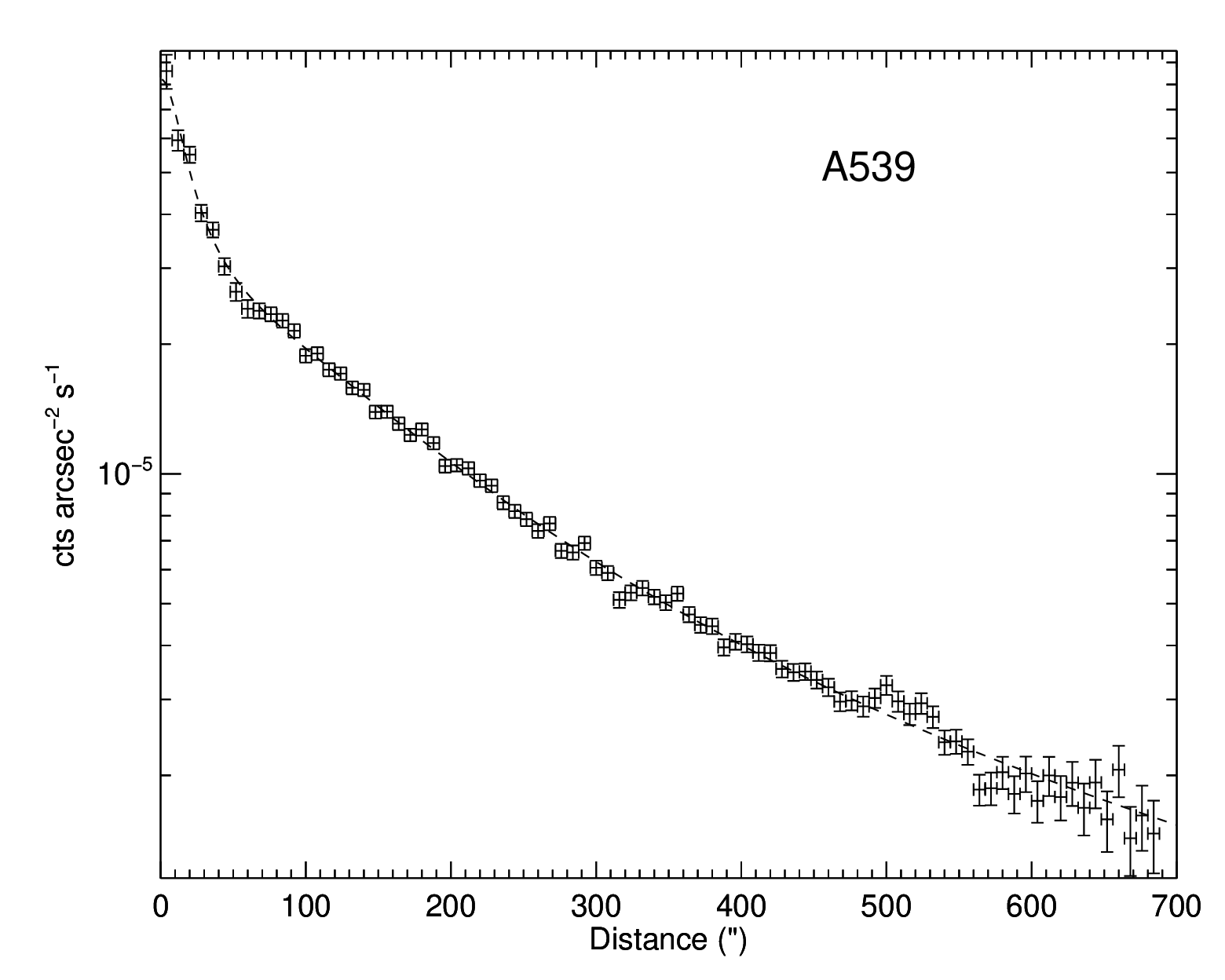}
\includegraphics[width=2.8in]{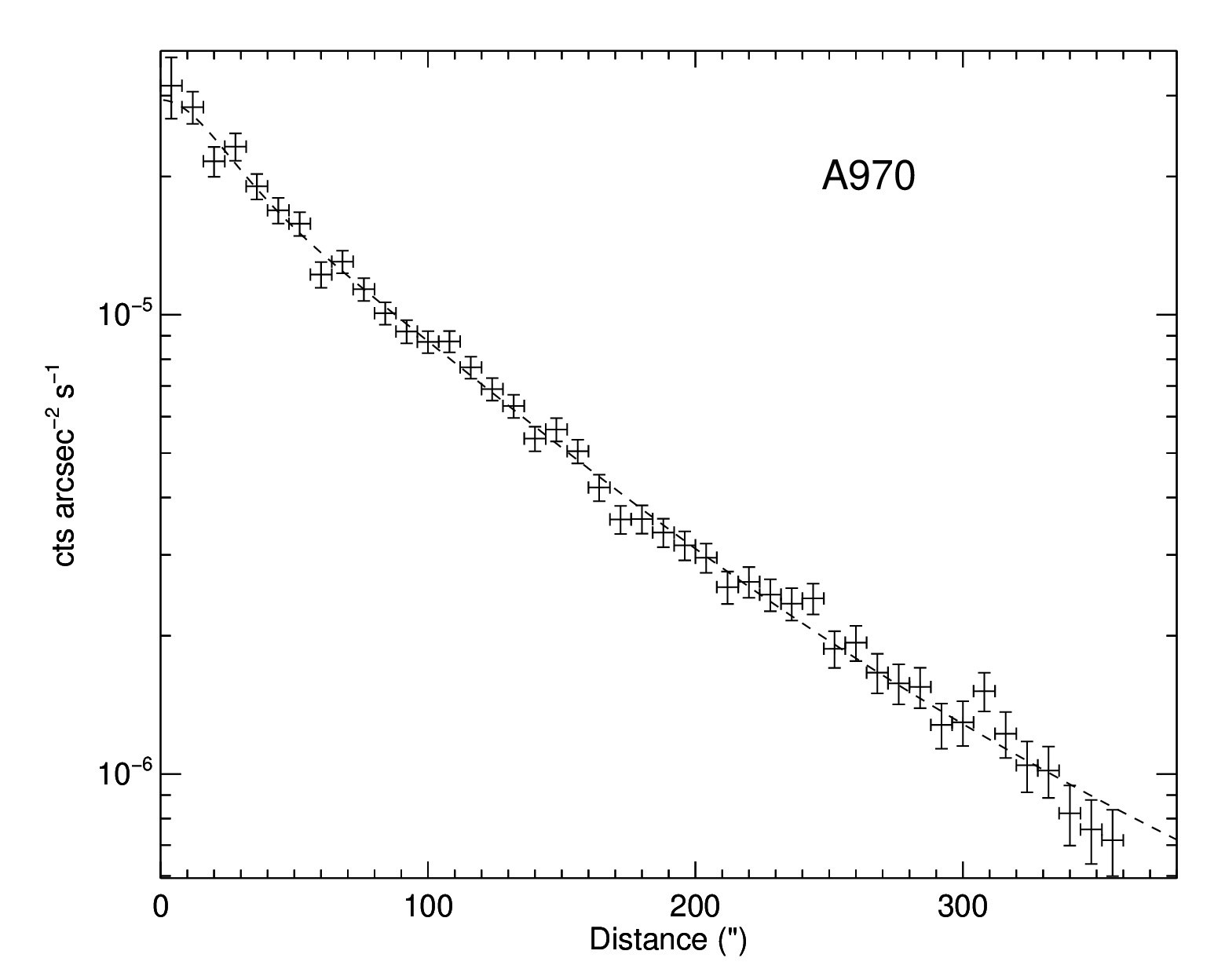}\\
\includegraphics[width=2.8in]{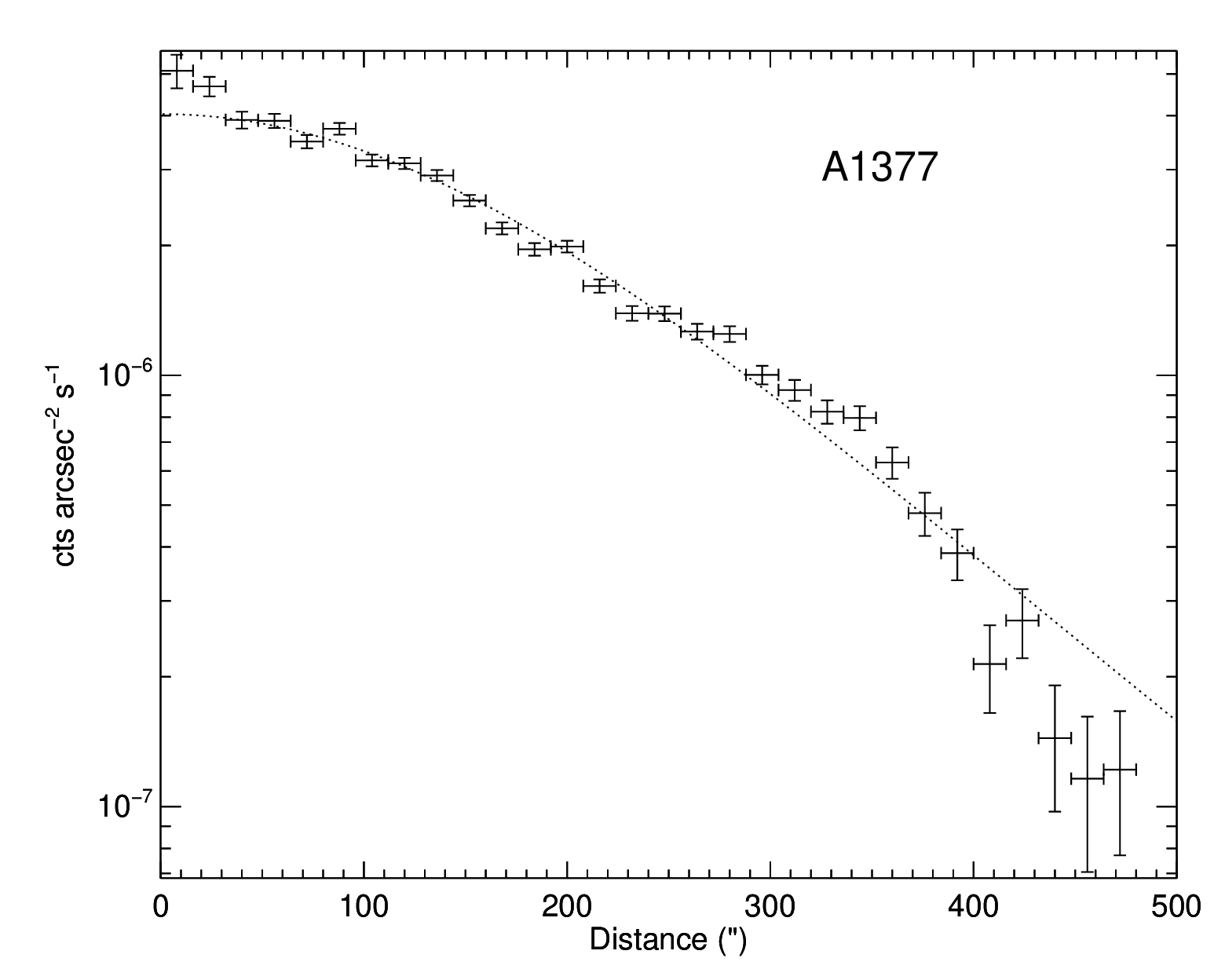}
\includegraphics[width=2.8in]{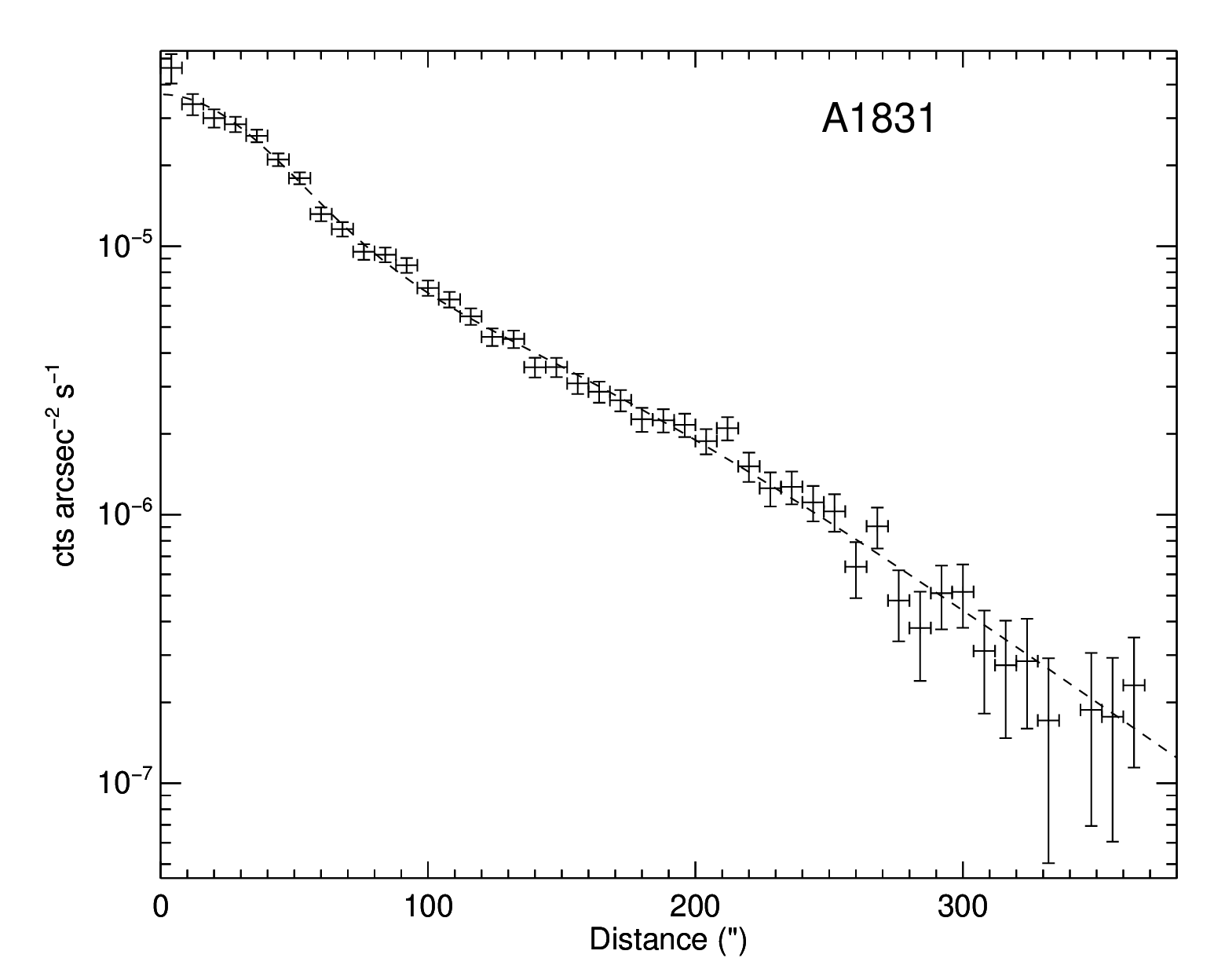}\\
\includegraphics[width=2.8in]{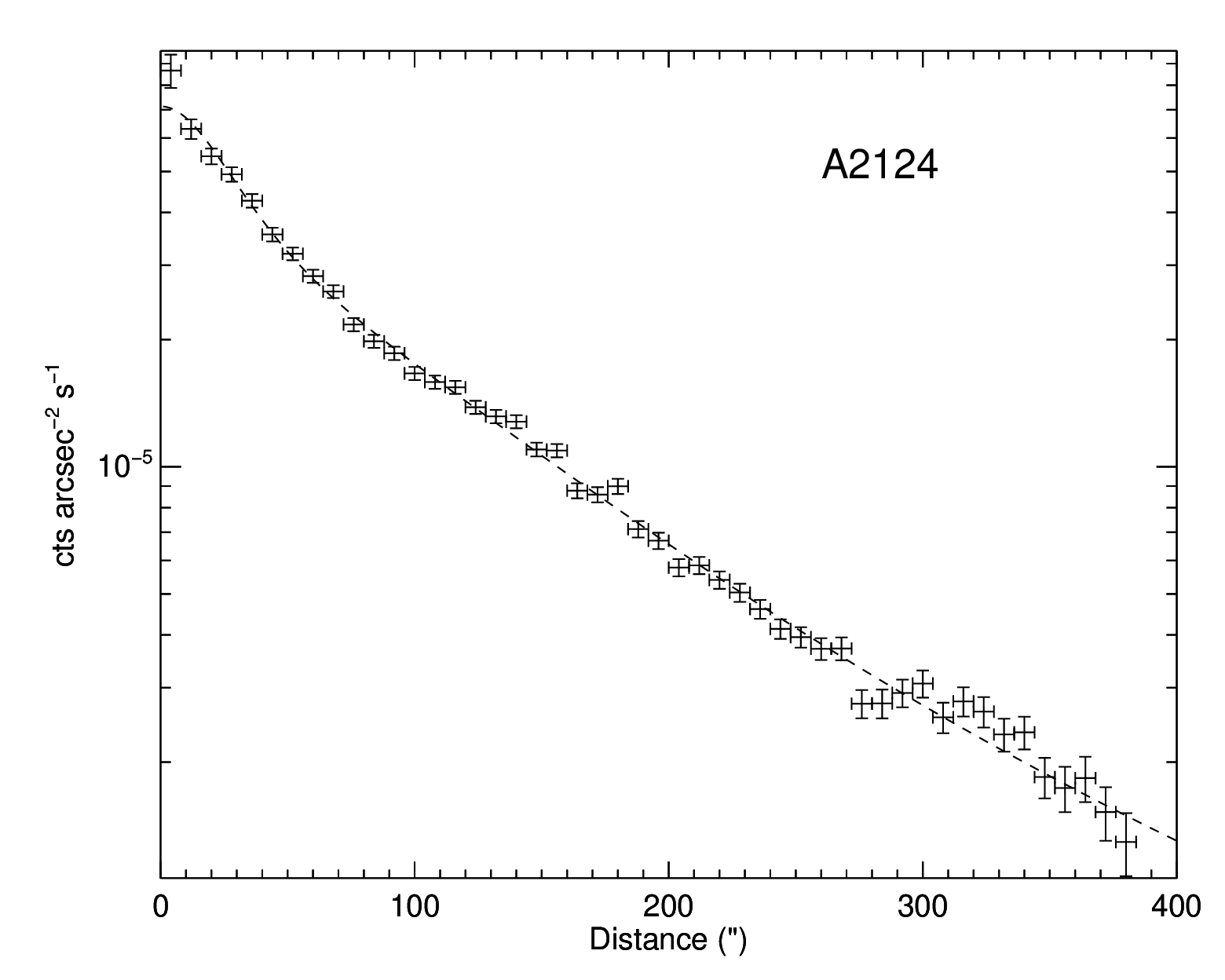}
\includegraphics[width=2.8in]{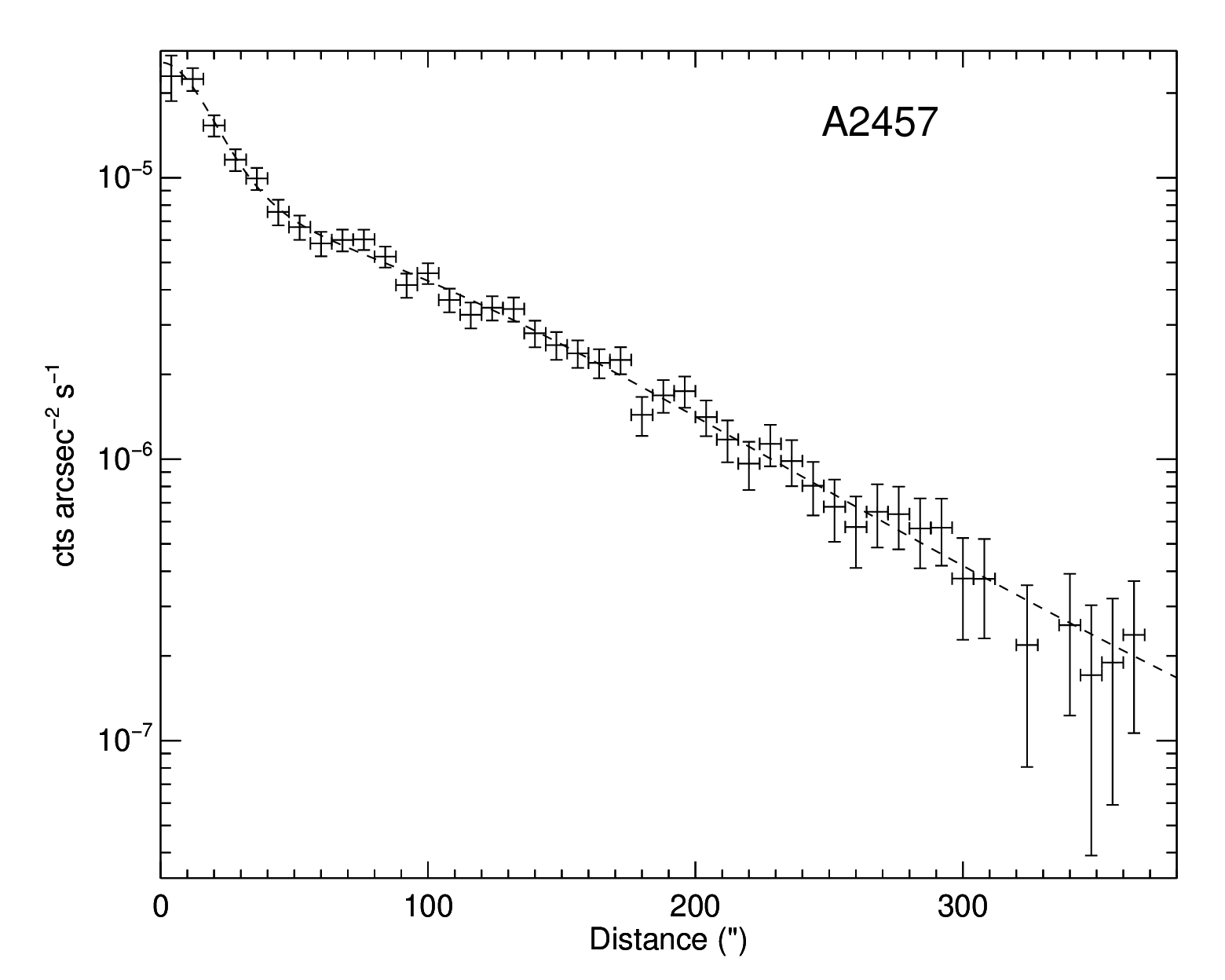}\\
\includegraphics[width=2.8in]{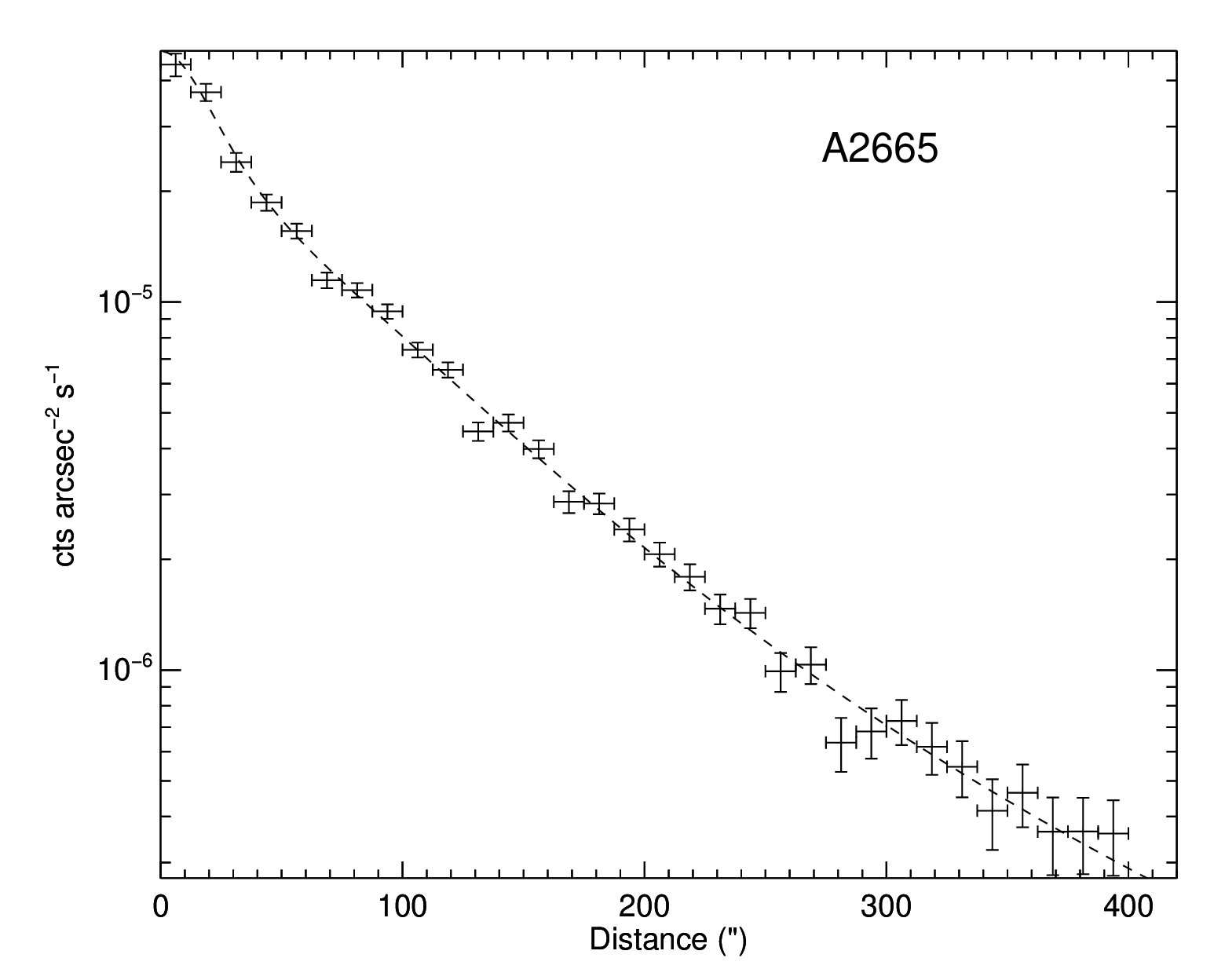}
\includegraphics[width=2.8in]{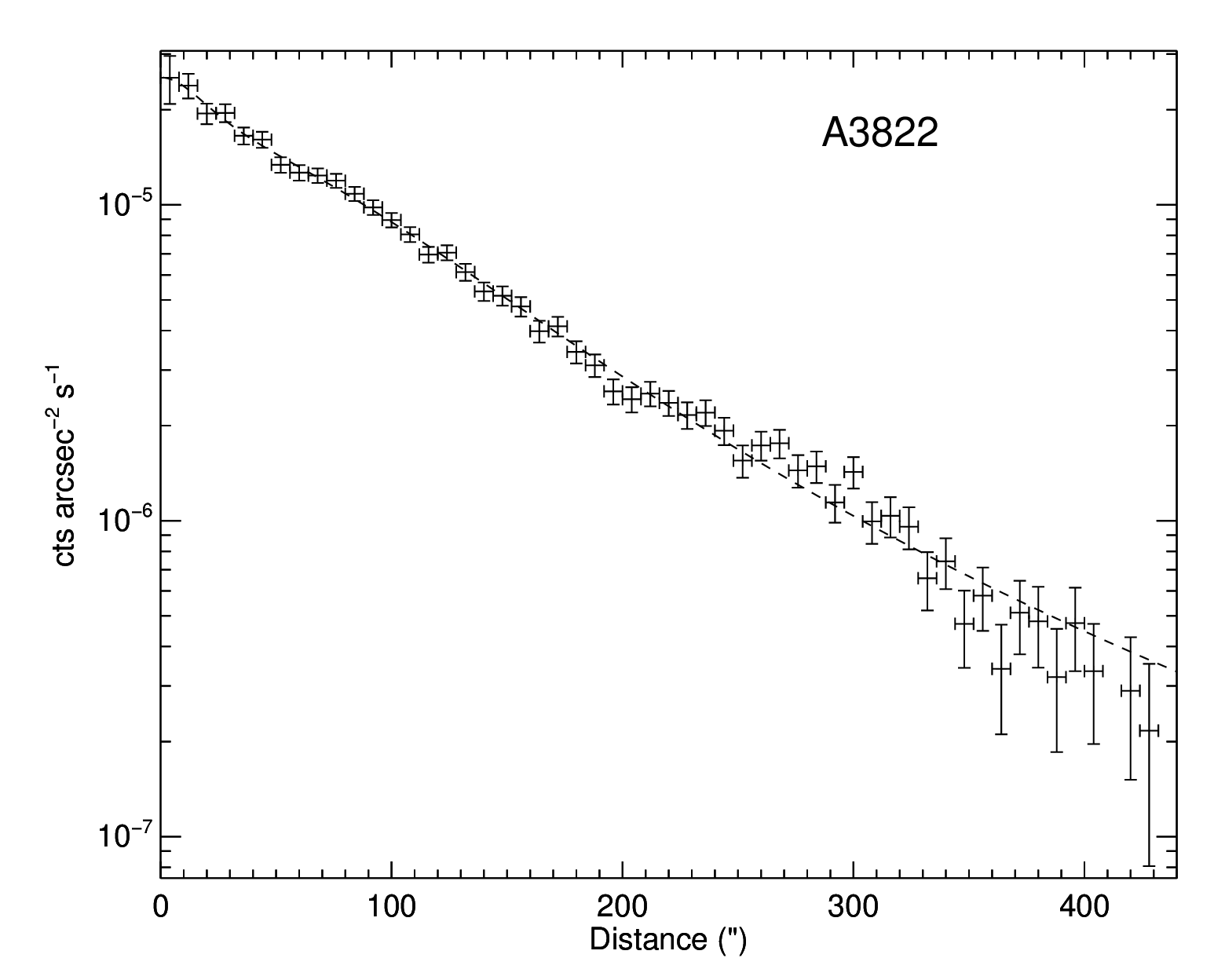}\\
 \end{longtable}%
\captionof{figure}{X-ray surface brightness profiles of the clusters A193, A376, A539, A970, A1831B, A2124, A2457, A2665, and A3822 fitted with 
double-$\beta$ model (single-$\beta$ for A1377), shown with the dashed line. The results of the fitting are provided in Table 
\ref{tab:xsb_db_fit_results}.
  \label{fig:xsb_prof_db_fit}}
\addtocounter{table}{-1}%
\end{center}

\begin{center}
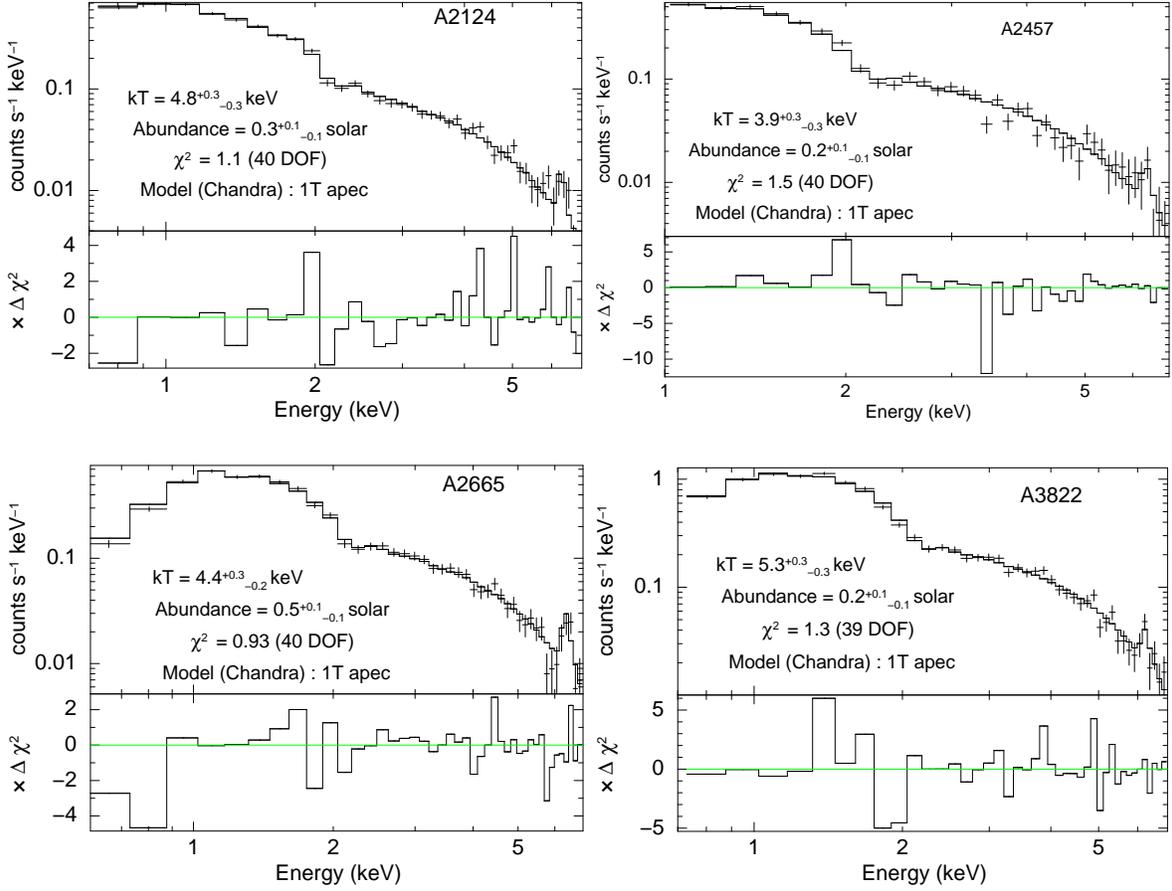

\begin{longtable}{cc}
\includegraphics[width=2.2in,angle=270]{f4a.eps}
\includegraphics[width=2.2in,angle=270]{f4b.eps}\\
\includegraphics[width=2.2in,angle=270]{f4c.eps}
\includegraphics[width=2.2in,angle=270]{f4d.eps}\\
\includegraphics[width=2.2in,angle=270]{f4e.eps}
\includegraphics[width=2.2in,angle=270]{f4f.eps}\\
\includegraphics[width=2.2in,angle=270]{f4g.eps}
\includegraphics[width=2.2in,angle=270]{f4h.eps}\\
\includegraphics[width=2.2in,angle=270]{f4i.eps}
\includegraphics[width=2.2in,angle=270]{f4j.eps}\\
 \end{longtable}%
\captionof{figure}{Average spectra of the clusters A193, A376, A539, A970, A1377, A1831B, A2124, A2457, A2665, and A3822 from the 
\textit{Chandra ACIS} detector. All the spectra have been fitted with \textit{wabs*apec} 
model shown as a histogram. Details of the spectral analysis are given in \S\ref{sec:Total-Spec-Analy}, and the 
best-fit parameters are shown here as insets. For A539 the two different colored spectra used correspond to the two 
\textit{Chandra} observations of the cluster used in this paper.}
  \addtocounter{table}{-1}%
\end{center}
\label{fig:tot_spec}

\clearpage

\begin{center}
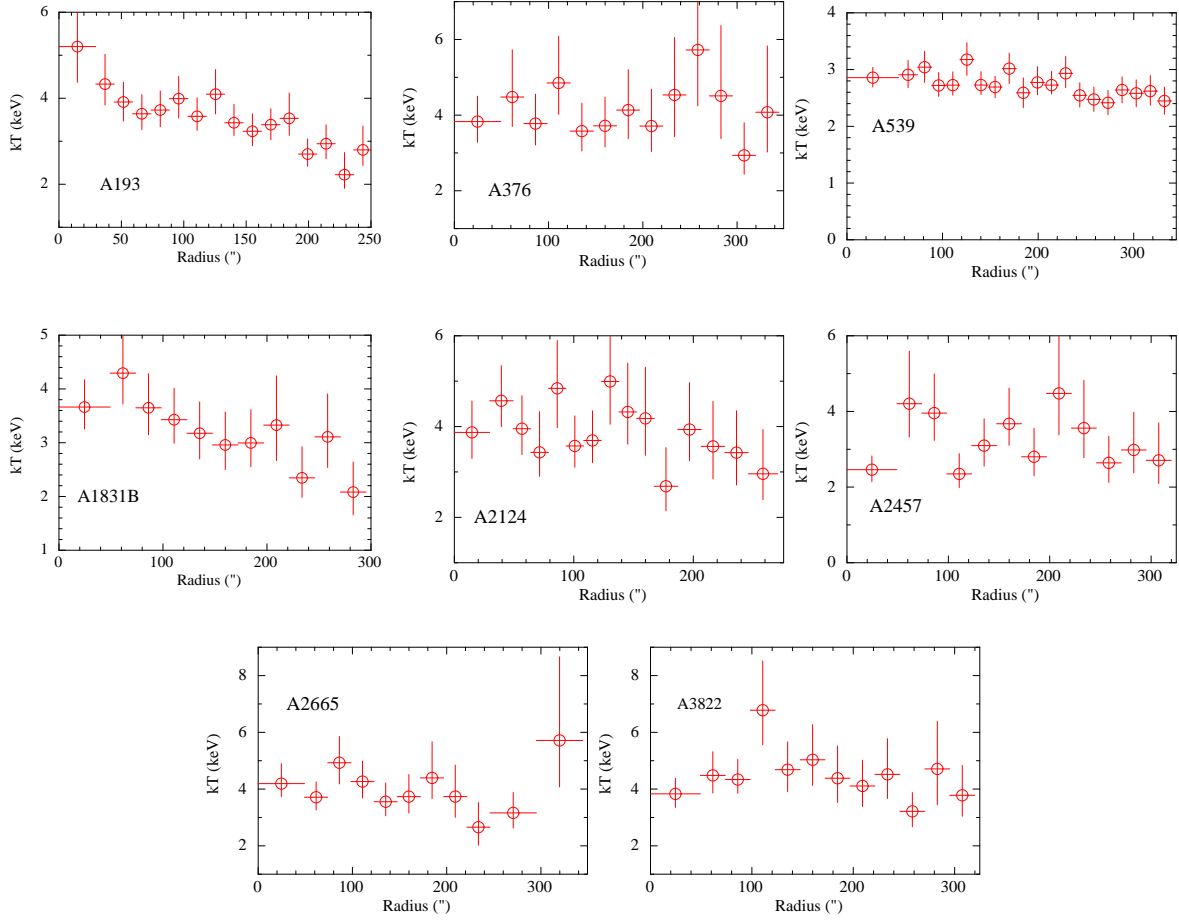

\begin{longtable}{ccc}
\includegraphics[height=2.0in,angle=270]{f5a.eps}
\includegraphics[height=2.0in,angle=270]{f5b.eps}
\includegraphics[height=2.0in,angle=270]{f5c.eps}\\
\includegraphics[height=2.0in,angle=270]{f5d.eps}
\includegraphics[height=2.0in,angle=270]{f5e.eps}
\includegraphics[height=2.0in,angle=270]{f5f.eps}\\
\includegraphics[height=2.0in,angle=270]{f5g.eps}
\includegraphics[height=2.0in,angle=270]{f5h.eps}
 \end{longtable}%
\captionof{figure}{Projected temperature (kT)  profiles obtained from the spectral analysis of circular annuli 
centred on the surface brightness peaks of the clusters A193, A376, A539, A1831B, A2124, A2457, A2665, and A3822, respectively. 
For each cluster, the value of elemental abundance was frozen to the average abundance 
obtained for the whole cluster.\label{fig:anu_temp_prof}}
\addtocounter{table}{-1}%
\end{center}

\begin{center}
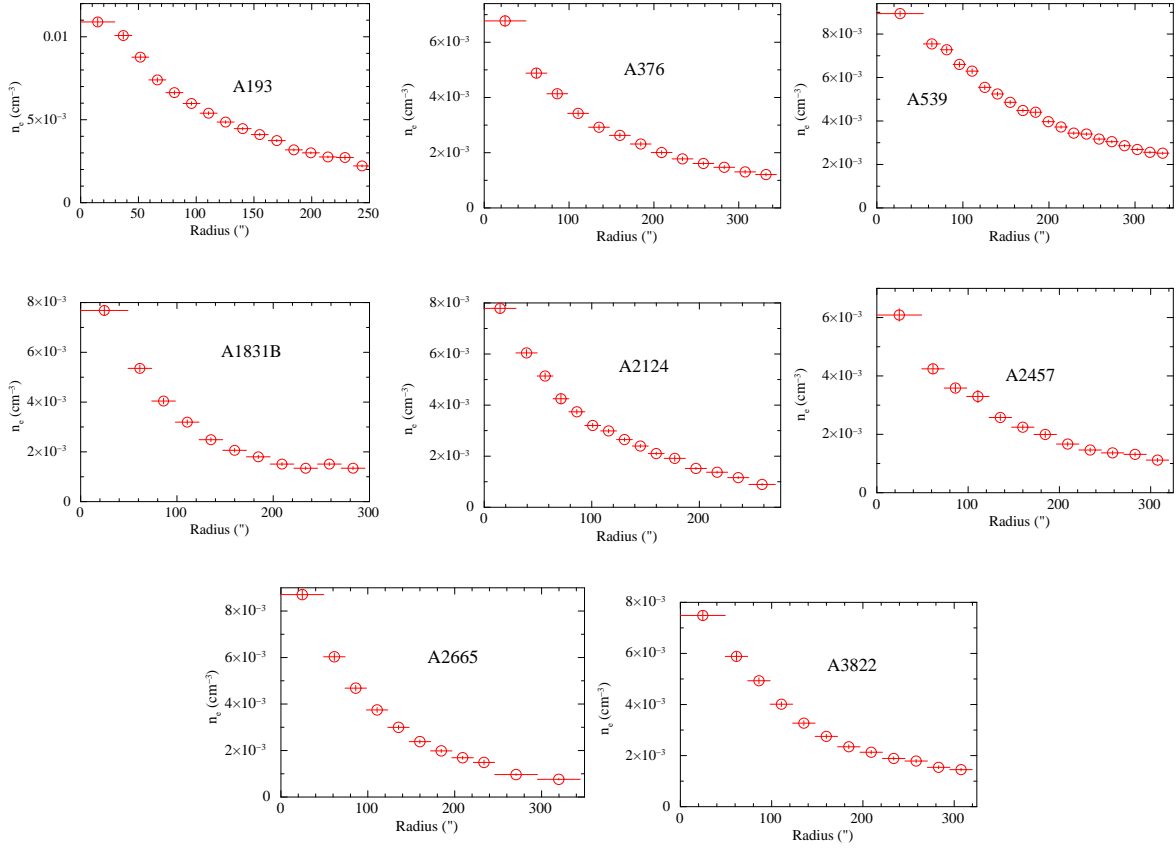

\begin{longtable}{ccc}
\includegraphics[height=2.0in,angle=270]{f6a_mod.eps}
\includegraphics[height=2.0in,angle=270]{f6b_mod.eps}
\includegraphics[height=2.0in,angle=270]{f6c_mod.eps}\\
\includegraphics[height=2.0in,angle=270]{f6d_mod.eps}
\includegraphics[height=2.0in,angle=270]{f6e_mod.eps}
\includegraphics[height=2.0in,angle=270]{f6f_mod.eps}\\
\includegraphics[height=2.0in,angle=270]{f6g_mod.eps}
\includegraphics[height=2.0in,angle=270]{f6h_mod.eps}
 \end{longtable}%
  \captionof{figure}{Projected electron number density (n$_{e}$)  profiles of the clusters A193, A376, A539, A1831B, A2124, A2457, 
A2665, and A3822, respectively.\label{fig:anu_dens_prof}}
\addtocounter{table}{-1}%
\end{center}

\clearpage

\begin{center}
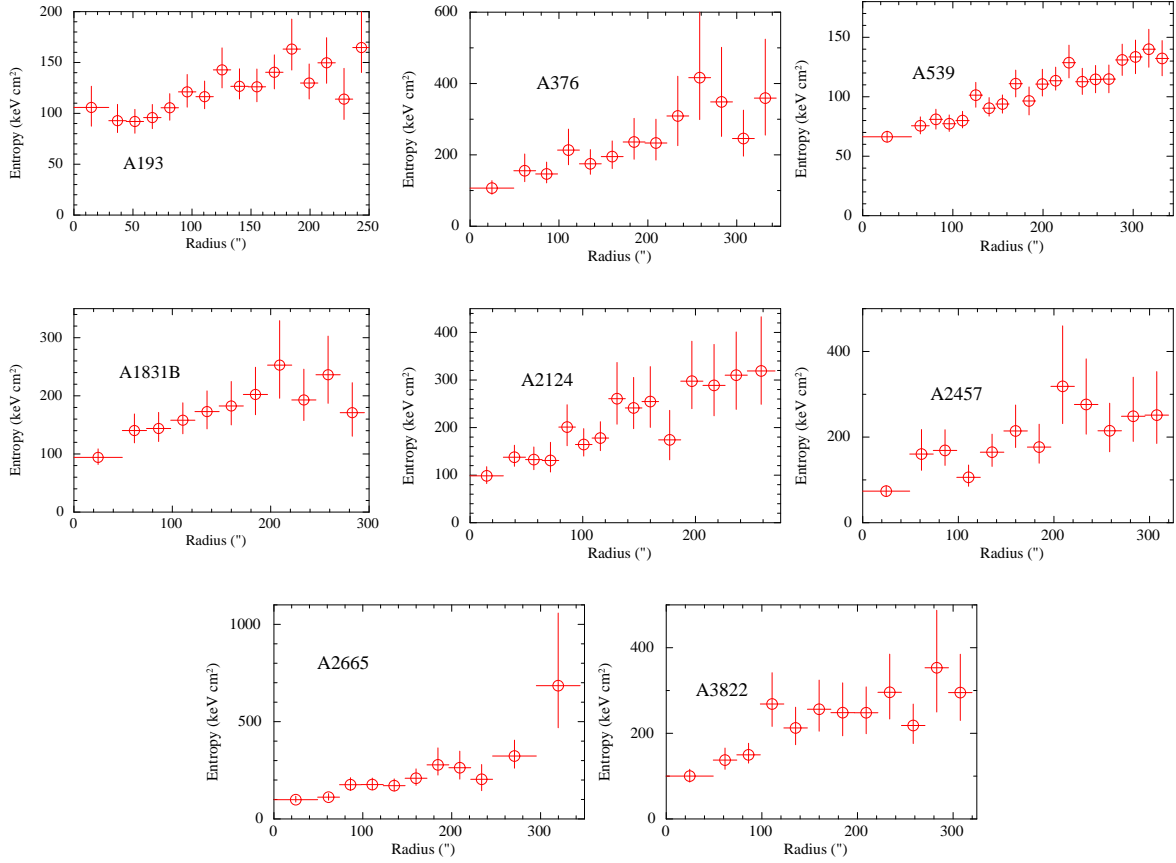

\begin{longtable}{ccc}
\includegraphics[height=2.0in,angle=270]{f7a_mod.eps}
\includegraphics[height=2.0in,angle=270]{f7b_mod.eps}
\includegraphics[height=2.0in,angle=270]{f7c_mod.eps}\\
\includegraphics[height=2.0in,angle=270]{f7d_mod.eps}
\includegraphics[height=2.0in,angle=270]{f7e_mod.eps}
\includegraphics[height=2.0in,angle=270]{f7f_mod.eps}\\
\includegraphics[height=2.0in,angle=270]{f7g_mod.eps}
\includegraphics[height=2.0in,angle=270]{f7h_mod.eps}
 \end{longtable}%
  \captionof{figure}{Projected entropy (S)  profiles of the clusters A193, A376, A539, A1831B, A2124, A2457, A2665, and A3822, 
respectively.\label{fig:anu_entr_prof}}
\addtocounter{table}{-1}%
\end{center}

\clearpage

\begin{center}
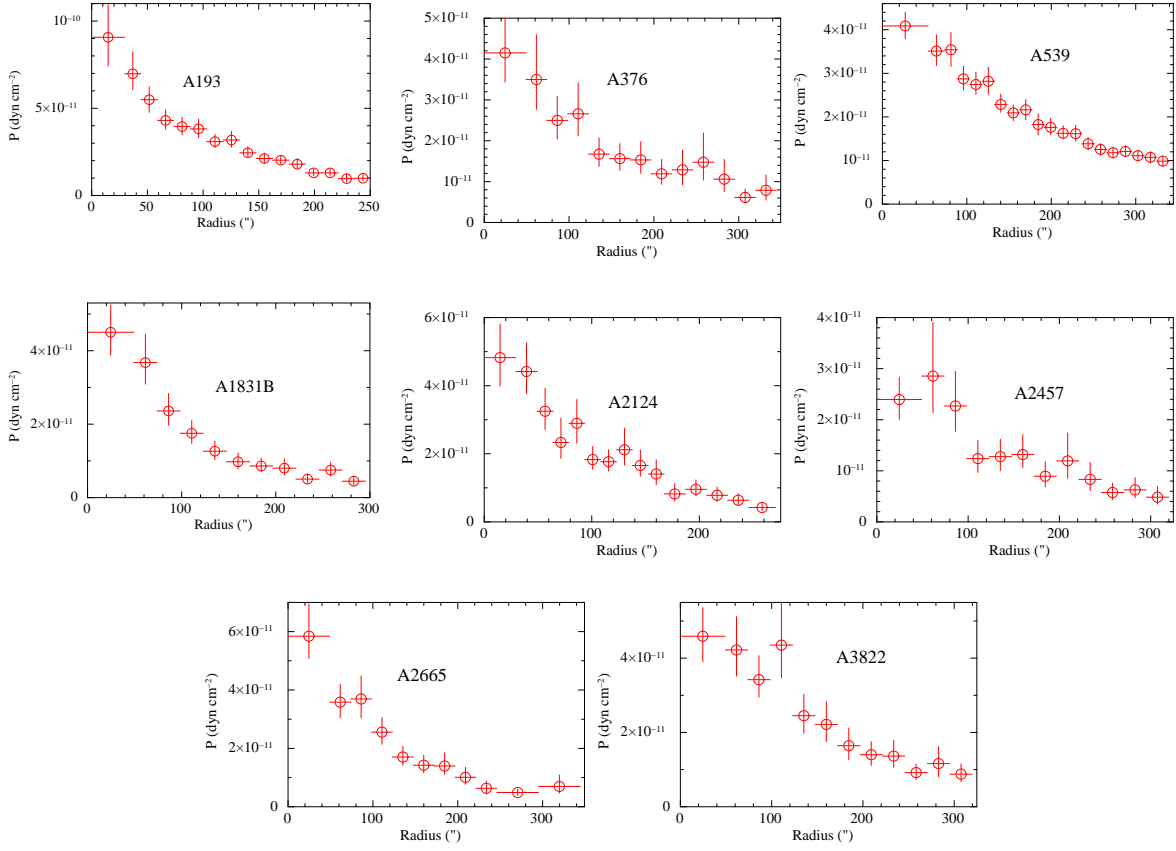

\begin{longtable}{ccc}
\includegraphics[height=2.0in,angle=270]{f8a_mod.eps}
\includegraphics[height=2.0in,angle=270]{f8b_mod.eps}
\includegraphics[height=2.0in,angle=270]{f8c_mod.eps}\\
\includegraphics[height=2.0in,angle=270]{f8d_mod.eps}
\includegraphics[height=2.0in,angle=270]{f8e_mod.eps}
\includegraphics[height=2.0in,angle=270]{f8f_mod.eps}\\
\includegraphics[height=2.0in,angle=270]{f8g_mod.eps}
\includegraphics[height=2.0in,angle=270]{f8h_mod.eps}
 \end{longtable}%
  \captionof{figure}{Projected pressure (P)  profiles of the clusters A193, A376, A539, A1831B, A2124, A2457, A2665, 
and A3822, respectively.\label{fig:anu_pres_prof}}
\addtocounter{table}{-1}%
\end{center}

\clearpage

\begin{center}
\begin{longtable}{c}
\includegraphics[height=4.0in]{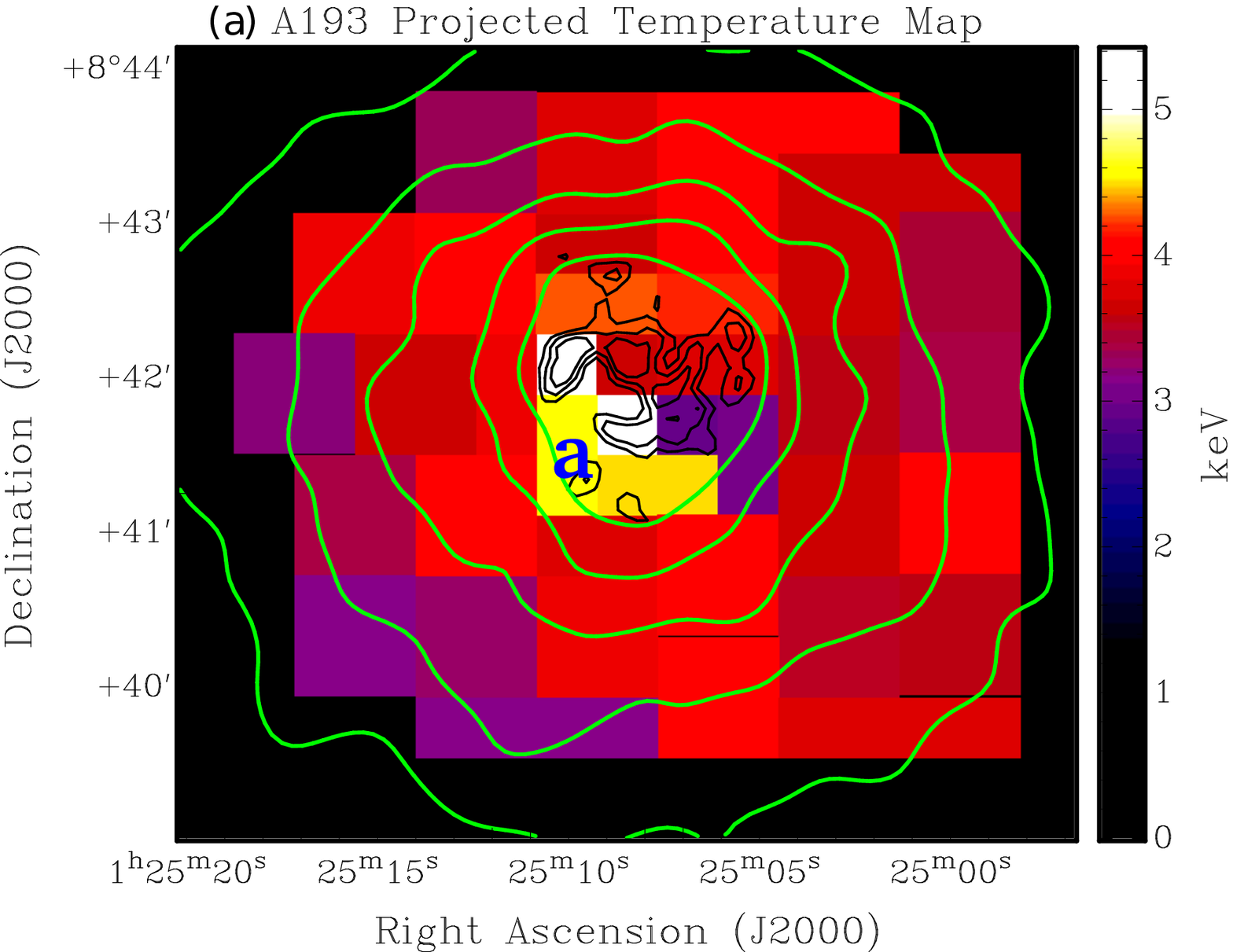}\\
\includegraphics[height=4.0in]{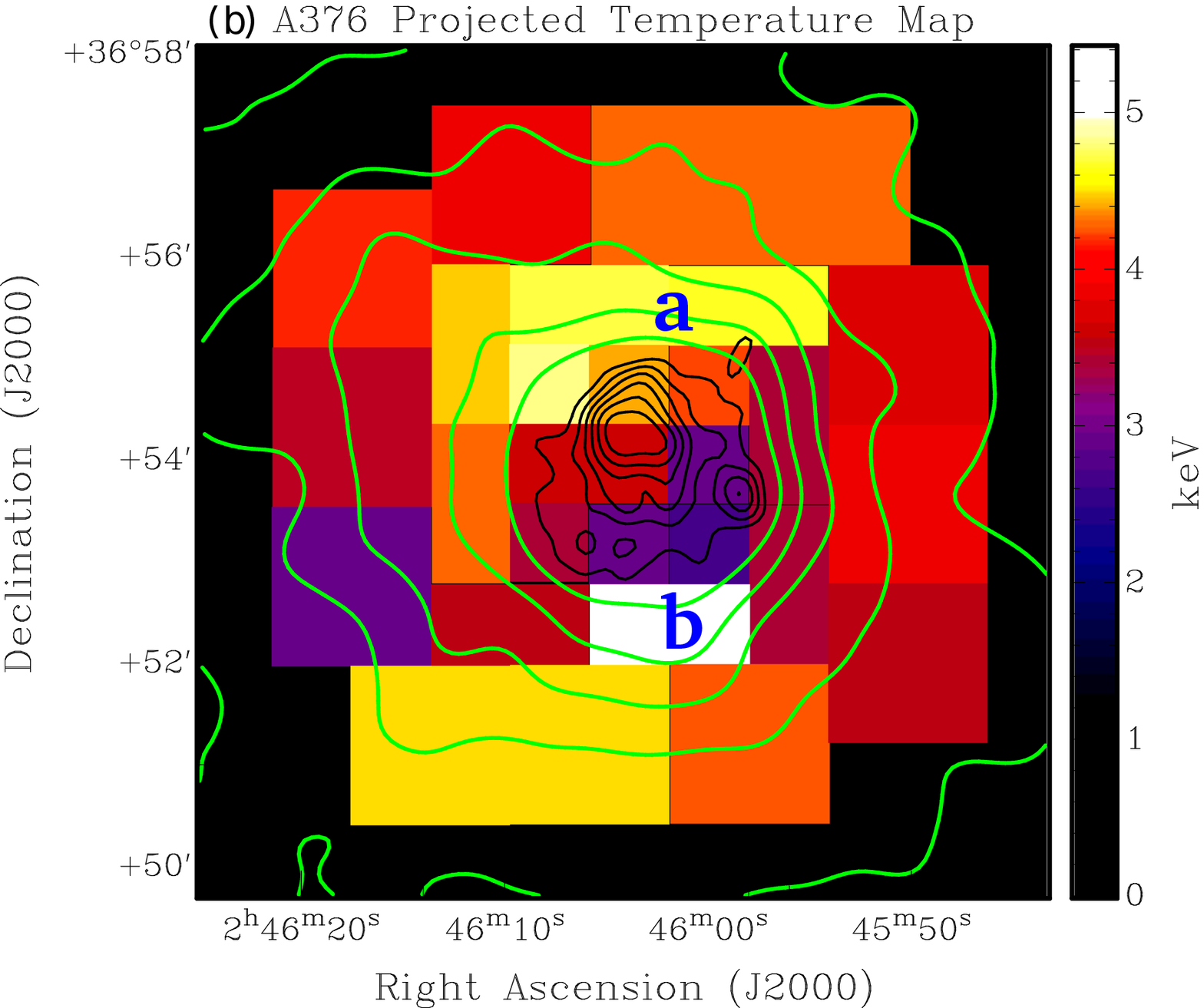}\\
\includegraphics[height=4.0in]{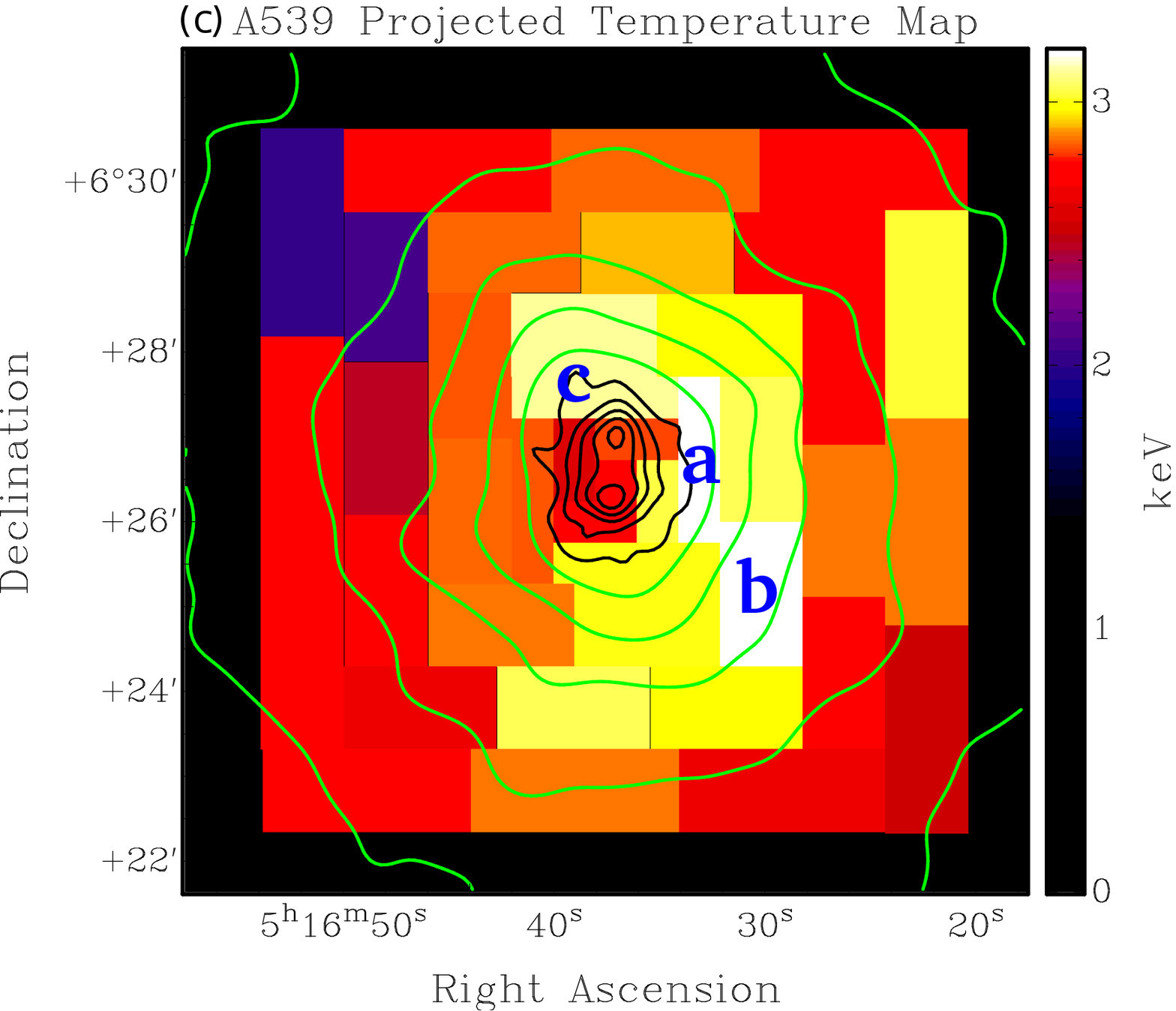}\\
\includegraphics[height=3.7in]{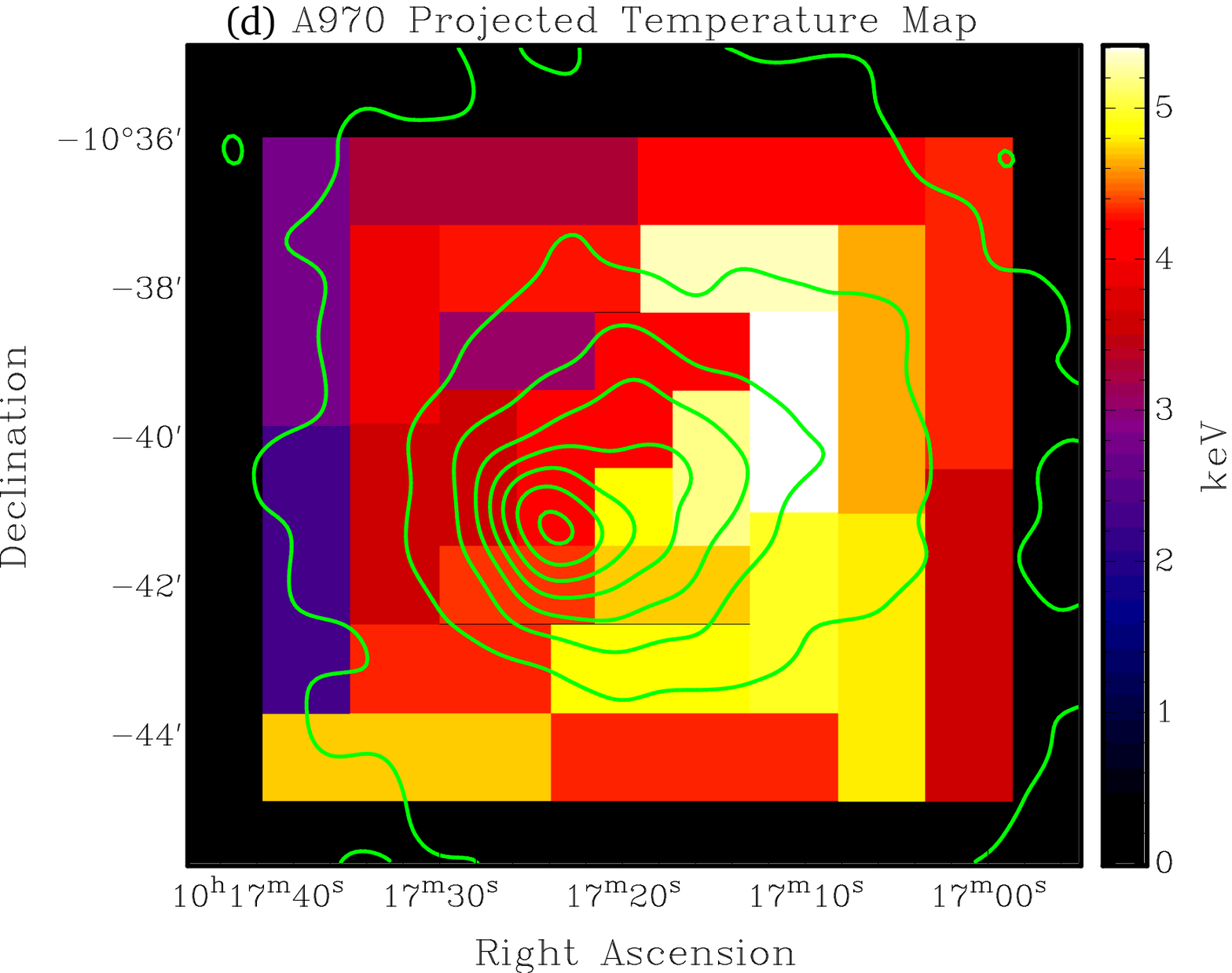}\\
\includegraphics[height=3.7in]{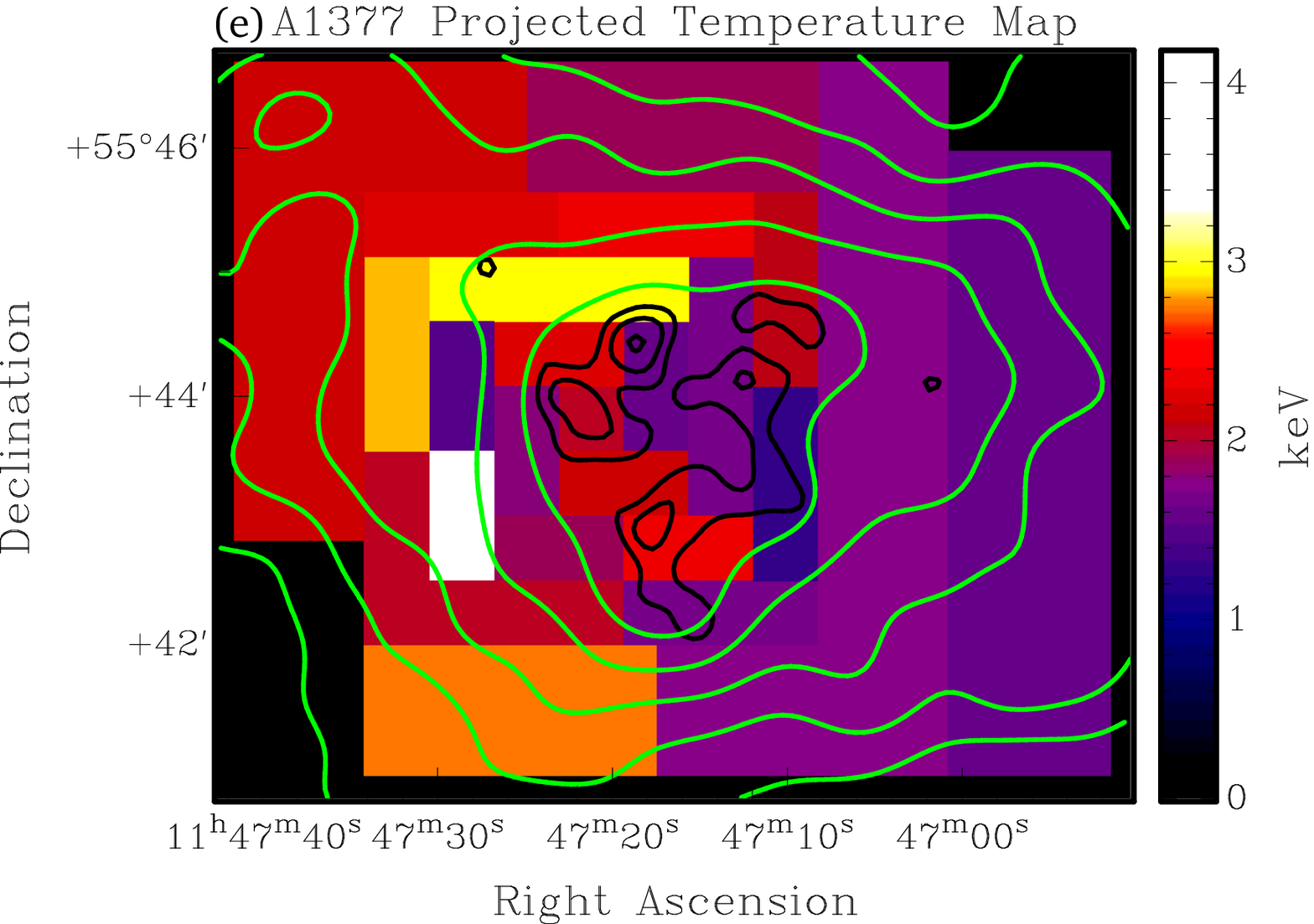}\\
\includegraphics[height=4.0in]{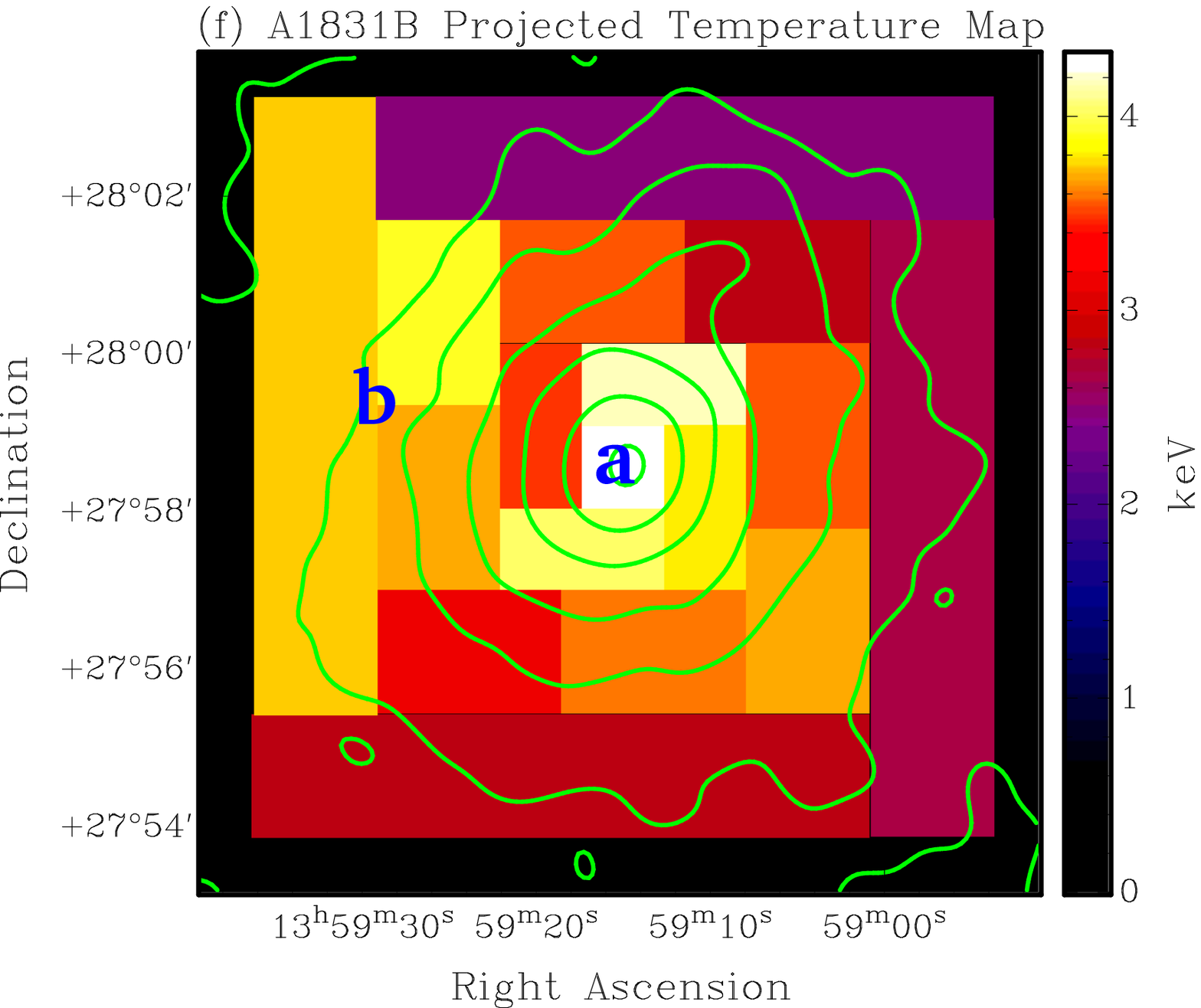}\\
\includegraphics[height=4.0in]{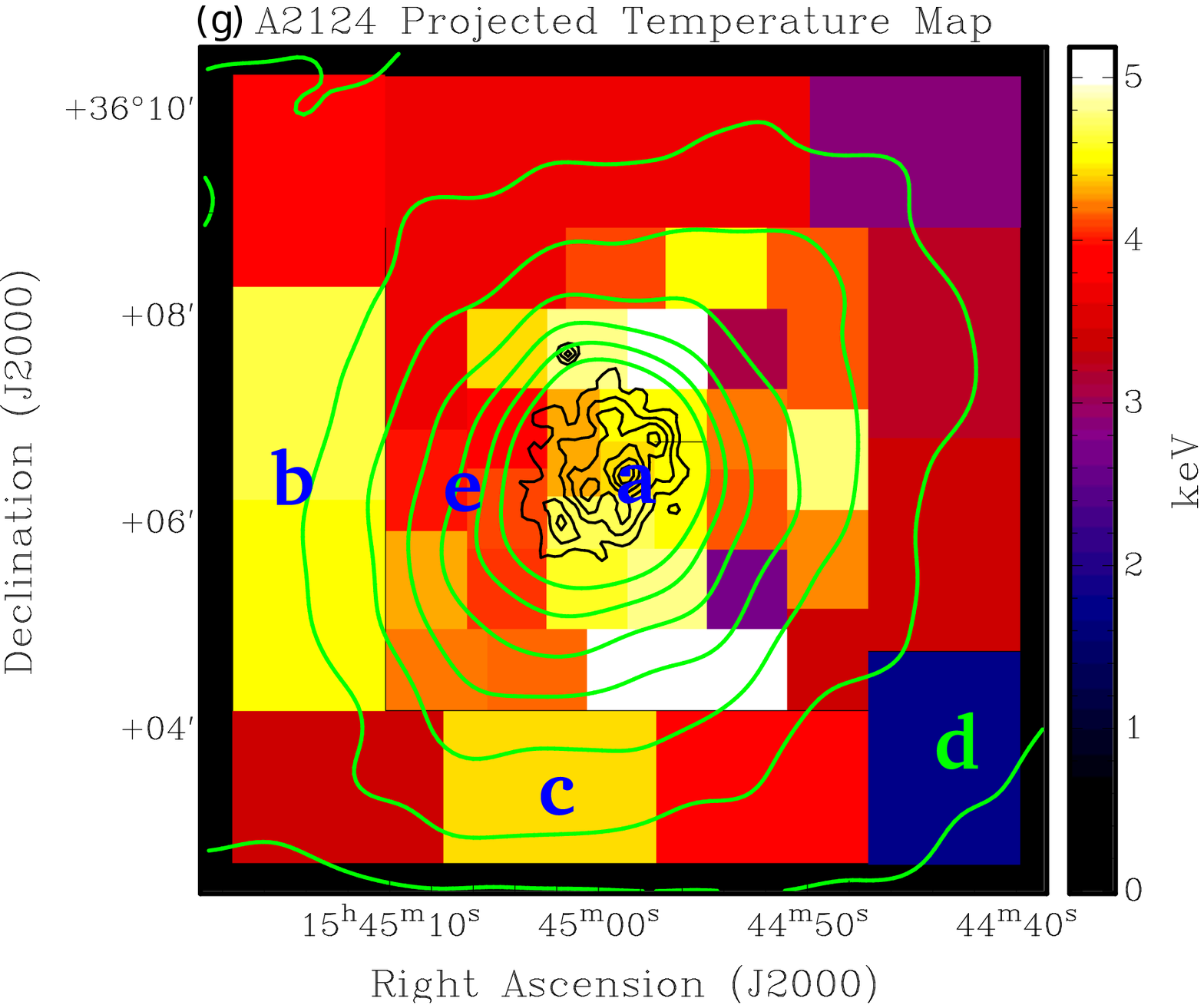}\\
\includegraphics[height=4.0in]{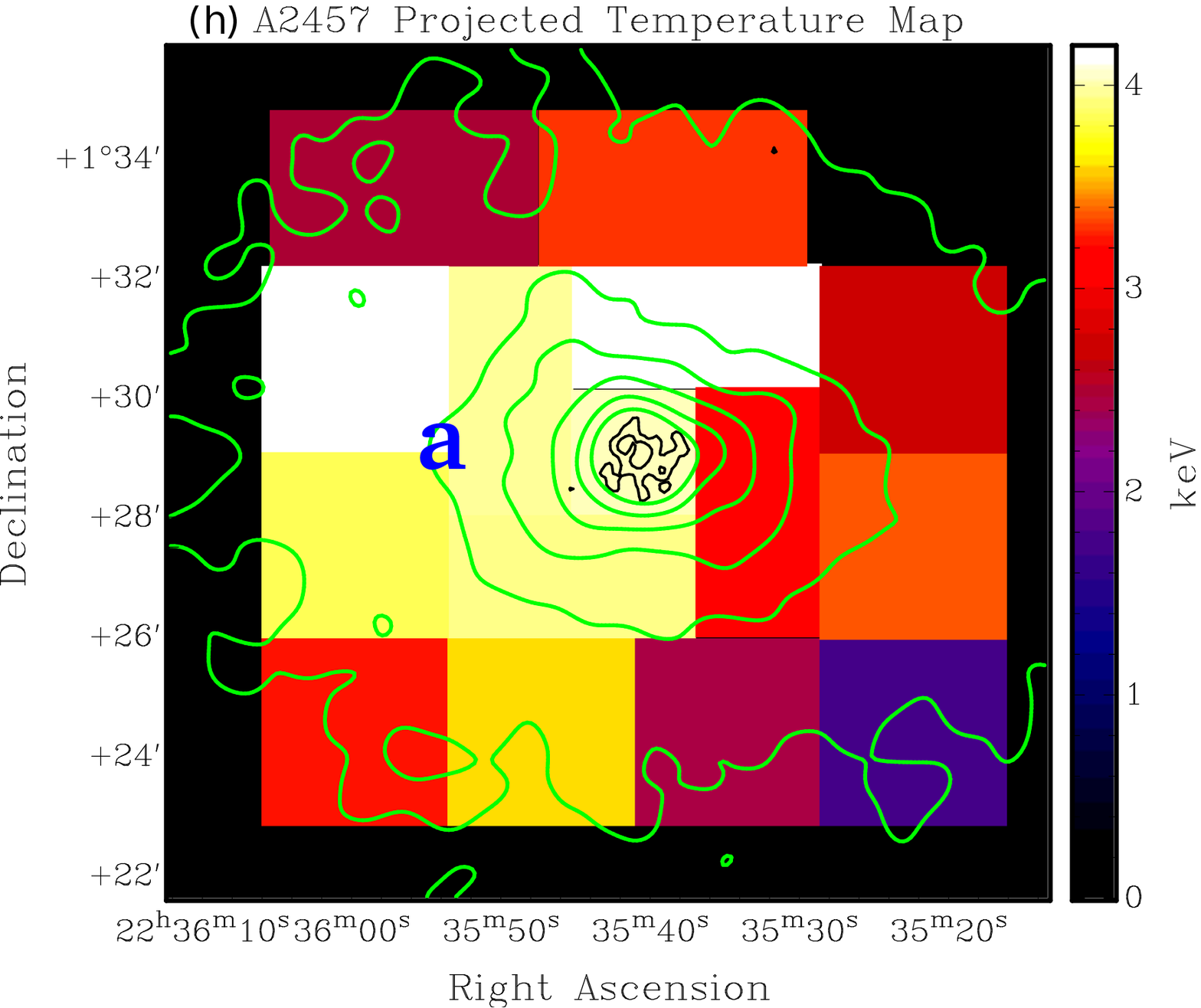}\\
\includegraphics[height=4.0in]{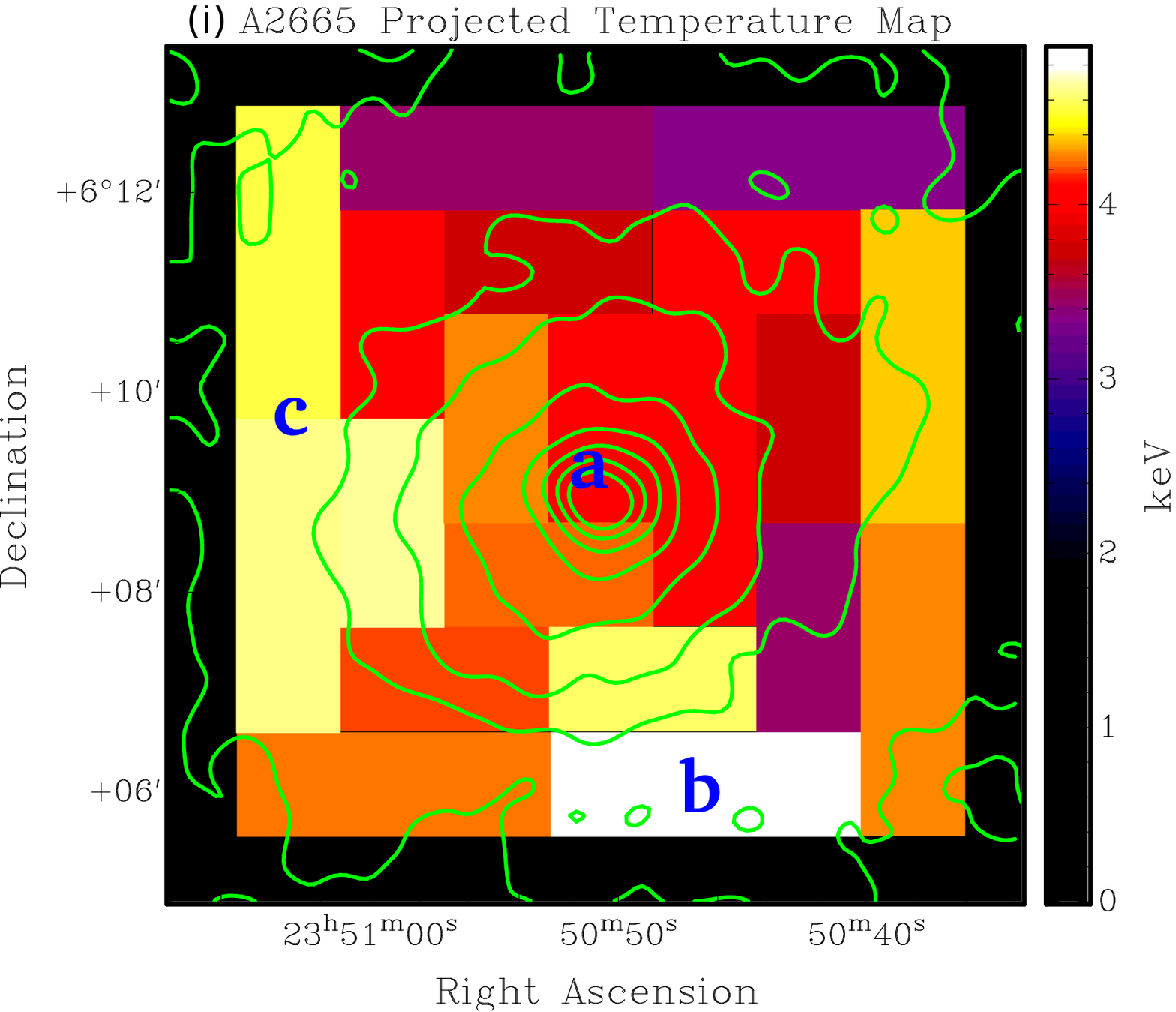}\\
\includegraphics[height=4.0in]{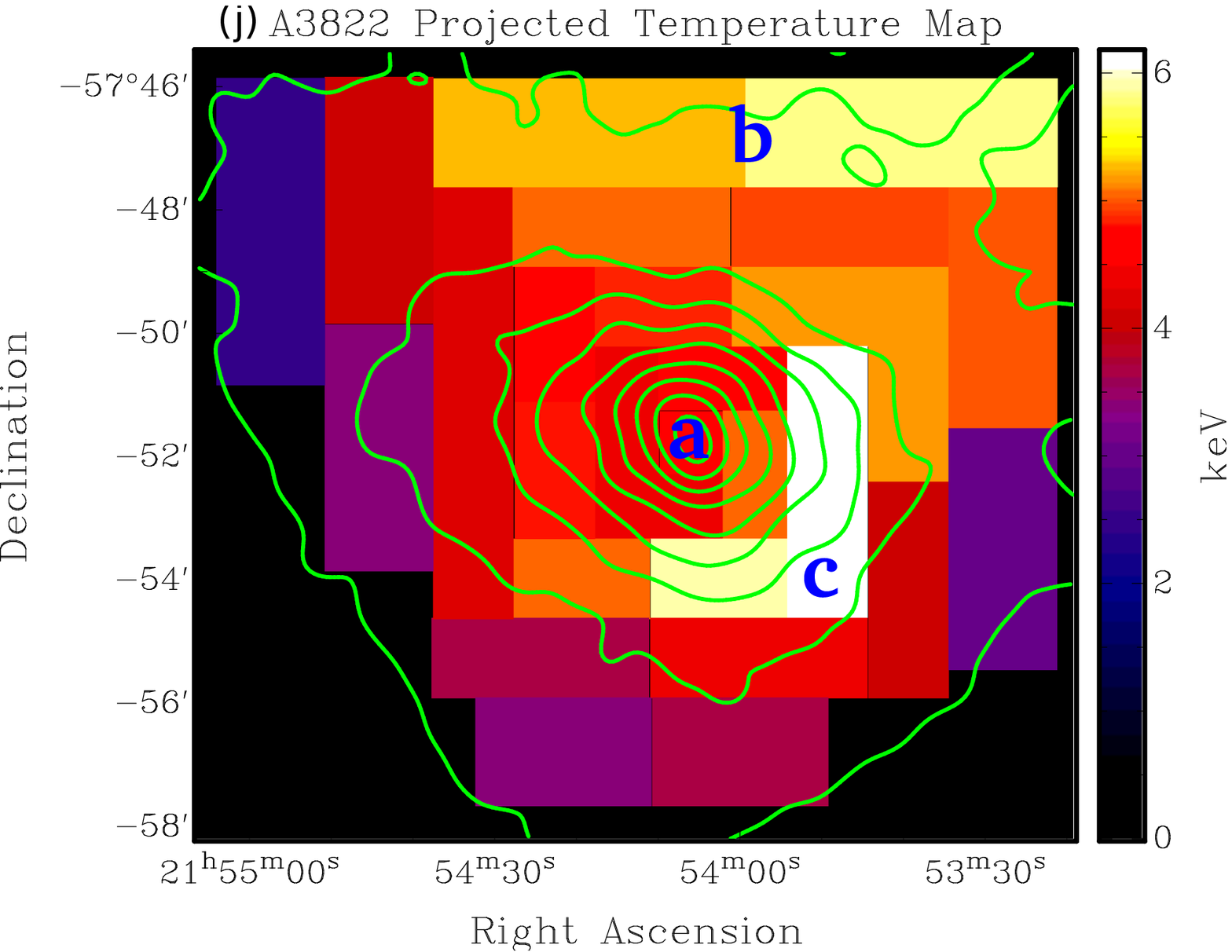}\\
 \end{longtable}%
  \captionof{figure}{(a)-(j): Projected temperature (kT) maps of the clusters A193, A376, A539, A970, A1377, A1831B, A2124, A2457, A2665, 
and A3822, respectively, obtained from the spectral analysis of box shaped regions (see \S\ref{sec:box_thermodynamic_maps}). 
The overlaid green contours for A193 and A2665 are from their \textit{Chandra} images smoothed with a Gaussian kernel of width 
12$^{\prime\prime}$ with intensity levels distributed from 14$\sigma$ to 43$\sigma$ and 3$\sigma$ to 
106$\sigma$ above the mean background level, respectively. The overlaid green contours for A376, A539, A970, A1377, A1831, A2124, A2457 
and A3822 are from their \textit{Chandra} images smoothed with a Gaussian kernel of width 20$^{\prime\prime}$ with intensity 
levels distributed from 5$\sigma$ to 17$\sigma$, 15$\sigma$ to 78$\sigma$, 2$\sigma$ to 115$\sigma$, 6$\sigma$ to 20$\sigma$, 2$\sigma$ to 
92$\sigma$, 2$\sigma$ to 23$\sigma$, 1$\sigma$ to 25$\sigma$ and 3$\sigma$ to 47$\sigma$ above the mean background level, 
respectively. The overlaid black 
contours for A193, A376, A539, A1377, A2124 and A2457 are from their \textit{Chandra} images smoothed with Gaussian kernels of width 
4$^{\prime\prime}$, 8$^{\prime\prime}$, 8$^{\prime\prime}$, 8$^{\prime\prime}$, 4$^{\prime\prime}$ and 4$^{\prime\prime}$, with intensity 
levels distributed from 22$\sigma$ to 25$\sigma$, 12$\sigma$ to 30$\sigma$, 27$\sigma$ to 89$\sigma$, 10$\sigma$ to 13$\sigma$, 
10$\sigma$ to 17$\sigma$ and 14$\sigma$ to 32$\sigma$ above the mean background level, respectively.\label{fig:2D_temp_map}}%
  \addtocounter{table}{-1}%
\end{center}

\clearpage

\begin{center}
\begin{longtable}{c}
\includegraphics[height=3.9in]{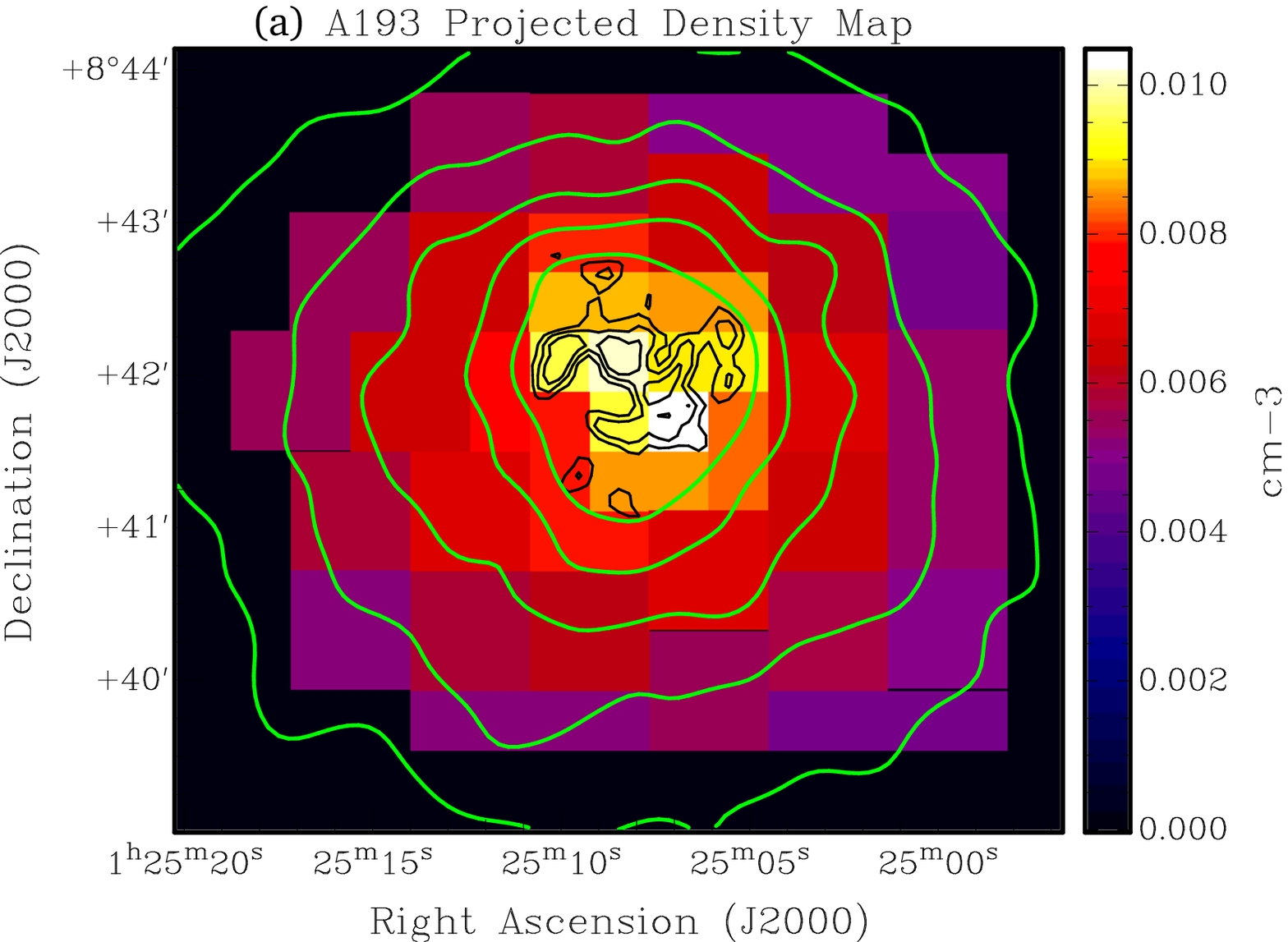}\\
\includegraphics[height=4.0in]{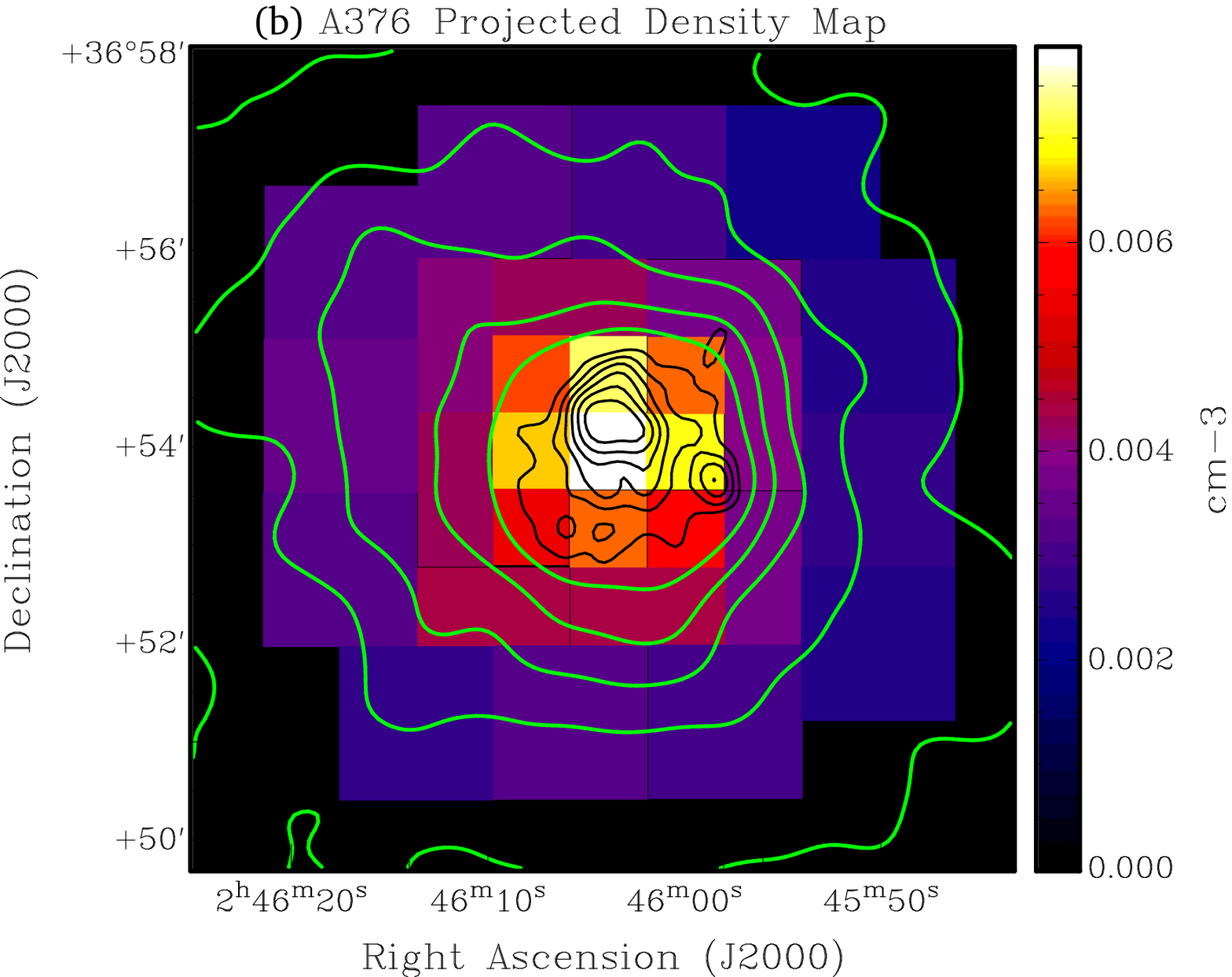}\\
\includegraphics[height=4.0in]{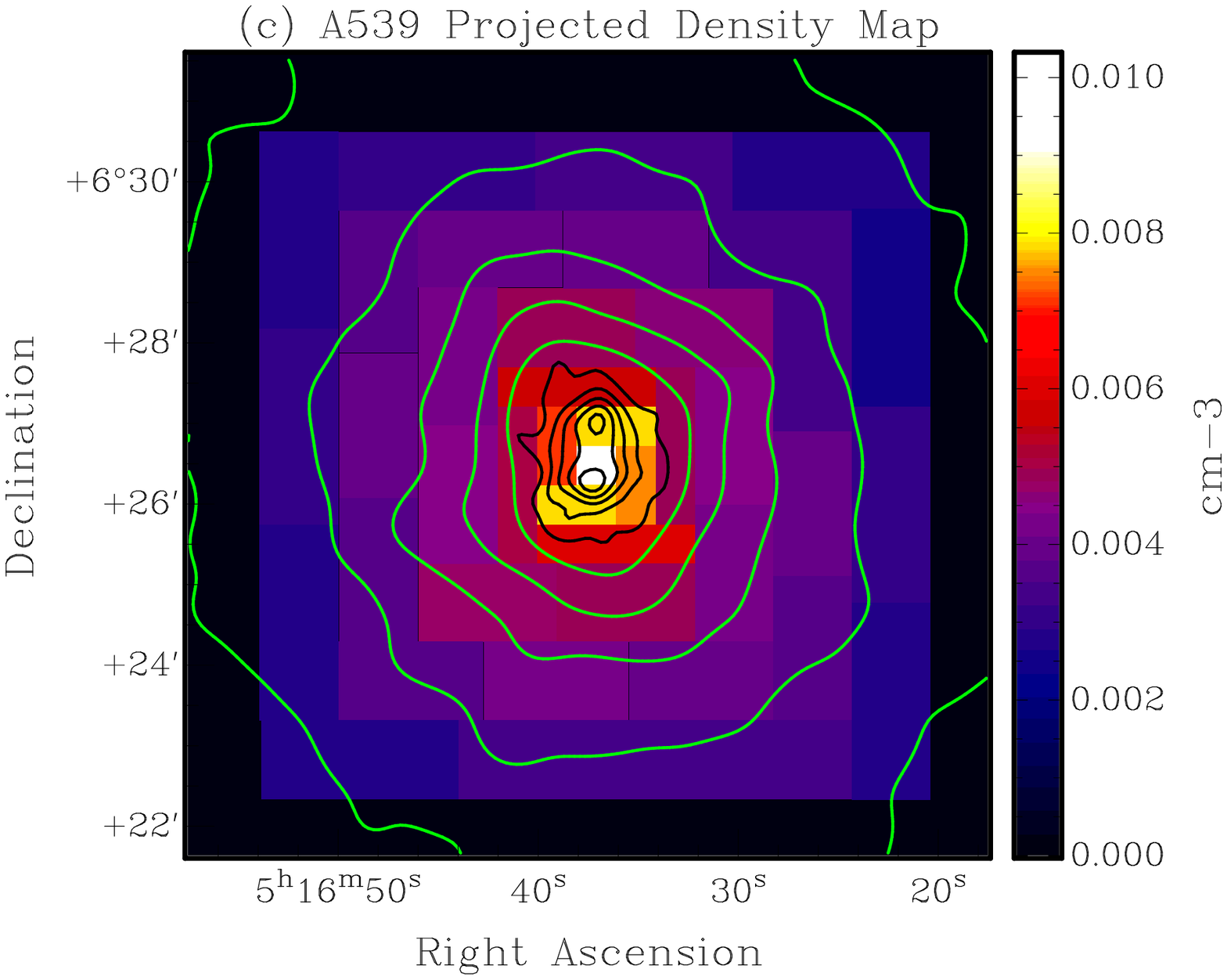}\\
\includegraphics[height=3.7in]{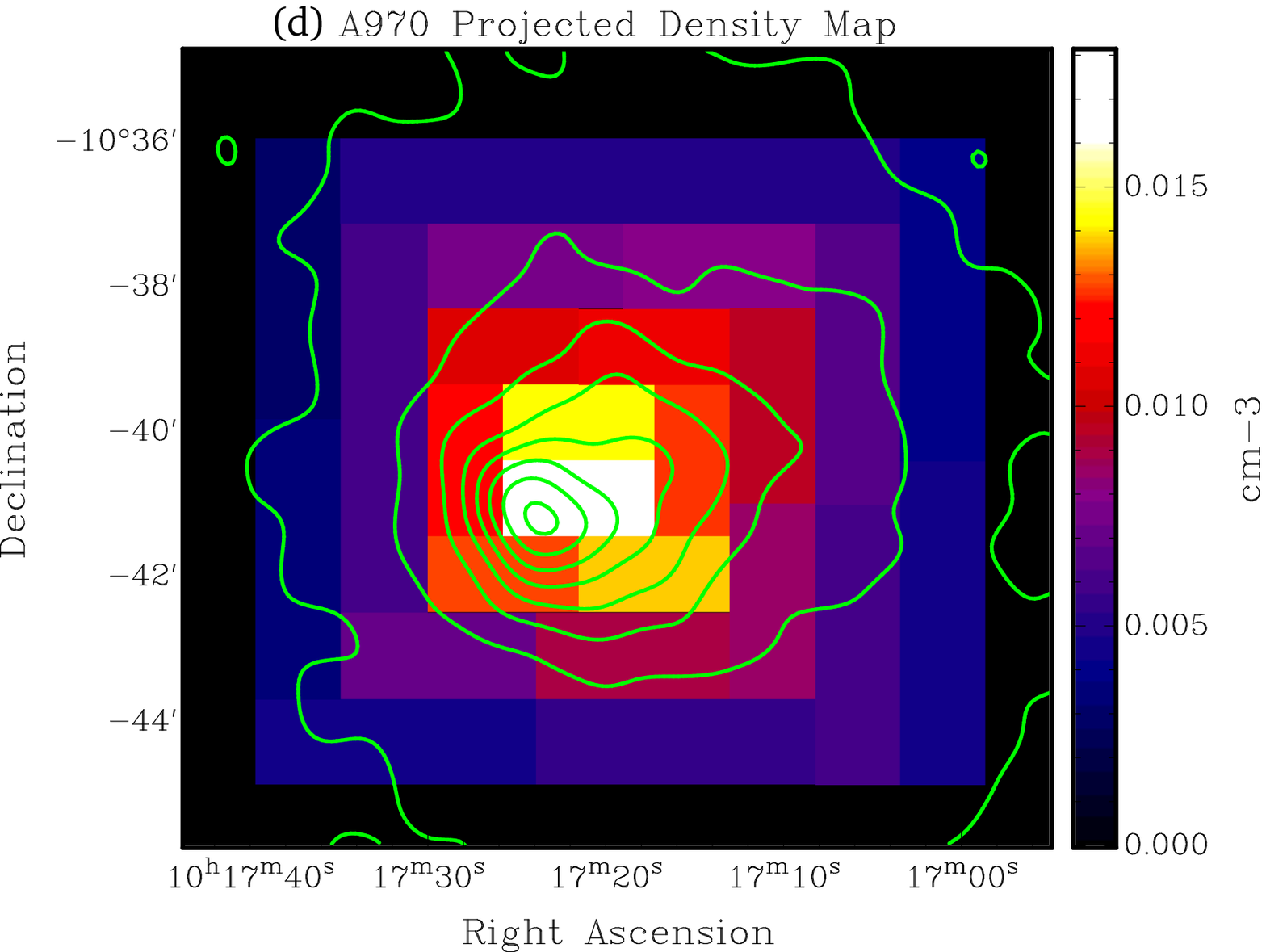}\\
\includegraphics[height=3.7in]{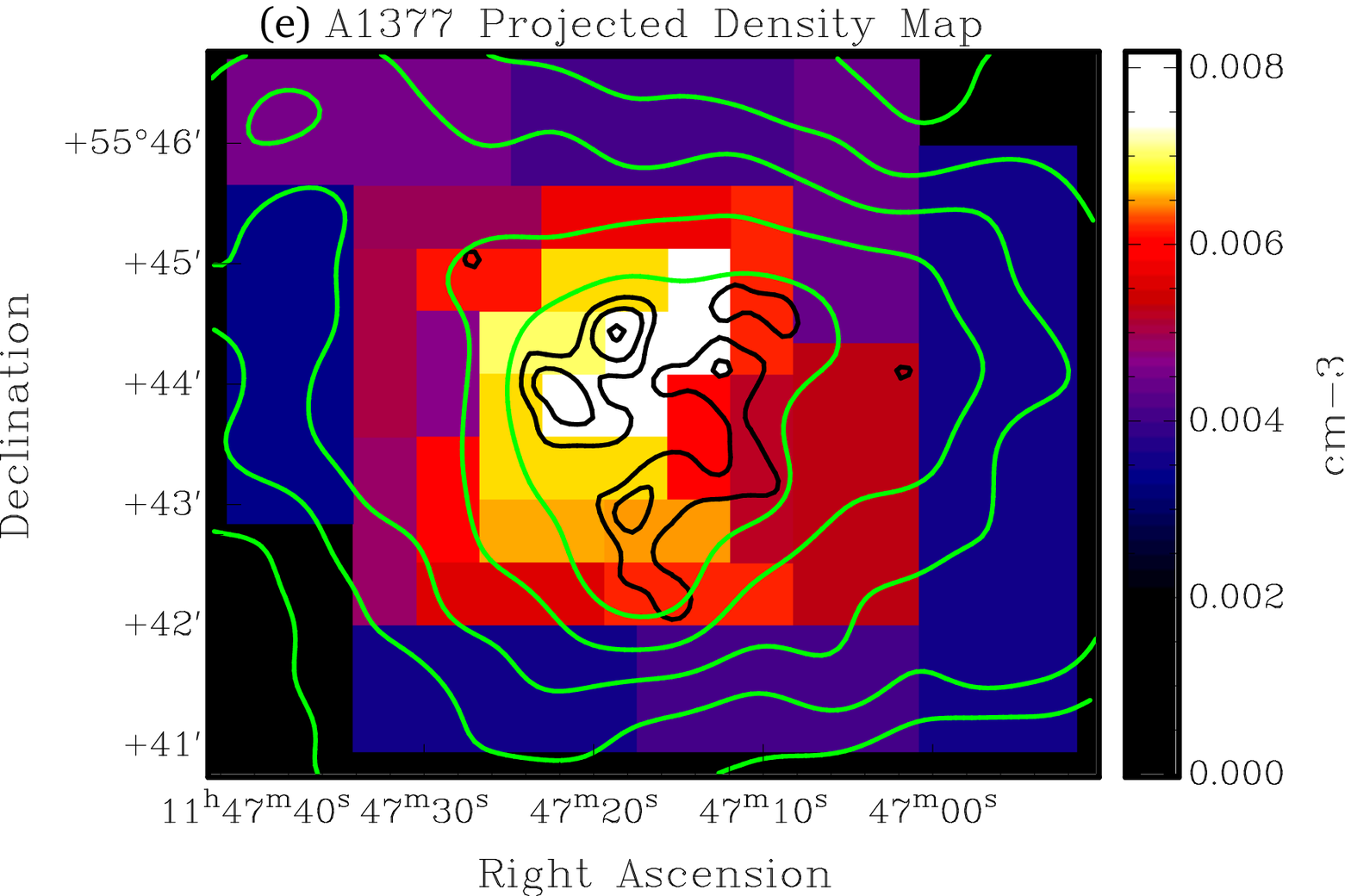}\\
\includegraphics[height=4.0in]{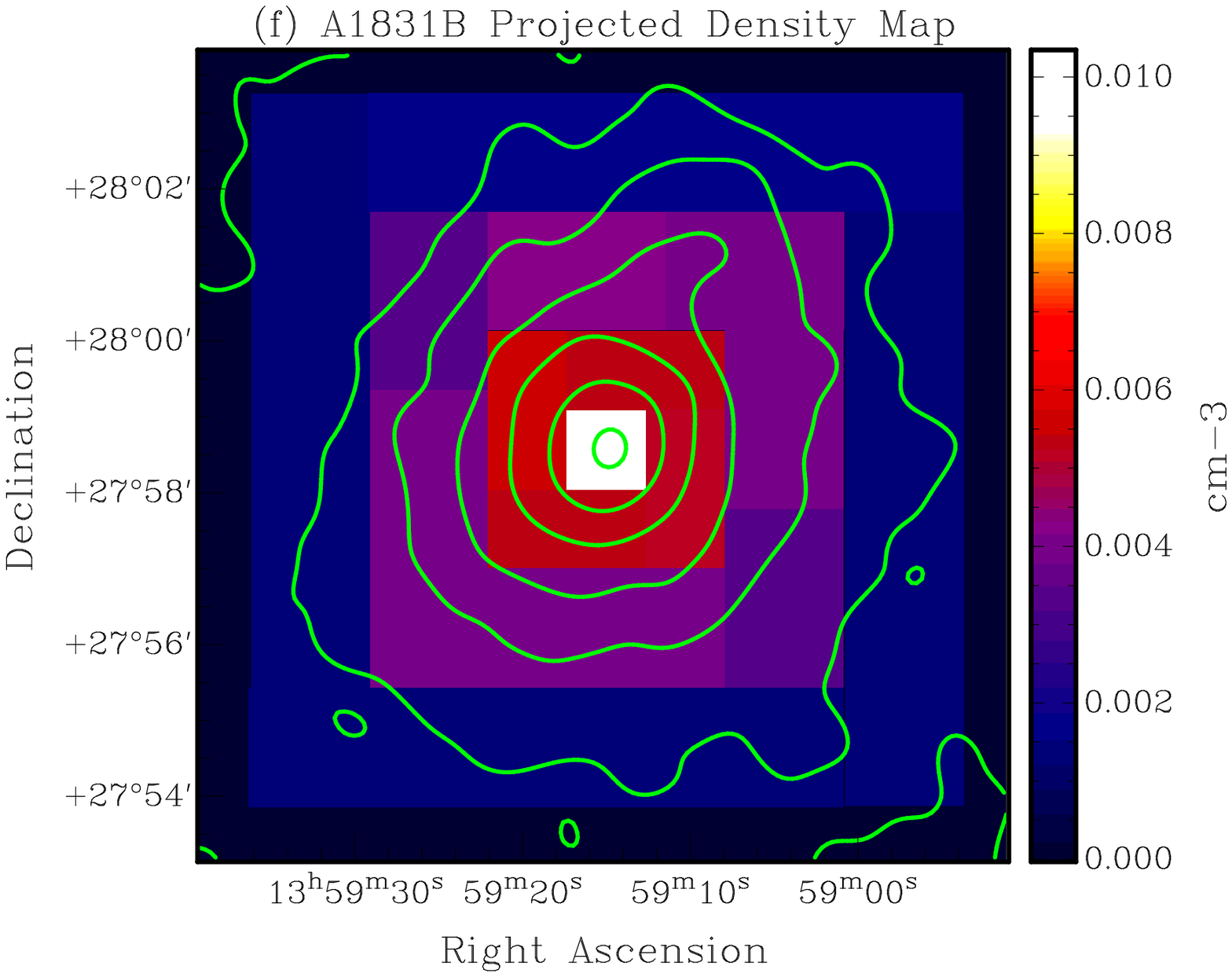}\\
\includegraphics[height=4.0in]{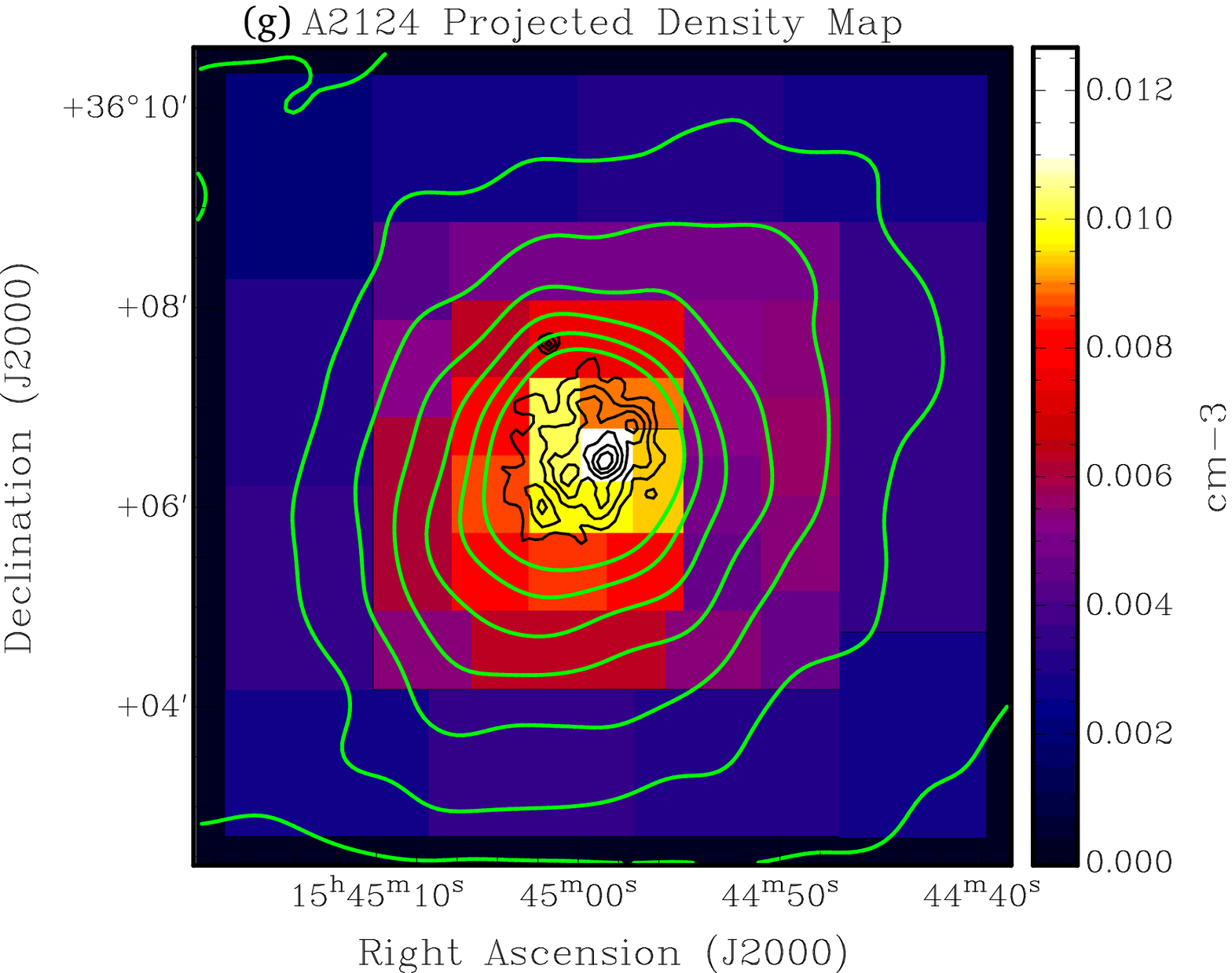}\\
\includegraphics[height=4.0in]{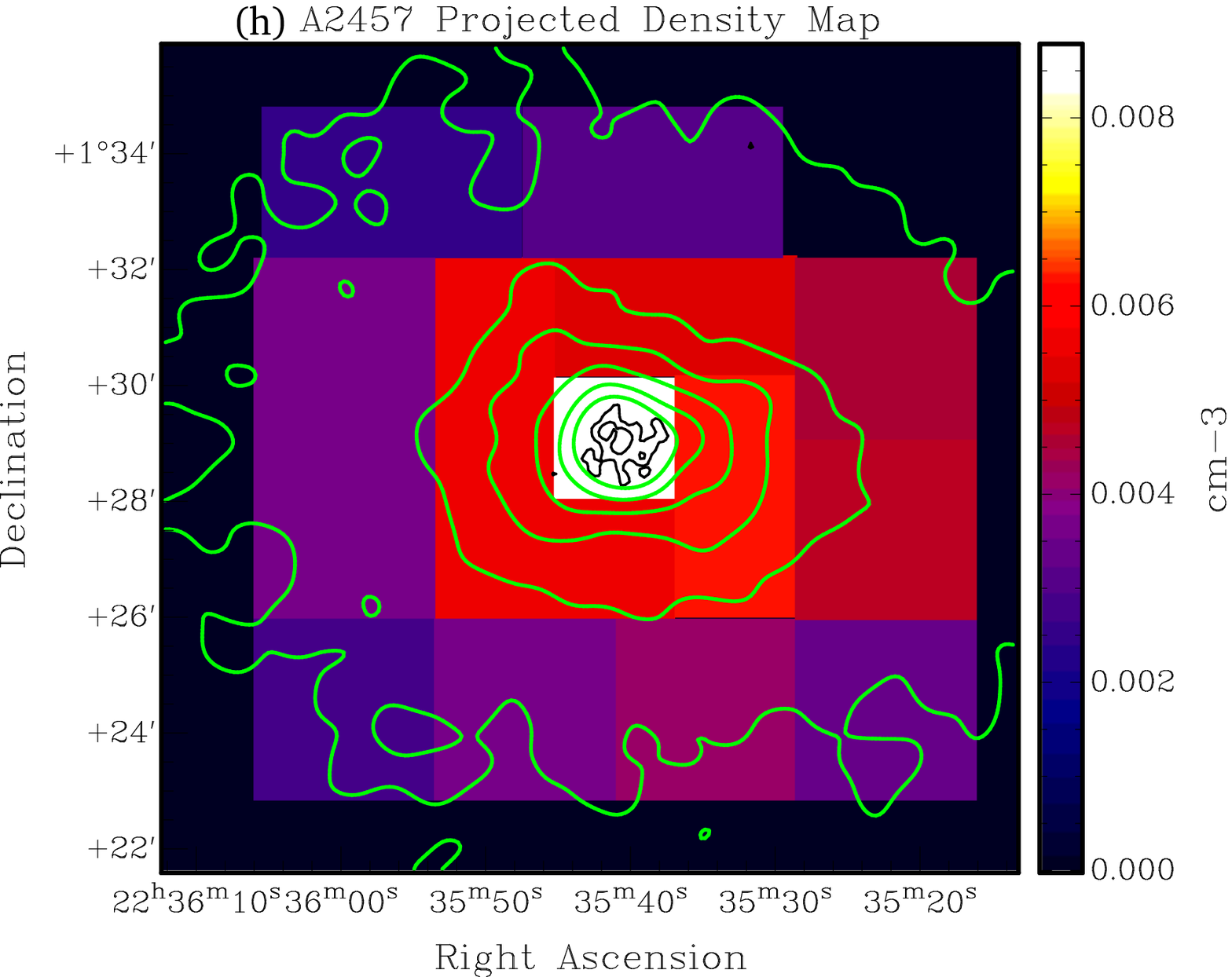}\\
\includegraphics[height=4.0in]{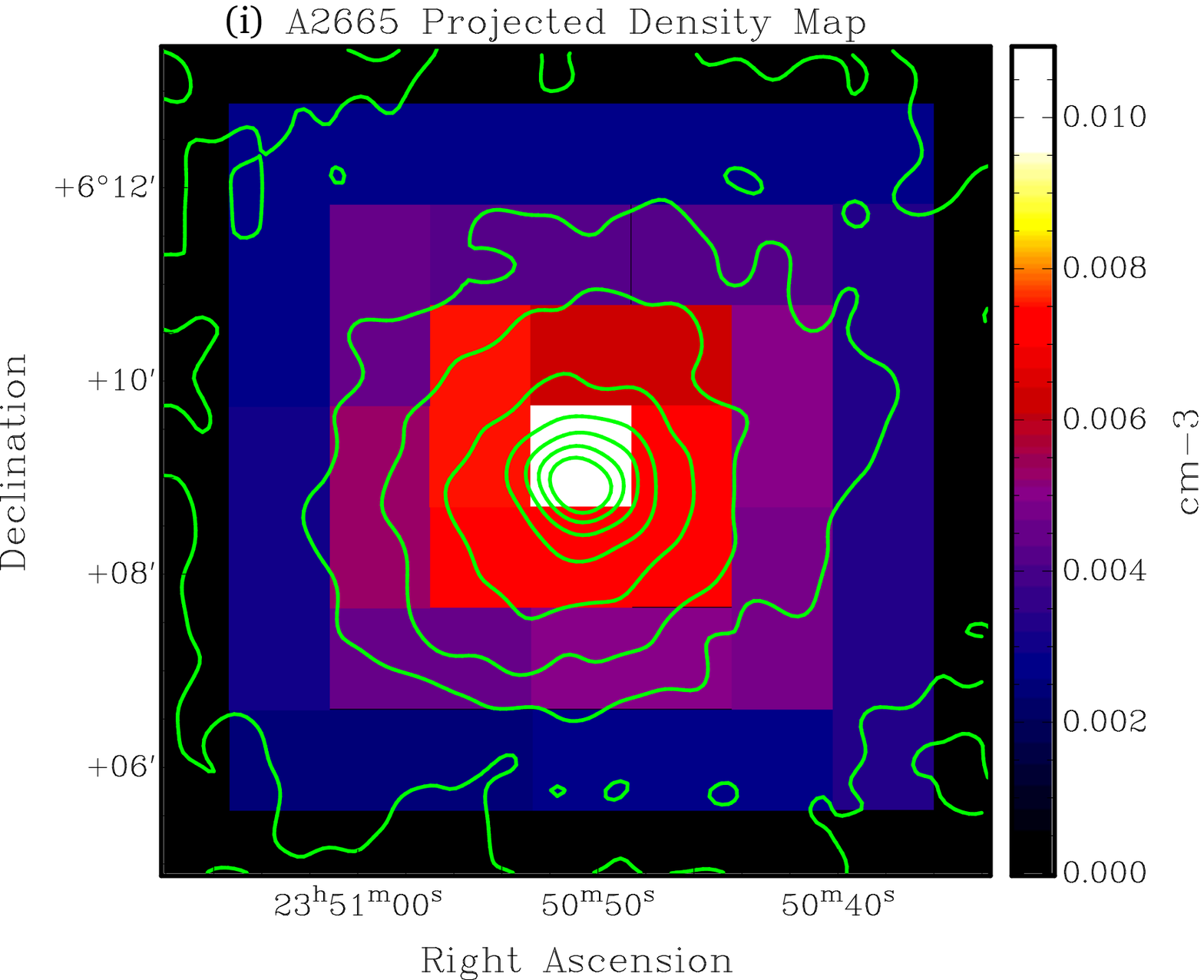}\\
\includegraphics[height=4.0in]{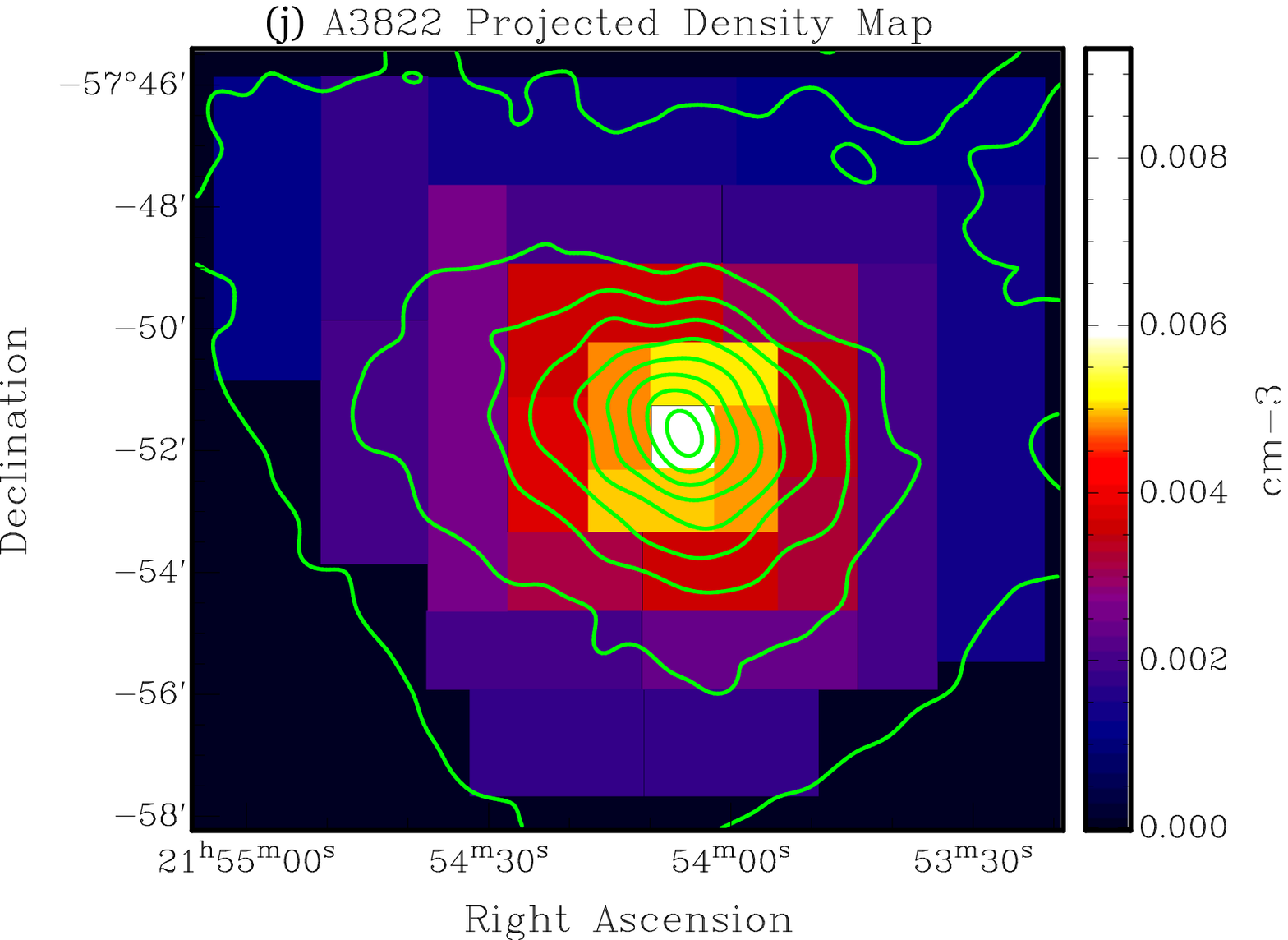}\\
 \end{longtable}%
   \captionof{figure}{(a)-(j): Projected electron number density (n$_{e}$) maps of the clusters A193, A376, A539, A970, A1377, A1831B, 
A2124, A2457, A2665 and A3822, respectively, obtained from the spectral analyses of box shaped regions (see 
\S\ref{sec:box_thermodynamic_maps}). The overlaid intensity contours are from the \textit{Chandra} images as explained in 
the caption of Figure \ref{fig:2D_temp_map}. \label{fig:2D_dens_map}}%
  \addtocounter{table}{-1}%
\end{center}

\clearpage

\begin{center}
\begin{longtable}{c}
\includegraphics[height=3.9in]{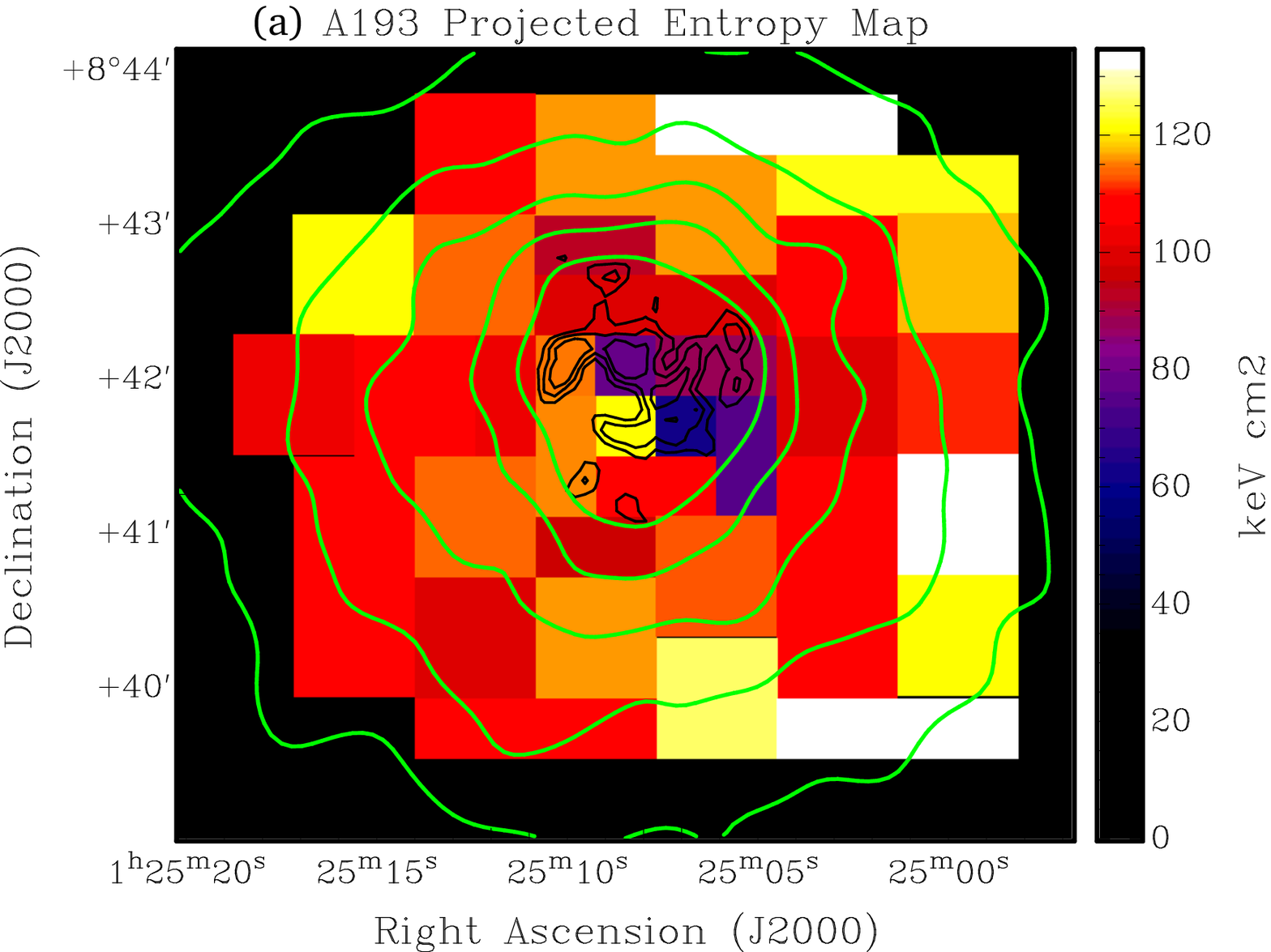}\\
\includegraphics[height=4.0in]{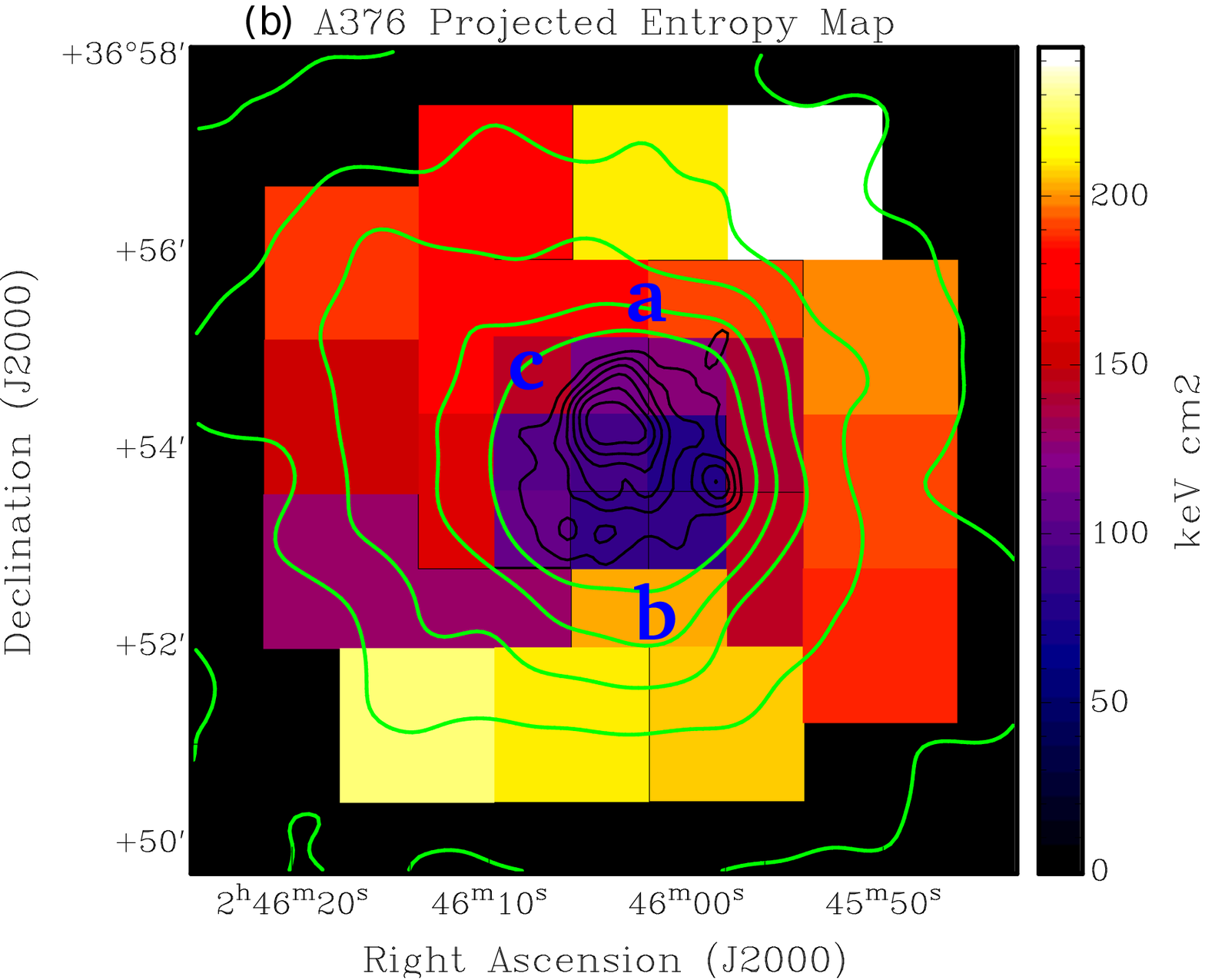}\\
\includegraphics[height=4.0in]{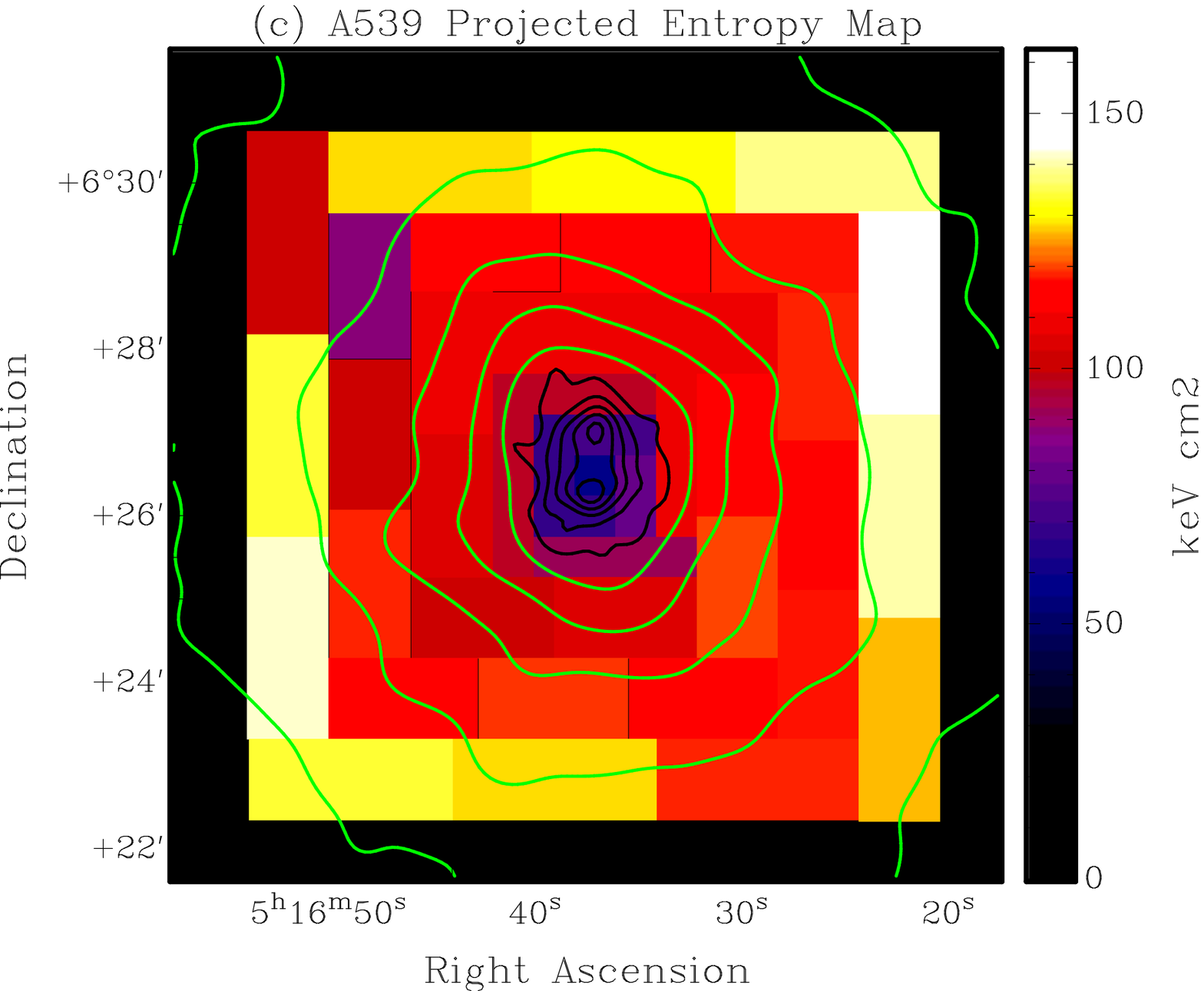}\\
\includegraphics[height=3.7in]{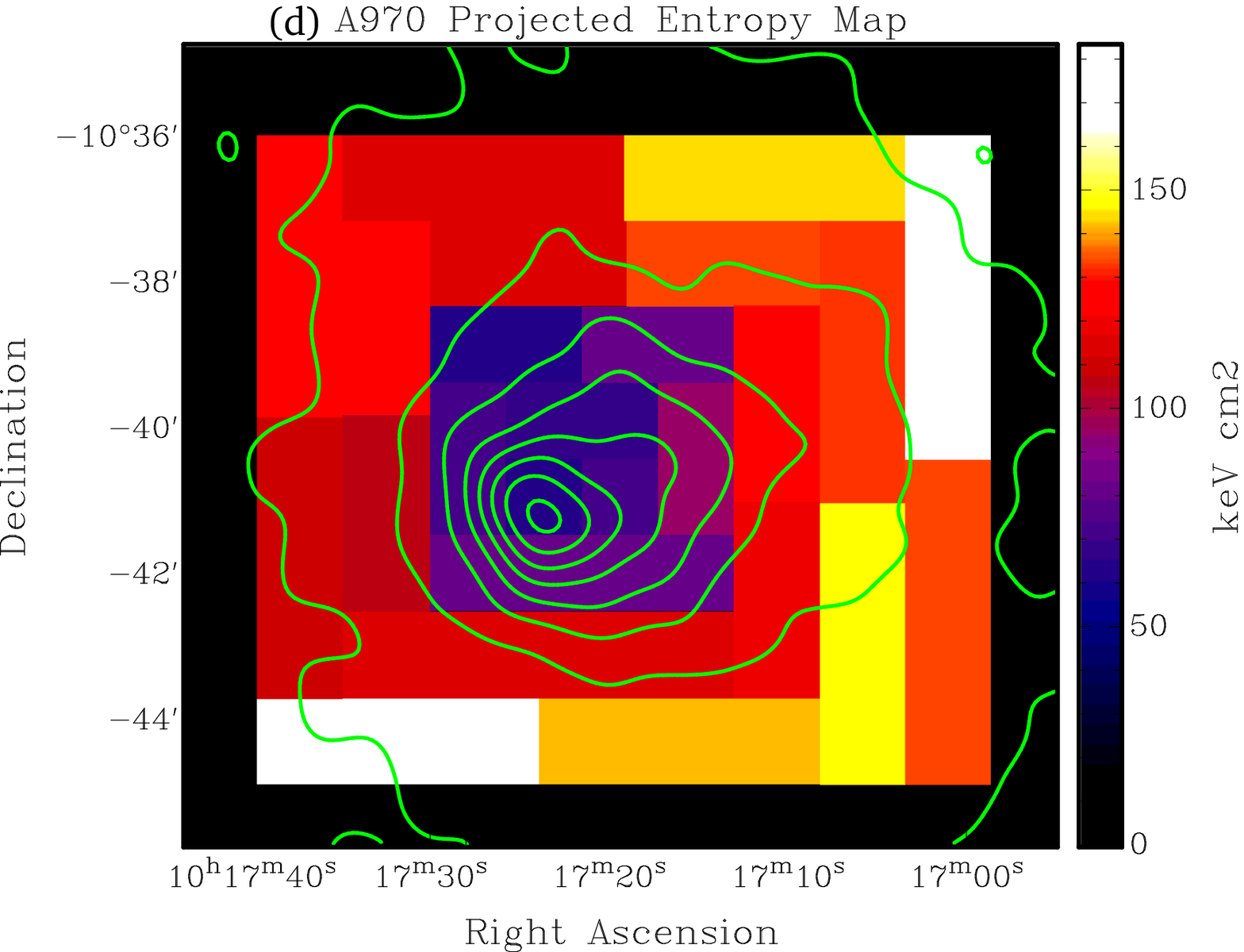}\\
\includegraphics[height=3.7in]{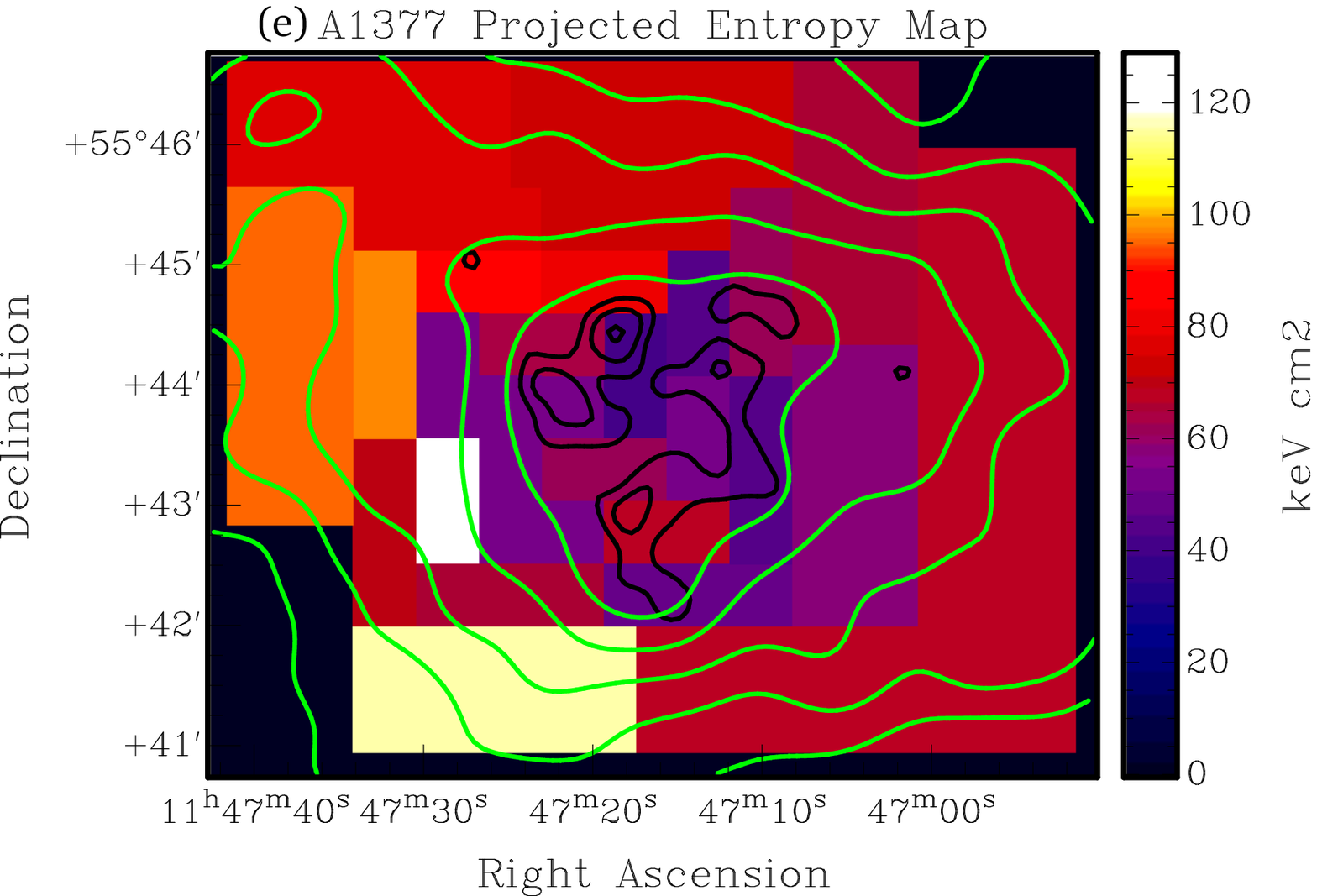}\\
\includegraphics[height=4.0in]{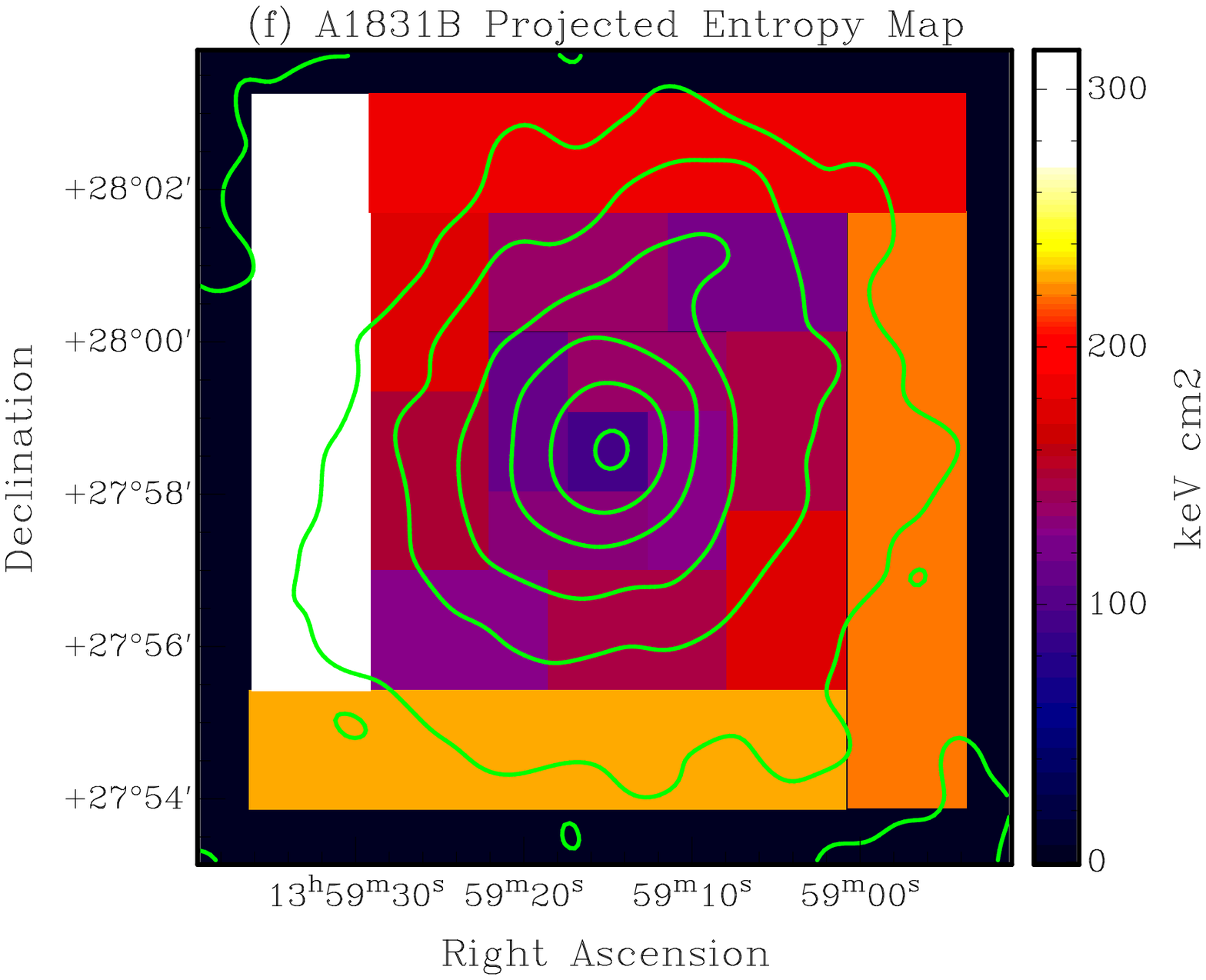}\\
\includegraphics[height=4.0in]{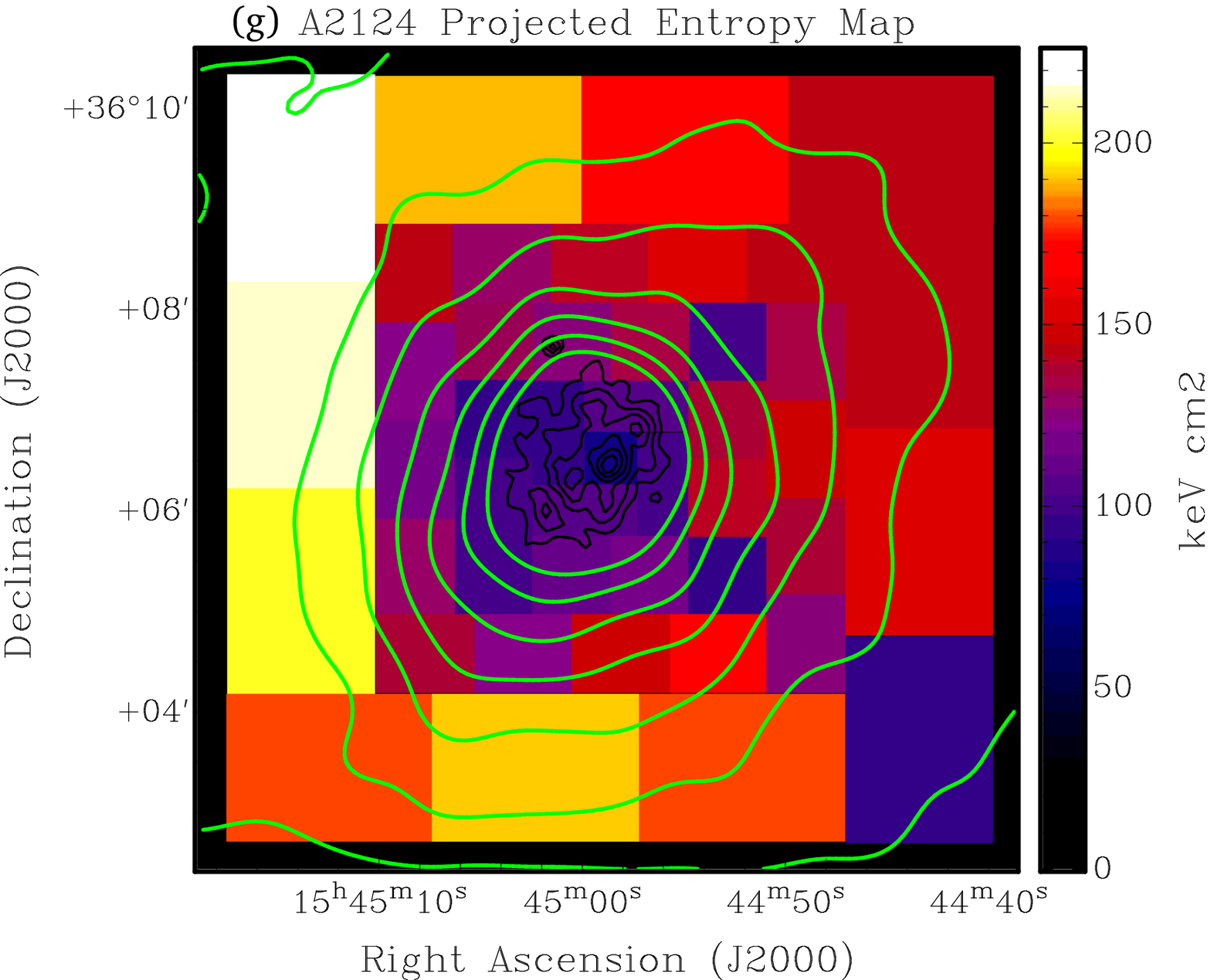}\\
\includegraphics[height=4.0in]{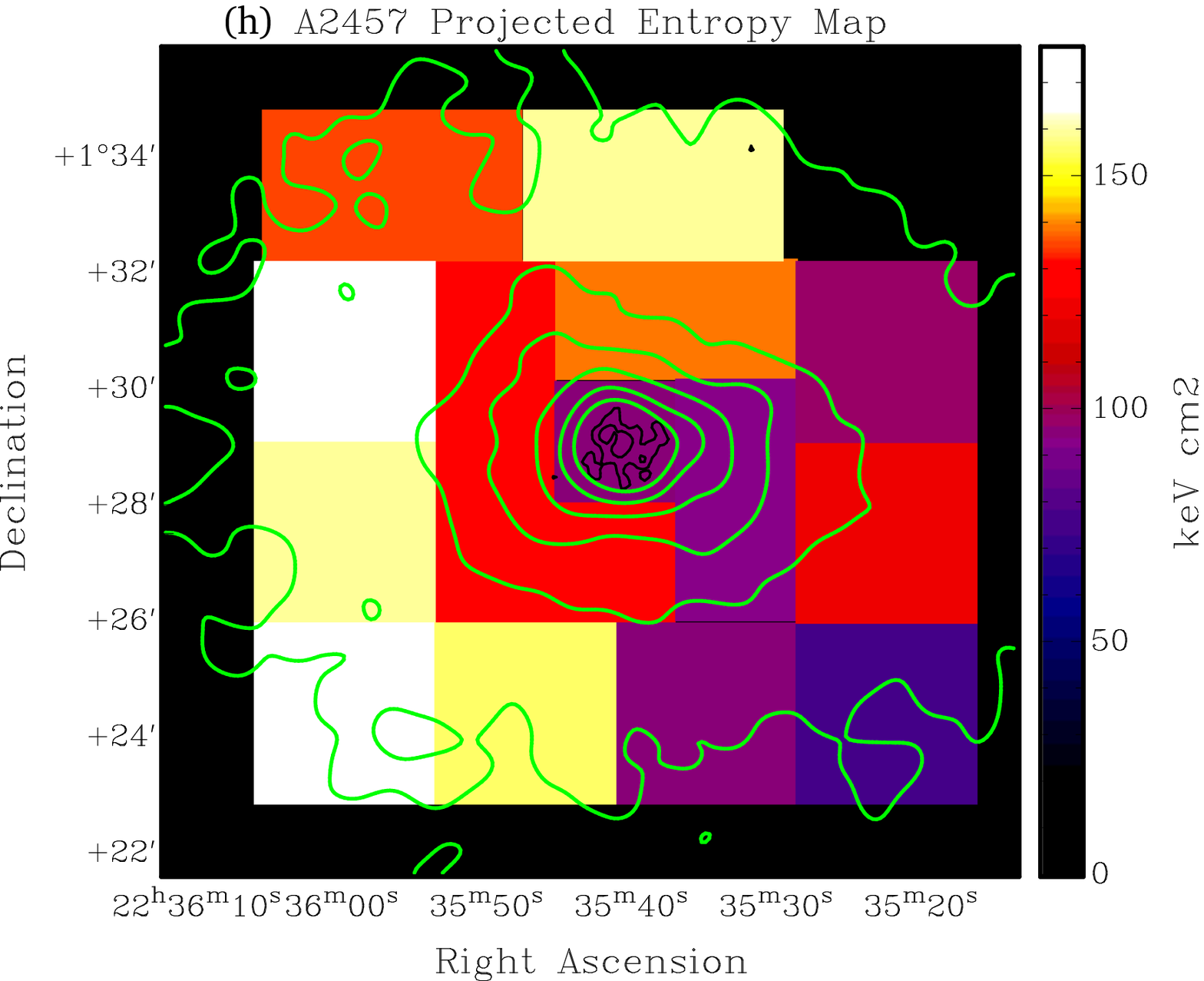}\\
\includegraphics[height=4.0in]{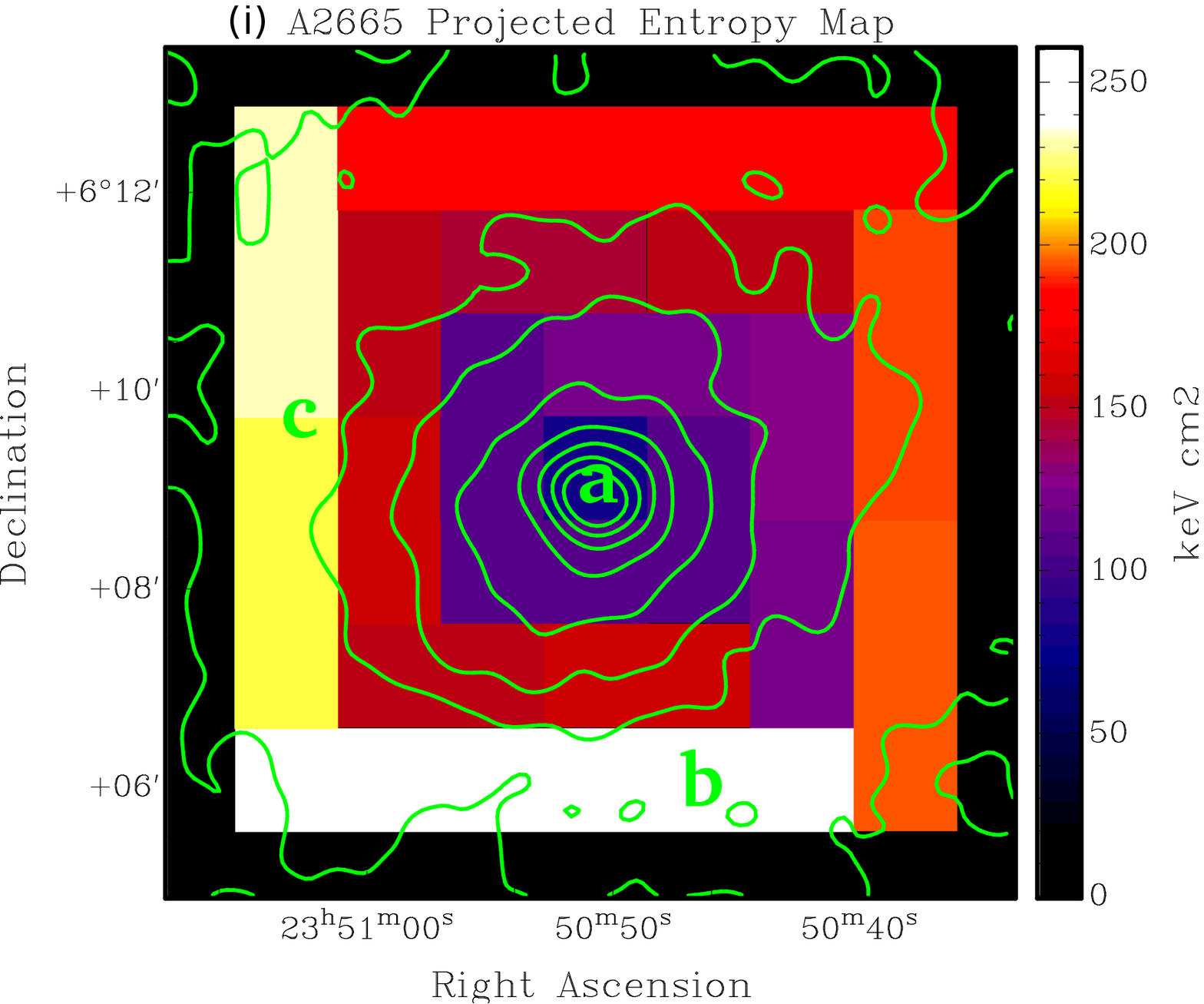}\\
\includegraphics[height=4.0in]{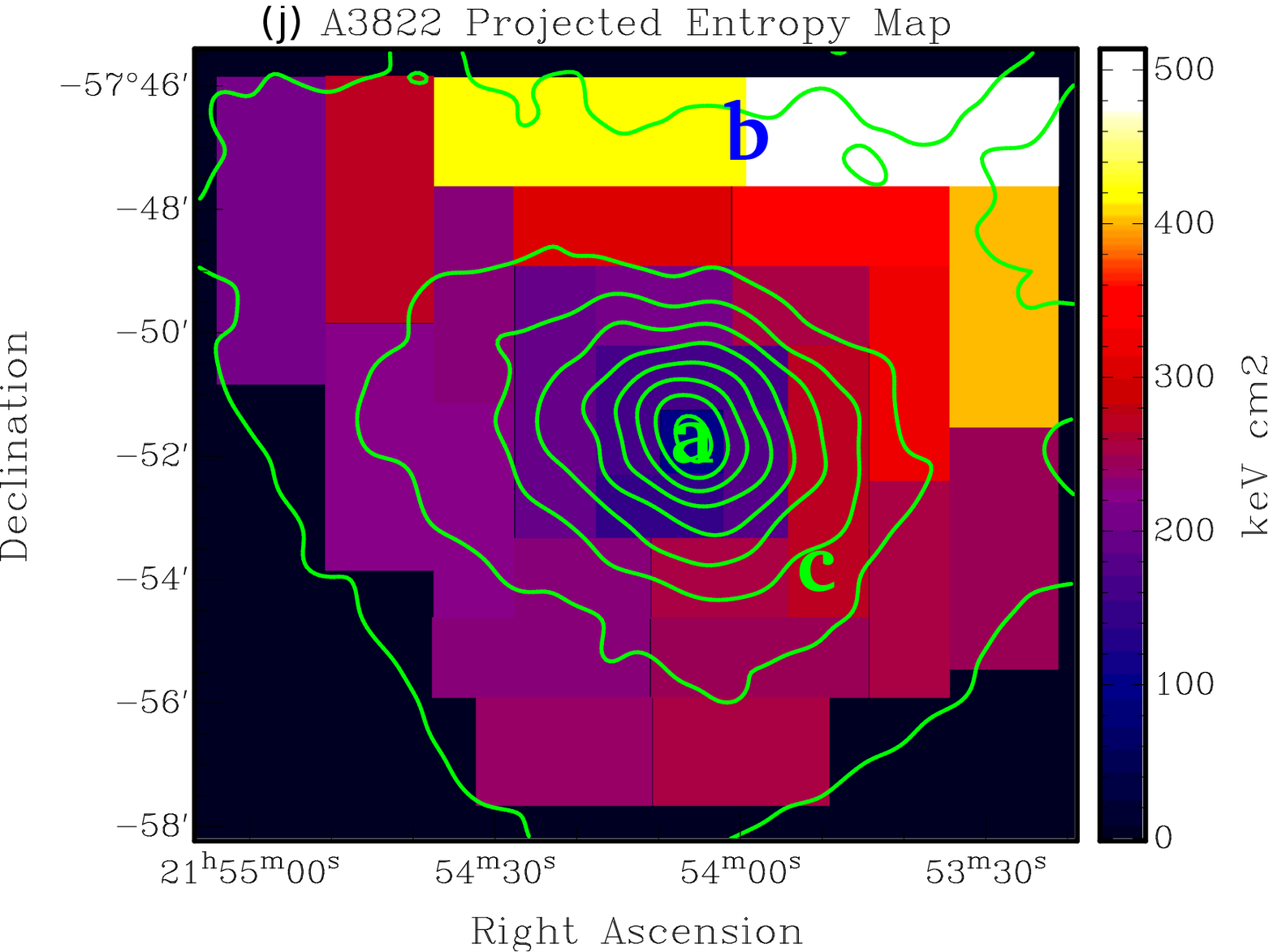}\\
 \end{longtable}%
   \captionof{figure}{(a)-(j): Projected entropy (S) maps of the clusters A193, A376, A539, A970, A1377, A1831B, A2124, A2457, A2665, and A3822, 
respectively, obtained from the spectral analyses of box shaped regions (details of spectral analyses are 
given in \S\ref{sec:box_thermodynamic_maps}). The overlaid intensity contours are from the \textit{Chandra} images as explained in 
the caption of Figure \ref{fig:2D_temp_map}. \label{fig:2D_entr_map}}%
  \addtocounter{table}{-1}%
\end{center}

\clearpage

\begin{center}
\begin{longtable}{c}
\includegraphics[height=3.9in]{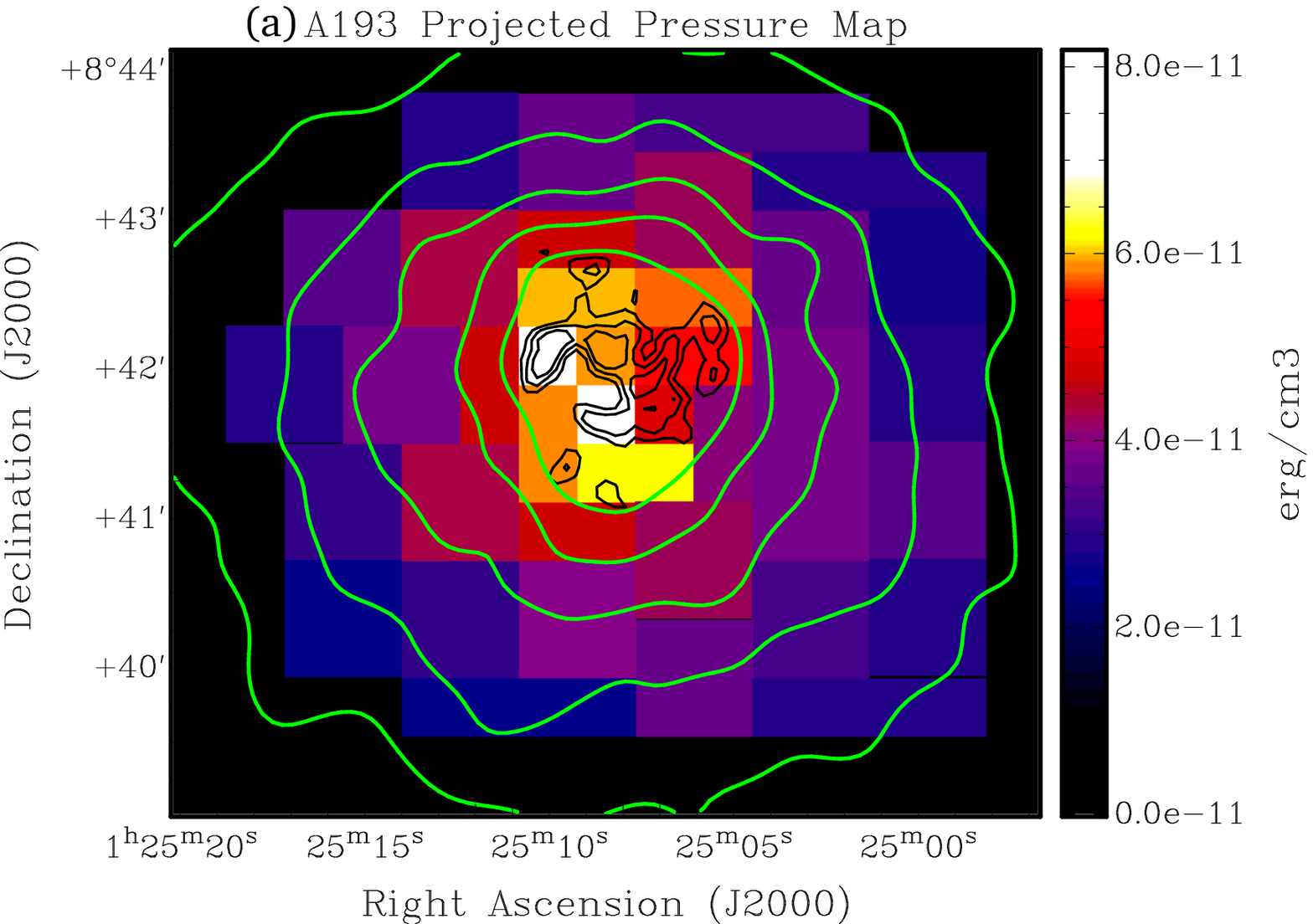}\\
\includegraphics[height=4.0in]{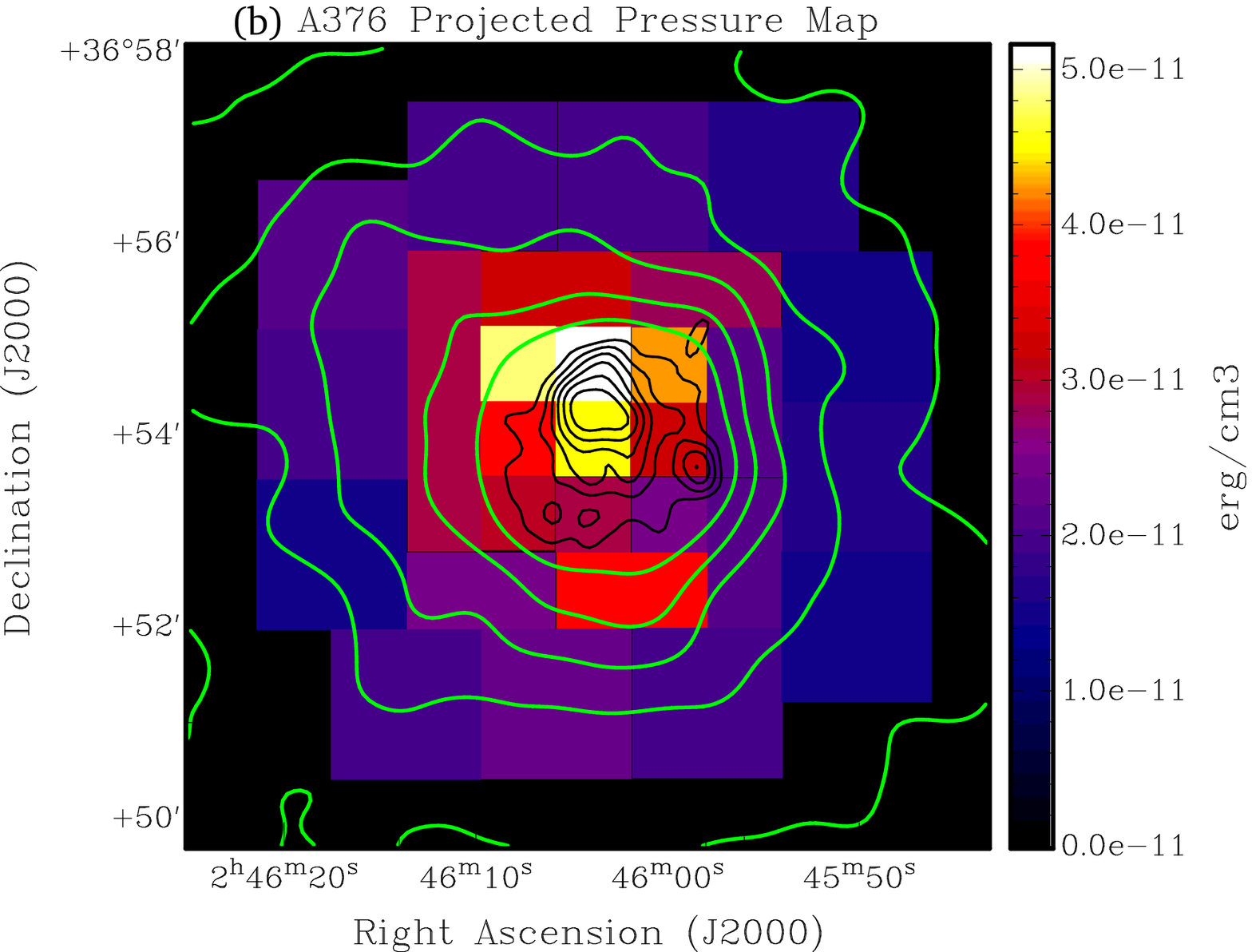}\\
\includegraphics[height=4.0in]{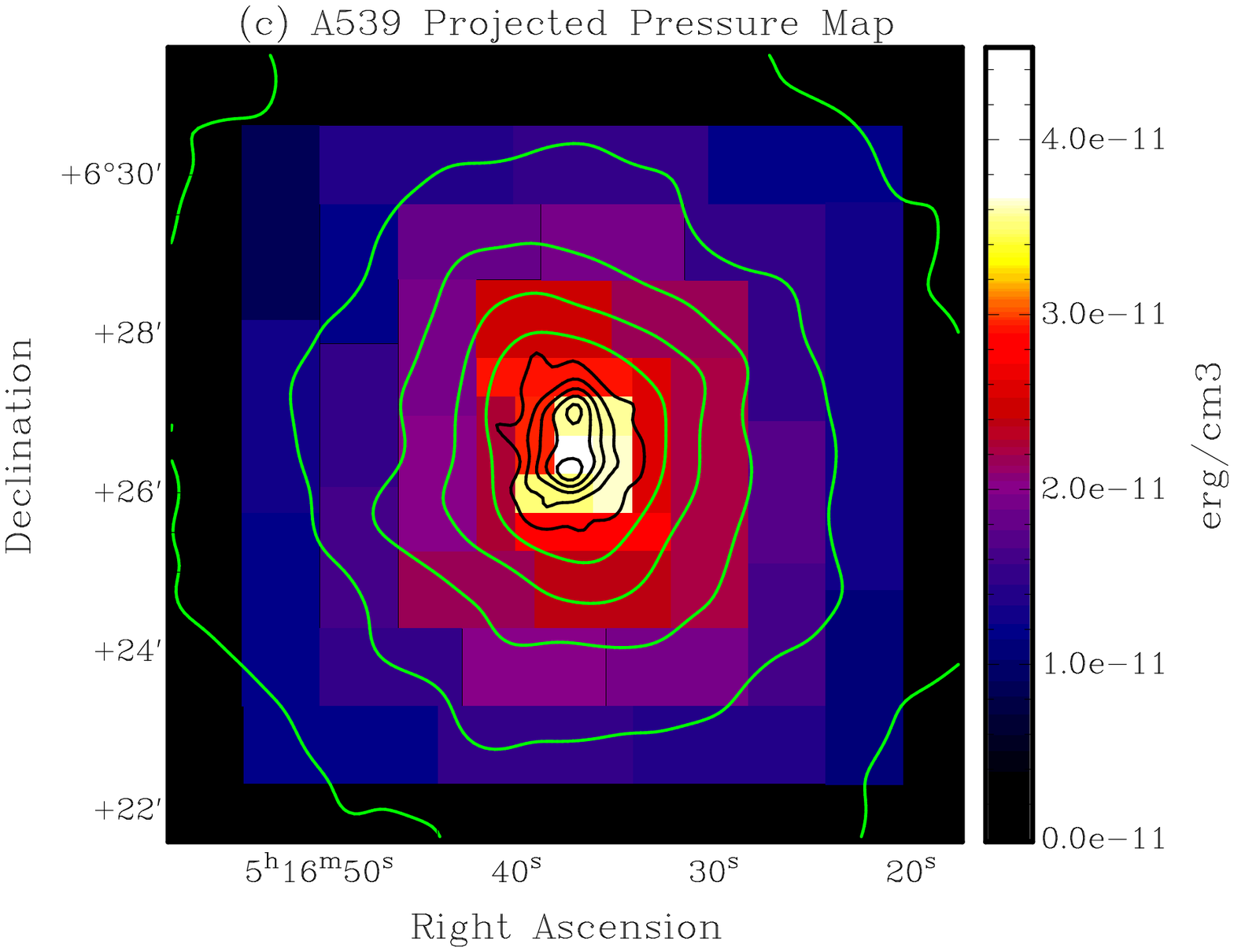}\\
\includegraphics[height=3.7in]{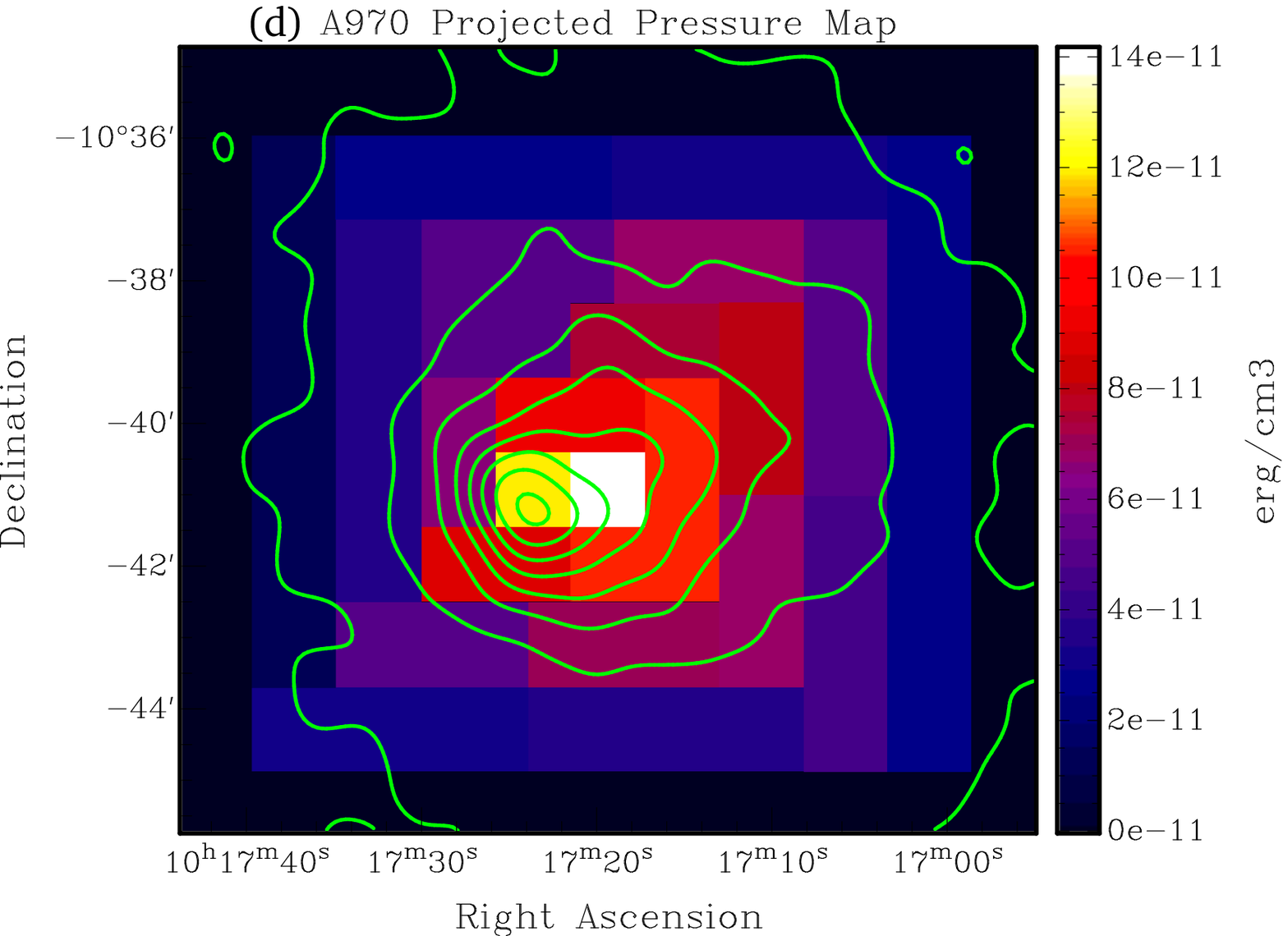}\\
\includegraphics[height=3.7in]{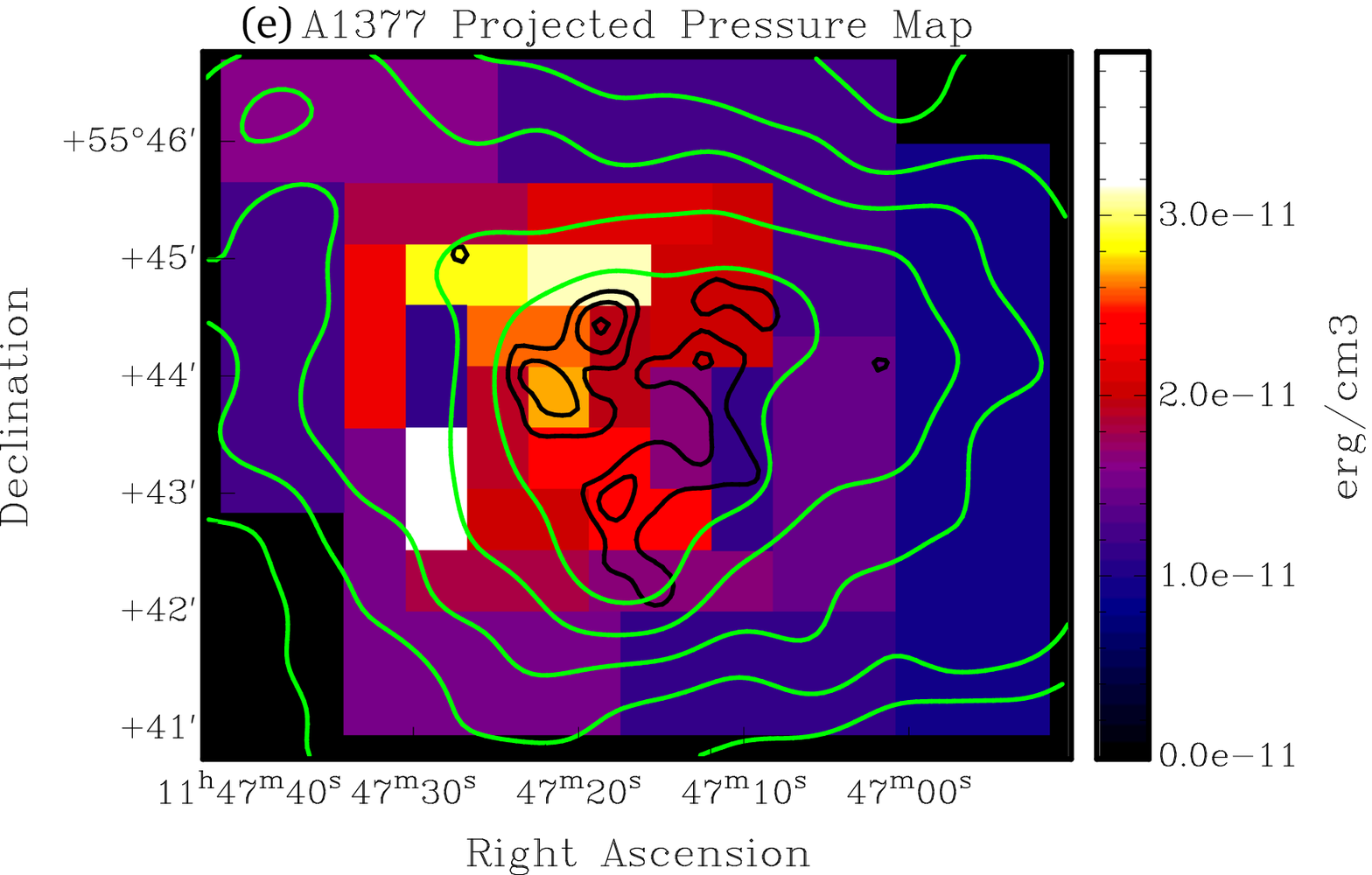}\\
\includegraphics[height=4.0in]{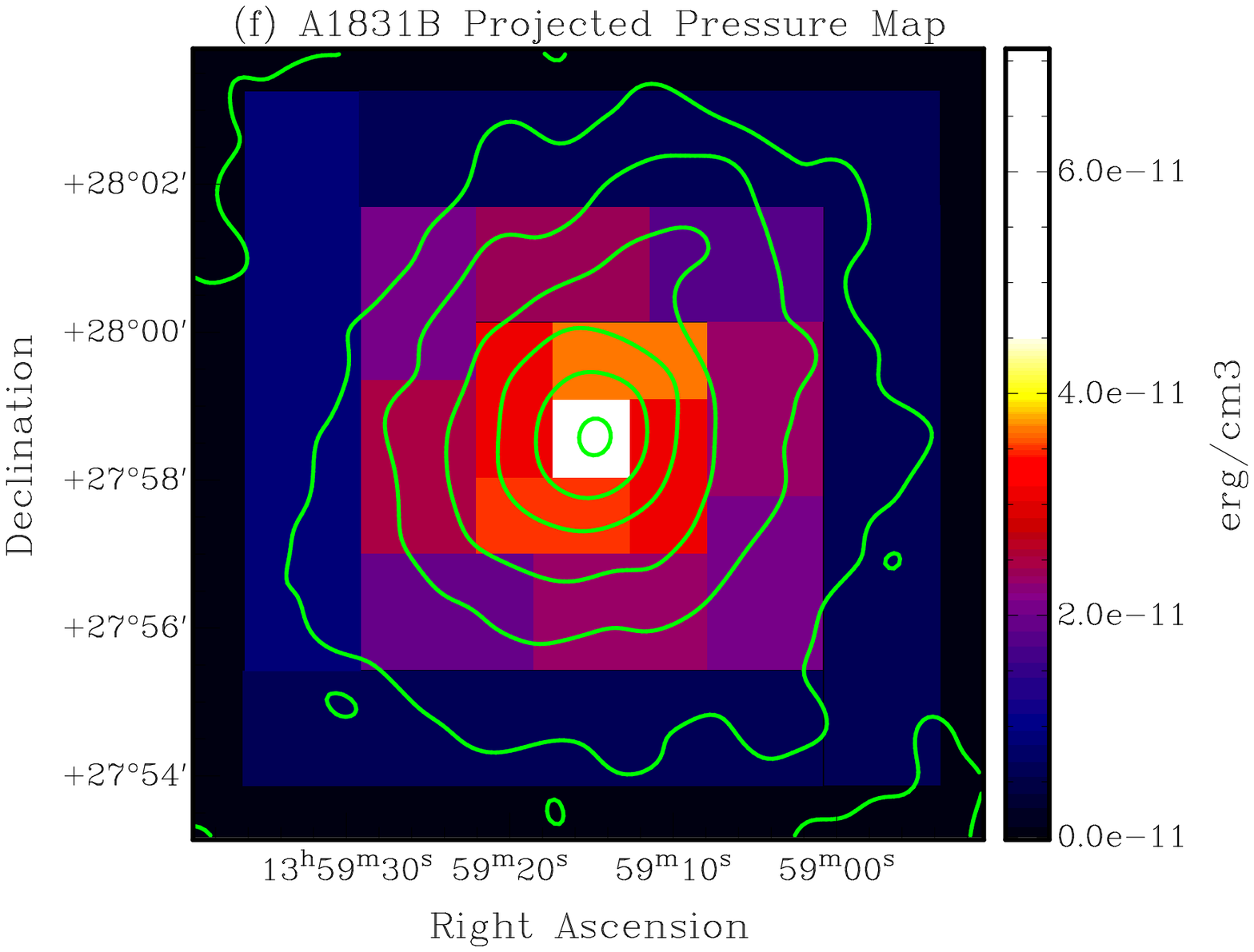}\\
\includegraphics[height=4.0in]{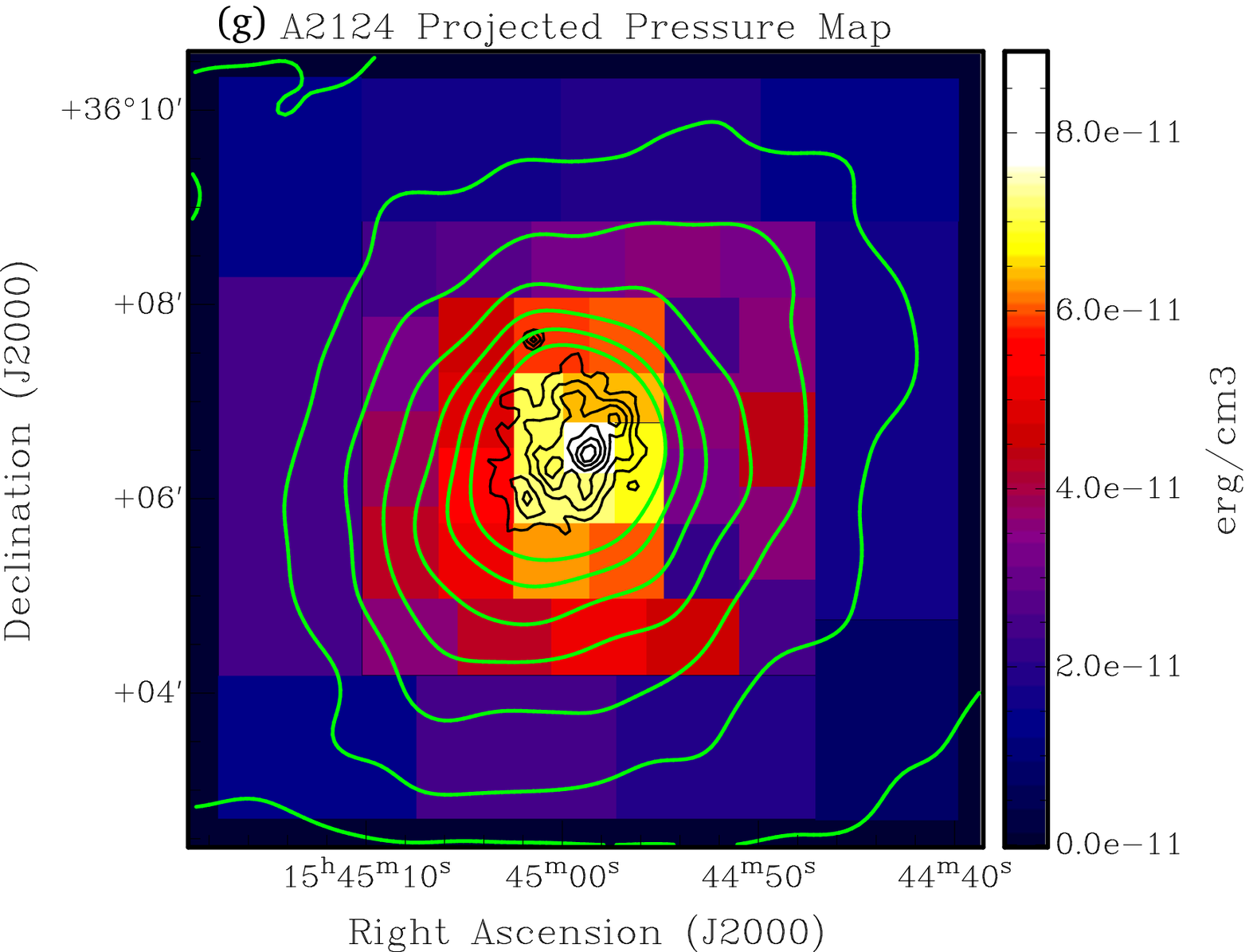}\\
\includegraphics[height=4.0in]{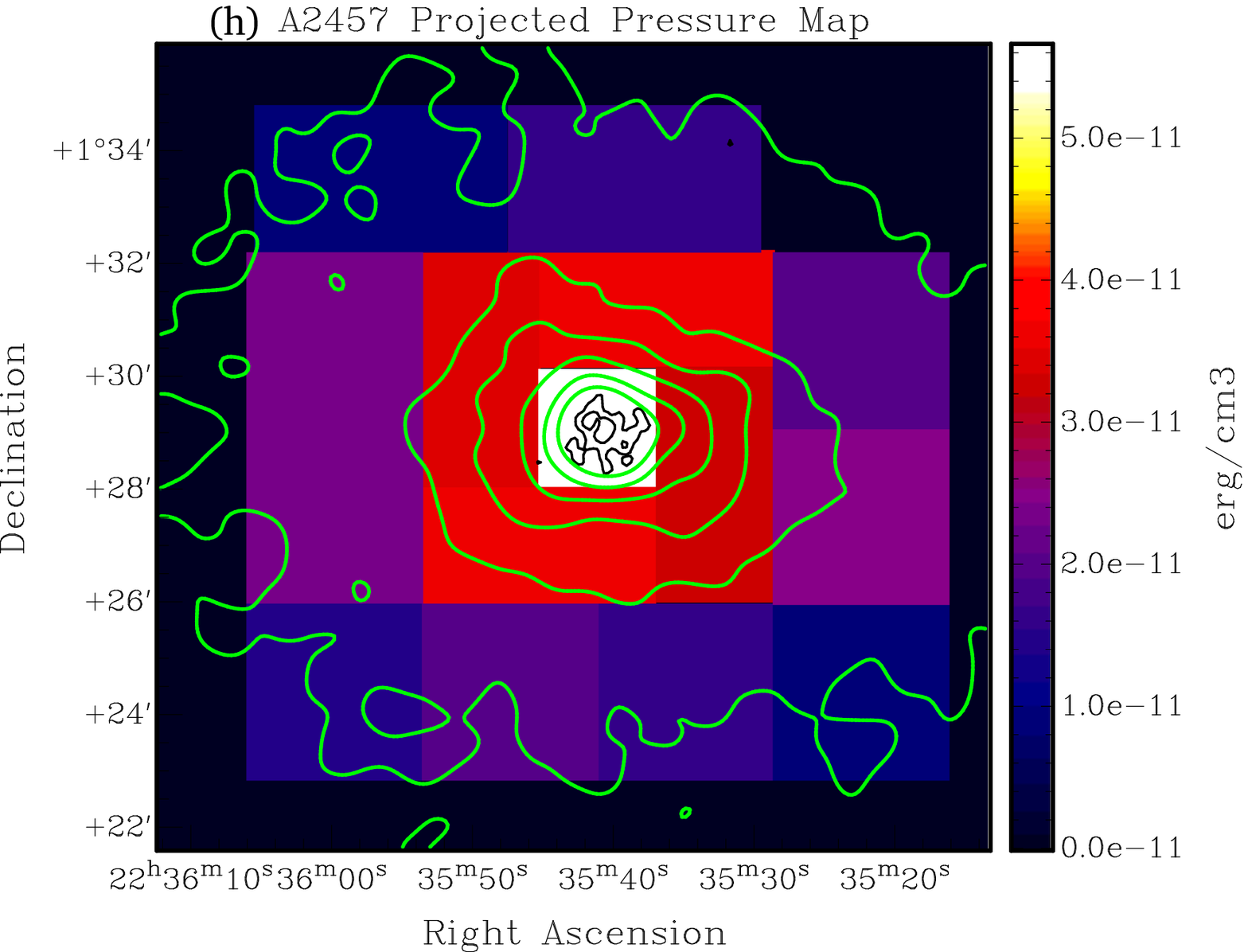}\\
\includegraphics[height=4.0in]{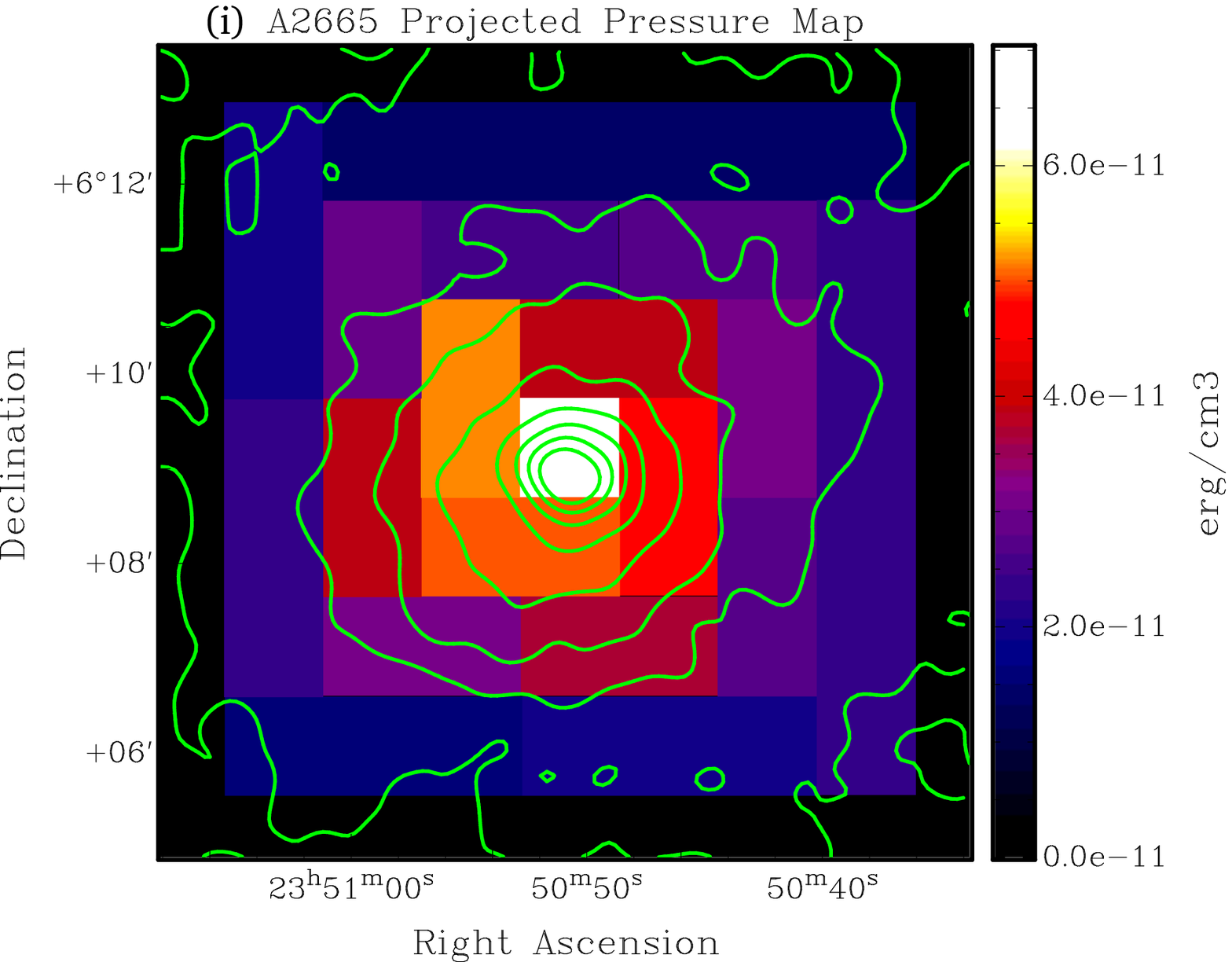}\\
\includegraphics[height=4.0in]{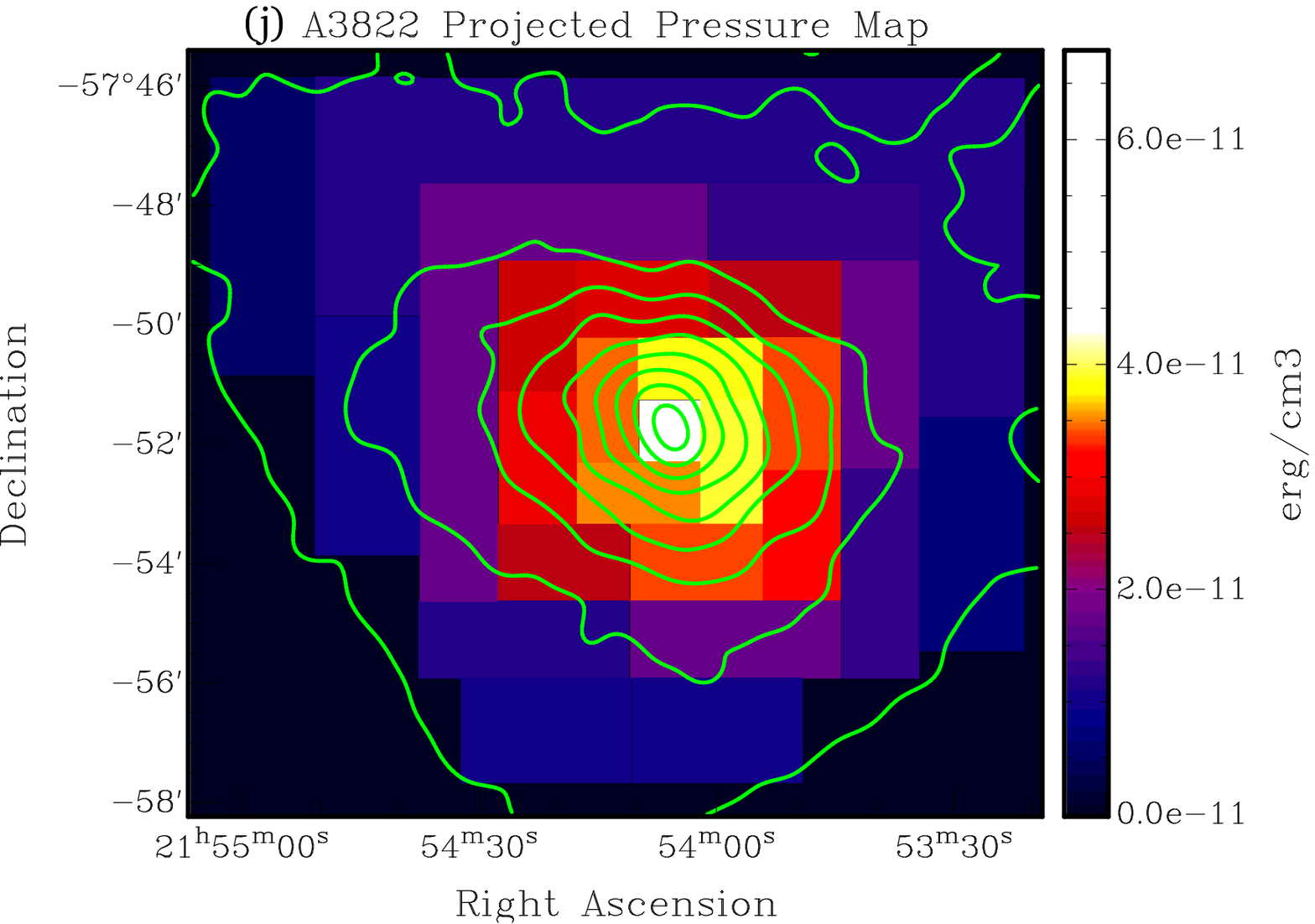}\\
 \end{longtable}%
   \captionof{figure}{(a)-(j): Projected pressure (P) maps of the clusters A193, A376, A539, A970, A1377, A1831B, A2124, A2457, A2665, 
and A3822, respectively, obtained from the spectral analysis of box shaped regions (see \S\ref{sec:box_thermodynamic_maps}). 
The overlaid intensity contours are from the \textit{Chandra} images as explained in the caption of Figure \ref{fig:2D_temp_map}. 
\label{fig:2D_pres_map}}%
  \addtocounter{table}{-1}%
\end{center}
  
 \clearpage

\end{document}